# Fast current-induced skyrmion motion in synthetic antiferromagnets

**Authors**: Van Tuong Pham[1,3,‡], Naveen Sisodia[1,5,†], Ilaria Di Manici[1,†], Joseba Urrestarazu-Larrañaga[1,†], Kaushik Bairagi[1], Johan Pelloux-Prayer[1], Rodrigo Guedas[1,6], Liliana D. Buda-Prejbeanu[1], Stéphane Auffret[1], Andrea Locatelli[2], Tevfik Onur Mentes[2], Stefania Pizzini[3], Pawan Kumar[4], Aurore Finco[4], Vincent Jacques[4], Gilles Gaudin[1] and Olivier Boulle[1,*]

[1] Univ. Grenoble Alpes, CNRS, CEA, SPINTEC; 38054 Grenoble, France
[2] Elettra-Sincrotrone Trieste S.C.p.A.; 34149 Basovizza, Trieste, Italy
[3] Univ. Grenoble Alpes, CNRS, Institut Néel; 38042 Grenoble, France
[4] Laboratoire Charles Coulomb, CNRS, Université de Montpellier; 34095 Montpellier, France
[5] Department of Physics, Indian Institute of Technology Gandhinagar, Gandhinagar-382355, Gujarat, India
[6] Instituto de Sistemas Optoelectrónicos y Microtecnología (ISOM), Universidad Politécnica de Madrid, Avda. Complutense 30, 28040 Madrid, Spain
* Corresponding author. e-mail: olivier.boulle@cea.fr
† These authors contributed equally to this work.
‡ Now at IMEC, Leuven, Belgium

**Abstract**
Magnetic skyrmions are topological magnetic textures with great promise as nanoscale bits of information in memory and logic devices. Although room temperature ferromagnetic skyrmions and their current-induced manipulation have been demonstrated, their velocity has been limited to about 100 m/s; in addition, their dynamics are perturbed by the skyrmion Hall effect, a motion transverse to the current direction caused by the skyrmion topological charge. Here we show that skyrmions in compensated synthetic antiferromagnets can be moved by current along the current direction at a velocity up to 900 m/s. This is explained by the cancellation of the net topological charge leading to a vanishing skyrmion Hall effect. Our results open an important path toward the realization of logic and memory devices based on the fast manipulation of skyrmions in tracks.

**Main Text**

Magnetic skyrmions have attracted a widespread interest in recent years owing to their rich physics at the interface of magnetism and topology as well as promising applications in the field of non-volatile memory and logic. Magnetic skyrmions are topologically non-trivial chiral whirling spin textures (*1*). They have a topological charge S=±1, *i.e* the skyrmion magnetization wraps the spin unit sphere once. Their topological stability, along with their small lateral dimensions, down to the nanometer scale, grant them particle-like properties. Skyrmions can also be moved efficiently by an electrical current (*2–6*). These unique properties can be exploited to store and compute information at the nanoscale in devices where skyrmions are the information carriers. The observation of magnetic skyrmions at room temperature in sputtered ultrathin films as well as their current-induced motion have lifted important roadblocks towards the practical realization of such devices (*3, 4, 6–9*).

However, ferromagnetic (FM) skyrmions suffer from several limitations in view of applications. Firstly, their dynamics in tracks are perturbed by the skyrmion Hall effect (*3*), *i.e.* a motion transverse to the driving force [see Fig.1A, (1) and (2)]. This effect results from a Magnus force $G\mathbf{z} \times \mathbf{v}$ acting on the moving skyrmion, a direct consequence of its topology (here G is the amplitude of the skyrmion gyrovector proportional to the skyrmion topological charge S and the layer magnetic moment, **v** is the skyrmion velocity vector, **z** is the unit vector normal to the film plane). This is an important issue for devices because it can cause skyrmions to move toward the track edges where they can be annihilated. Secondly, the current-induced skyrmion velocity barely exceeds 100 m/s (*3–6*, *10*, *11*). According to the Thiele equation (*3*, *12*, *13*), the skyrmion velocity scales as $v \sim \frac{1}{\sqrt{\alpha^2 + \alpha_T^2}} \sim \cos\theta_{SkHE}$ and the skyrmion Hall angle $\theta_{SkHE}$ as $\tan\theta_{SkHE} = \frac{v_y}{v_x} = \frac{\alpha_T}{\alpha}$, with $\alpha$ the magnetic damping and $\alpha_T = G/D$, where D is the dissipation constant. Thus, the skyrmion Hall effect limits the skyrmion velocity, via the introduction of an additional topological magnetic damping $\alpha_T$ (*14*). These limitations could be circumvented by considering skyrmions in antiferromagnets (*15–18*, *18–24*). In these materials, antiferromagnetically coupled skyrmions reside in different sublattices have opposite core polarities and hence opposite topological charges. Thus, the coupled skyrmions experience opposite Magnus forces resulting in a cancellation of the skyrmion Hall effect (see Fig. 1B). This leads to an increased velocity with predicted current induced motions over 1000 m/s (*15–18*, *18–24*). Although the stabilization of antiferromagnetic (AF) topological spin textures has been recently shown in bulk $MnSc_2S_4$(*25*) and $Fe_2O_3$(*26*), in IrMn antiferromagnets using exchange bias(*27*), and in synthetic antiferromagnets (*28*, *29*), demonstrating their current-nduced dynamics remains challenging. However, recent experiments in ferrimagnets revealed the impact of the angular momentum compensation on the skyrmion dynamics : for instance, a vanishing skyrmion Hall effect was demonstrated for elongated pinned bubbles at the angular momentum compensation of the ferrimagnets (*30*); reduced skyrmion Hall angle (0-25°) and enhanced velocities were also reported in ferrimagnetic alloys and synthetic ferrimagnets (*31–33*).

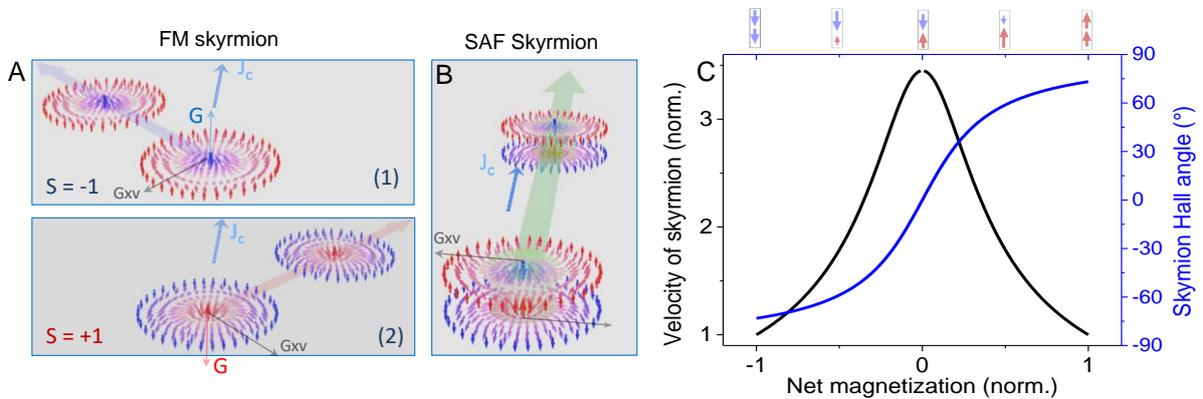

**Fig. 1 Compensation of skyrmion Hall effect and enhanced velocity in SAF skyrmions (A)** Skyrmions with a core magnetization pointing down (resp. up) hold a topological charge of S=-1 (resp. S=+1) and have a gyrovector G along +z (resp –z). When moving, they are deflected to the left (resp. right) by the Magnus force $F_M = G \times v$ where **v** is the velocity of the skyrmion as illustrated in (1) and (2). **(B)** In SAF, AF coupled skyrmions with opposite topological charges experience opposite Magnus forces. The two forces are canceled out and the SAF skyrmion move along the current direction. **(C)** Normalized skyrmion velocity (black line) and skyrmion Hall angle (blue line) as a function of the net magnetization in a SAF calculated using experimentally obtained parameters (*36*).

Compensated synthetic antiferromagnets (SAFs) combine key advantages for studying the dynamics of AF skyrmions and utilizing them in devices. SAFs are composed of ultrathin ferromagnetic layers separated by a non-magnetic spacer and AF-coupled via the Ruderman-Kittel-Kasuya-Yosida (RKKY)-type interlayer interaction(*34*). The stabilization of AF skyrmions at room temperature was recently demonstrated (*28*, *29*) in compensated Pt/FM/Ru based SAFs as well as their nucleation using local current injection or ultrafast laser pulses(*29*). These stacks combine the necessary ingredients to stabilize skyrmions and move them efficiently with current: a large perpendicular magnetic anisotropy (PMA), the Dzyaloshinskii-Moriya interaction (DMI) and spin orbit torques (SOT). Owing to the cancellation of the Magnus force, a significant enhancement of the skyrmion velocity is expected in compensated SAFs compared to ferromagnetic stacks(*14*) (see Fig.1B). In addition, SAFs are deposited by sputtering, a technique widely used in the microelectronics industry, and are compatible with CMOS (complementary metal oxide semi-conductor) technology. They can also be integrated in magnetic tunnel junctions, which can be employed for the electrical readout of the skyrmions (*35*).

Here we show that in optimized compensated synthetic antiferromagnets, skyrmions can be driven by spin orbit torque at velocities up to 900 m/s along the current direction, *i.e.* without skyrmion Hall effect. This is explained by the vanishing net topological charge in the SAF. Our results are substantiated by analytical and micromagnetic simulations using experimentally derived material parameters.

**AF skyrmions in SAF at room temperature**

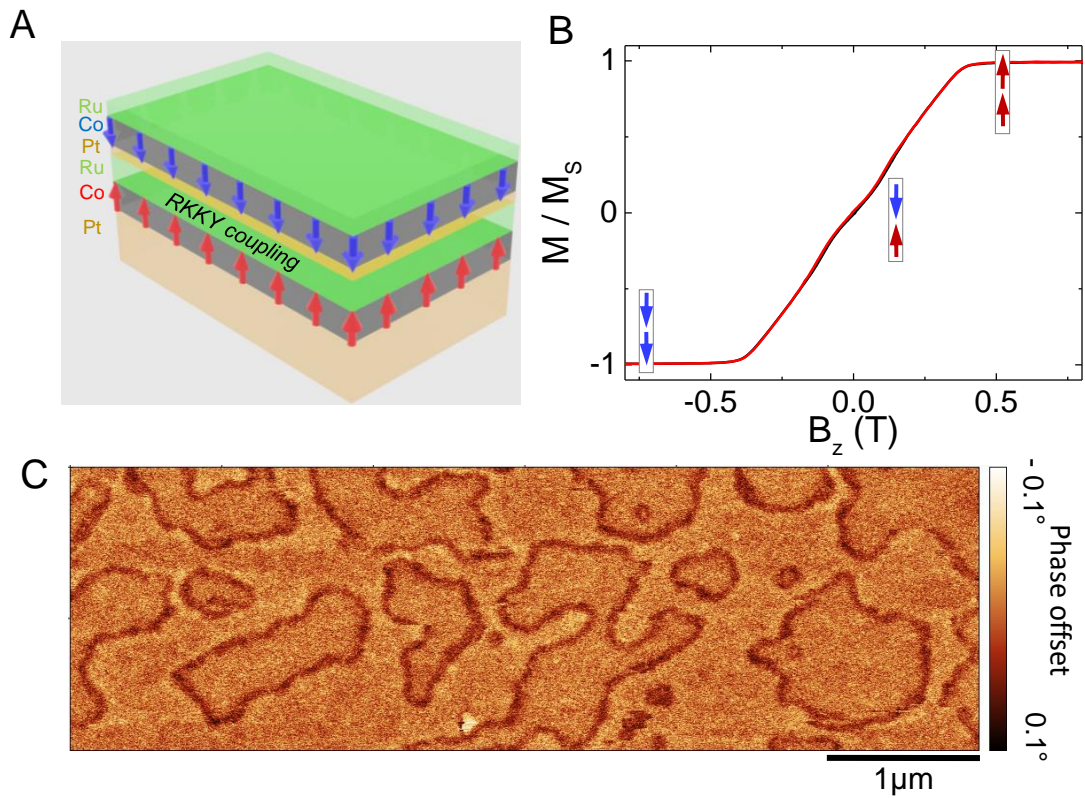

**Fig. 2 Skyrmions in optimized compensated SAF (A)** Composition of the SAF stack (see text for thicknesses). **(B)** Normalized magnetic moment as the function of the out-of-plane external magnetic field measured by vibrating sample magnetometry. Black and red lines corresponds to decreasing and increasing magnetic field respectively. **(C)** MFM images measured at zero applied magnetic field after applying sequentially an in-plane field of around 400 mT followed by a perpendicular field of 170 mT.

The SAF stack is composed of a Pt(3)/Co(1.58)/Ru(0.85)/Pt(0.5)/Co(1.58)/Ru(0.85) (nominal thickness in nm) multilayer capped with a thin Ta(1.5 nm) layer, deposited by sputtering deposition (Fig. 2A). We carefully optimized its composition in order to stabilize skyrmions and move them efficiently with currents(*5*): the Pt/Co interfaces provides the PMA and DMI necessary to stabilize the skyrmions and the spin orbit torque to move them with current (*36*). The Ru thickness was optimized to achieve AF RKKY interlayer exchange coupling with a coupling field of $\mu_0 H_{RKKY}=205$ mT. The thickness of the Co layers was adjusted such that the layers are perpendicularly magnetized but close to the in-plane/out-of-plane spin reorientation transition, leading to a linear and reversible hysteresis loop indicating a multidomain state (Fig. 2B). The lower effective PMA combined with the DMI reduces the domain wall energy and favors the nucleation of domains as well AF skyrmions (*29*). Isolated skyrmions can be nucleated using external magnetic field sequences (*29*, *36*). Figure 2C shows an example of a magnetic force microscopy (MFM) image at zero field displaying isolated skyrmions and larger domains nucleated after applying sequentially a large in-plane field followed by a perpendicular magnetic field (170 mT) (for other examples see (*36*)). The skyrmions have different shapes and their size ranges between 100 nm and 700 nm with an average of 215 ± 102 nm (*36*). This large dispersion can be explained by the deformation induced by the pinning of the skyrmions on material inhomogeneities as is also observed in ferromagnetic ultrathin films (*37*, *38*). In contrast to FM skyrmions, the stray field energy plays little role in the size of AF skyrmions owing to their vanishing moment, and their diameter d results from a fine balance between the DMI, exchange and anisotropy energies. Models (*29*, *39*) predict a skyrmion diameter $d \approx 2.7\Delta \frac{\left(\frac{D}{D_c}\right)^2}{\sqrt{1-\left(\frac{D}{D_c}\right)^2}}$ with $\Delta = \sqrt{\frac{A}{K_{eff}}}$ the domain wall width, D the DMI constant, $D_c = \frac{4\sqrt{AK_{eff}}}{\pi}$ the critical DMI for which the single domain state is no longer stable, where A is the exchange constant and $K_{eff}$ the effective PMA (*29*, *39*). Using experimental parameters, a diameter d = 197 nm is predicted, in line with our results. Quantitative measurements of the stray field of the SAF skyrmions using scanning nitrogen vacancy (NV) center magnetometry confirm that our SAF is fully compensated and that the skyrmion magnetizations in the two layers are AF-aligned, in agreement with micromagnetic simulations (*36*). X-ray magnetic circular dichroism photo-emission electron microscopy (XCMD-PEEM) experiments as well as micromagnetic simulations confirm the left handed chirality of the skyrmions expected from the DMI at the Pt/Co interface (*36*).

**Fast current induced dynamics of SAF skyrmions**

To study the current induced dynamics, the stack was patterned in 3 µm wide tracks in contact with metallic electrodes for current injection using a standard nanofabrication process (Fig S6, (*36*). In Fig. 3, sequences of MFM images display characteristic current induced motion experiments. Between each MFM image acquisition, a single current pulse is injected in the track. In Fig. 3A, the consecutive injections of 1 ns current pulses lead to the motion of a SAF skyrmion in the current direction. The distance and trajectory induced by the current pulse vary from one pulse to another, which can be explained by the effect of pinning on inhomogeneities in the material (*3*, *5*, *6*, *40*). However, on average, the skyrmion moves with minimal deviation from the current direction (~1° ). Note that the observed direction of motion, *i.e* along the current, is consistent with the sign of the spin Hall effect in Pt and the left handed chirality of the skyrmions arising from the DMI at the Pt/Co interface (*5*, *8*) and is in agreement with previous experiments in Pt/Co based stacks (*4*, *5*).

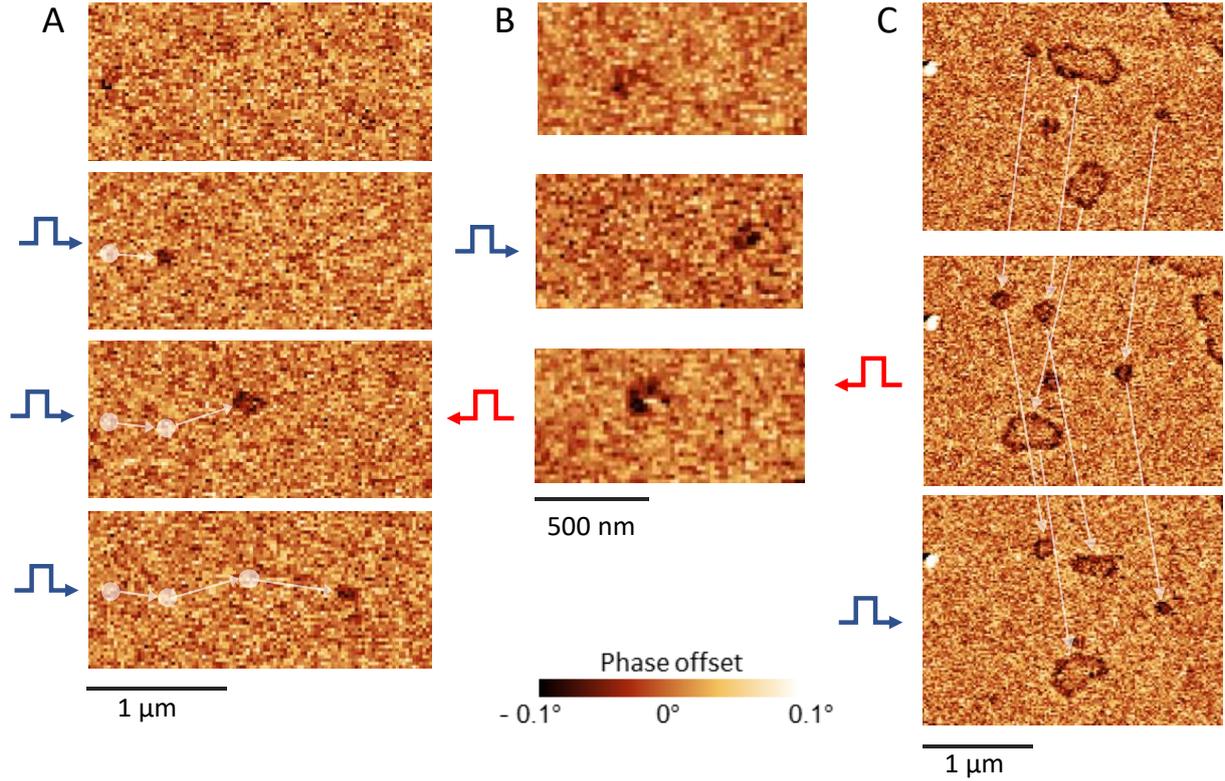

**Fig. 3. Current induced motion of SAF skyrmions** Current induced motion of SAF skyrmions observed by magnetic force microscopy (MFM). Plotted is the phase offset in color scale measured in lift mode. (A) Sequence of MFM images measured after the consecutive injection of a 1.160 ns, 1.18 ns, and 0.9 ns current pulses with densities J =8.0 × 1011 A/m² , 8.0 × 1011 A/m² and 7.6 × 1011 A/m² respectively (velocity of the skyrmion of 280 m/s, 400 m/s, 640 m/s respectively). (B) Sequence of MFM images measured before (top), after the consecutive injection of a positive current pulse (middle) and a negative current pulse (bottom) with width 0.6 ns and density 8.5 ×1011 A/m² and -7.9 ×1011A/m² respectively. (v=900 m/s and 720 m/s respectively) (C) Sequence of MFM images measured before (top), after the injection of a negative current pulse (middle) and after a positive current pulse (bottom) with widths 0.55 ns and density -8.2 × 1011 A/m² and 7.9 × 1011 A/m² respectively. All measurements are performed at zero applied magnetic field and room temperature. The white and colored dashed lines in A, B, C connect the positions of the skyrmions before and after the current pulses. Drift was corrected using a defect on the topography images.

Skyrmions can also be moved back and forth along the current direction by consecutive current pulses with opposite polarities as shown in the sequences of Fig. 3, B and C and in Fig. S24-S30 (*36*) . For instance, in Fig. 3B, a skyrmion is moved along the current direction by successive positive and negative 0.6 ns current pulses with velocities of 900 m/s and 720 m/s respectively. In Fig. 3C, several skyrmions with different sizes move in the direction of the applied current pulses. However, some skyrmions change their shape and size when moving and one of them does not move. This behavior is also observed for ferromagnetic skyrmions in ultrathin films and can be explained by the pinning of the skyrmion textures on inhomogeneities (*5*).

Despite these irregular dynamics, the average velocity $\bar{v}_{Sk}$ and skyrmion Hall angle $\bar{\theta}_{Sk}$ of the moving skyrmions can be obtained from systematic current-induced motion experiments (see Fig. 4 and (*36*)). Figure 4A, black squares, displays the dependence of $\bar{v}_{Sk}$ on the current density. The skyrmion velocity is calculated by dividing the distance travelled by the skyrmion over the current pulse width (between 0.5 ns and 1.3 ns). $\bar{v}_{Sk}$ increases with the current density with a low mobility regime at low current density (5×1011A/m²<J<6.5×1011A/m²) with 250<$\bar{v}_{Sk}$ <300 m/s and a larger mobility at higher current density with $\bar{v}_{Sk}$ reaching a maximum of 895 ± 285 m/s for a current density of 8.9×10¹¹ A/m². Our experiments reveal a vanishing skyrmion Hall effect: $\bar{\theta}_{Sk}$ = 6.4° ± 16.4° and does not show any dependence on the skyrmion velocity (Fig.4B, black squares). To study the impact of the magnetic compensation on the skyrmion dynamics, we have measured the same quantities in a Pt/Co/Ru trilayer (Fig. 4, blue dots) as well as a synthetic ferromagnetic (SF) stack where both Co layers are ferromagnetically coupled by RKKY using a thicker Ru interlayer (1.4 nm instead of 0.85 nm)(*36*) (Fig. 4, red triangles) . Note that data are for one sample for each material stack and the error bars denote the standard deviation across different skyrmions plus experimental error. In contrast to the SAF stack, a maximum velocity of around 120 m/s for the single Co layer stack and 80 m/s for the SF stack is observed. These values are in line with the one measured in previous current induced motion experiments of ferromagnetic skyrmions in ultrathin films (*4, 5*). The skyrmion Hall angle is found to increase with the velocity and eventually saturates at larger velocity to around 60° for the single layer and 85° for the SF stacks (*36*). This behavior was already reported in FM stacks and can be attributed to the interaction of the skyrmions with material disorder, which decreases the skyrmion Hall angle at lower drives/velocities (*3, 5, 10, 41*). These experiments demonstrate that the velocity enhancement and cancellation of the skyrmion Hall effect observed in the SAF stack are related to the SAF magnetic compensation.

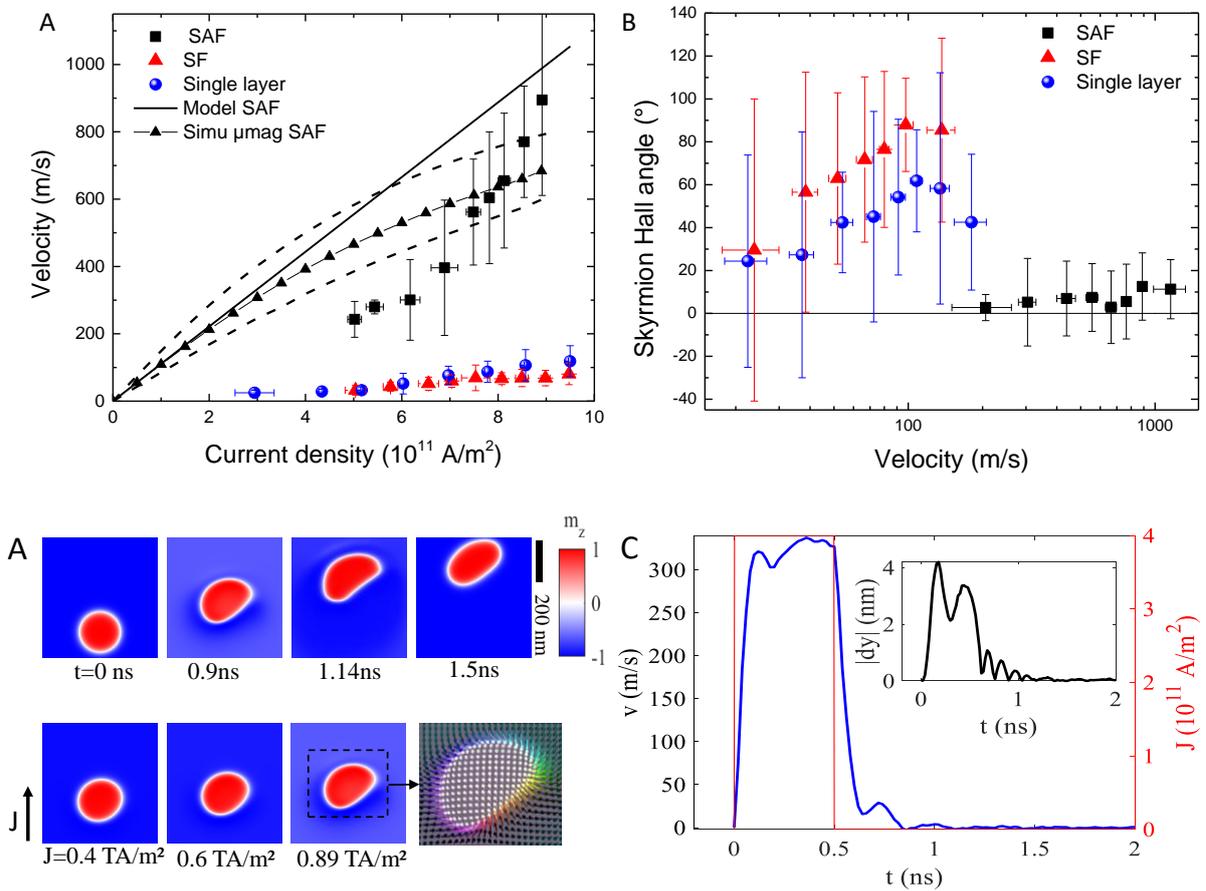

**Fig. 4. Fast current induced dynamics of SAF skyrmions with vanishing skyrmion Hall effect** (A) Average skyrmion velocity as a function of the current density measured experimentally in the SAF stack (black squares), synthetic ferromagnetic (SF) (red triangles) and single ferromagnetic layer stacks (blue dots). The black line corresponds to the prediction of the Thiele model for the SAF stack using experimentally obtained material parameters (see table S2 and section 2.7 in Ref (*36*)). The black triangles correspond to the velocity obtained from the micromagnetic simulations (α=0.14). The upper and lower dotted lines corresponds to the velocity obtained from micromagnetic simulations for α=0,1 and α=0,18 respectively. (B) Skyrmion Hall angle as a function of the velocity of the SAF skyrmion (black squares), SF skyrmions (red triangles) and single layer skyrmion (blue circles). Each point is an average over 3 to 47 measurements (*36*). Note that shown data are for one sample per material stack and the error bars denote the standard deviation across different skyrmions for the same sample plus the experimental errors.

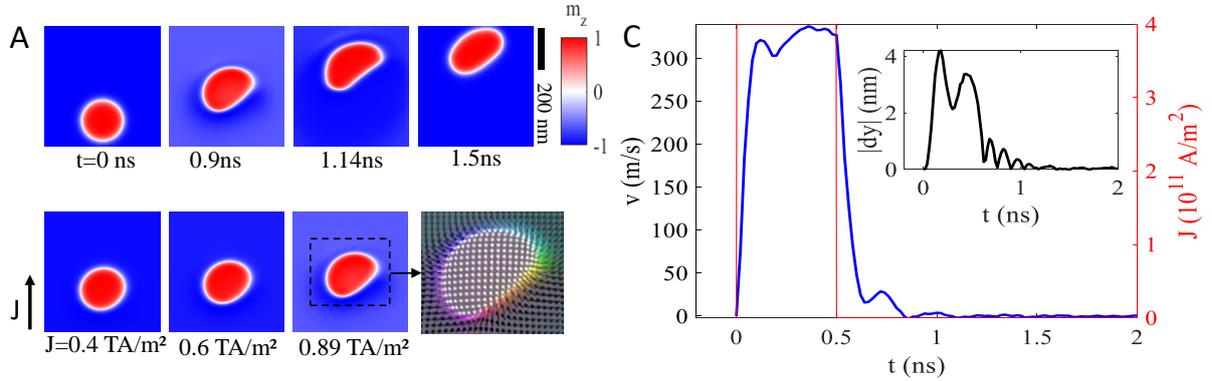

**Fig. 5 Simulations of the current-induced skyrmion deformation and inertial dynamics** (A-B) Deformation of a skyrmion (A) with time during a 0.5 ns Gaussian current pulse for J=8.9×1011A/m2 and (B) for various current densities at the maximum of the 0.5 ns Gaussian pulse. The panels plot the z-axis magnetization in color scale. The last panel in the bottom row shows a zoom of the spin texture for J=8.9x1011 A/m² (Black/white contrast indicates the magnetisation pointing downward/upward and the colou=r wheel illustrates the orientation of the in-plane magnetisation).C) Time dependence of the instantaneous skyrmion velocity (blue line) induced by a 0.50 ns square current pulse of density J=4×1011 A/m2 (red line). The inset shows the time dependence of the distance between the AF coupled skyrmions, dy, in a direction perpendicular to the current.

**Micromagnetic simulations and analytical modelling**
To better understand these results, we carried out analytical modeling as well as micromagnetic simulations using experimentally obtained magnetic and transport parameters, including magnetic damping, SOT and DMI (*36*). The experimental results are first compared to the Thiele equation, which is known to capture the skyrmion dynamics well (*5*, *42*). Under the assumption that the skyrmion spin texture is rigid during its motion, the steady state velocity **v** results from a balance between the different forces on the skyrmion $\boldsymbol{F}_{SOT} + \boldsymbol{G} \times \boldsymbol{v} + \alpha[\boldsymbol{D}]\boldsymbol{v} = 0$ where **F**$_{SOT}$ is the SOT force along the current direction and [**D**] the dissipation tensor. For a SAF skyrmion, the net gyrovector **G** = 0 owing to the opposite topological charges of the AF coupled skyrmions composing the SAF. Thus, SAF skyrmions move along the current direction with no skyrmion Hall effect, in agreement with the experimental observations. From the Thiele equation, the velocity can be expressed as (*36*) $v = \frac{\pi B_{SOT} J R \gamma}{2\alpha\left(\frac{R}{\Delta}+\frac{\Delta}{R}\right)}$ where R is the skyrmion radius, Δ the domain wall width, B$_{SOT}$ is the effective damping like SOT field and γ the gyromagnetic constant. The dependence of the velocity on the current density predicted by the model is plotted in

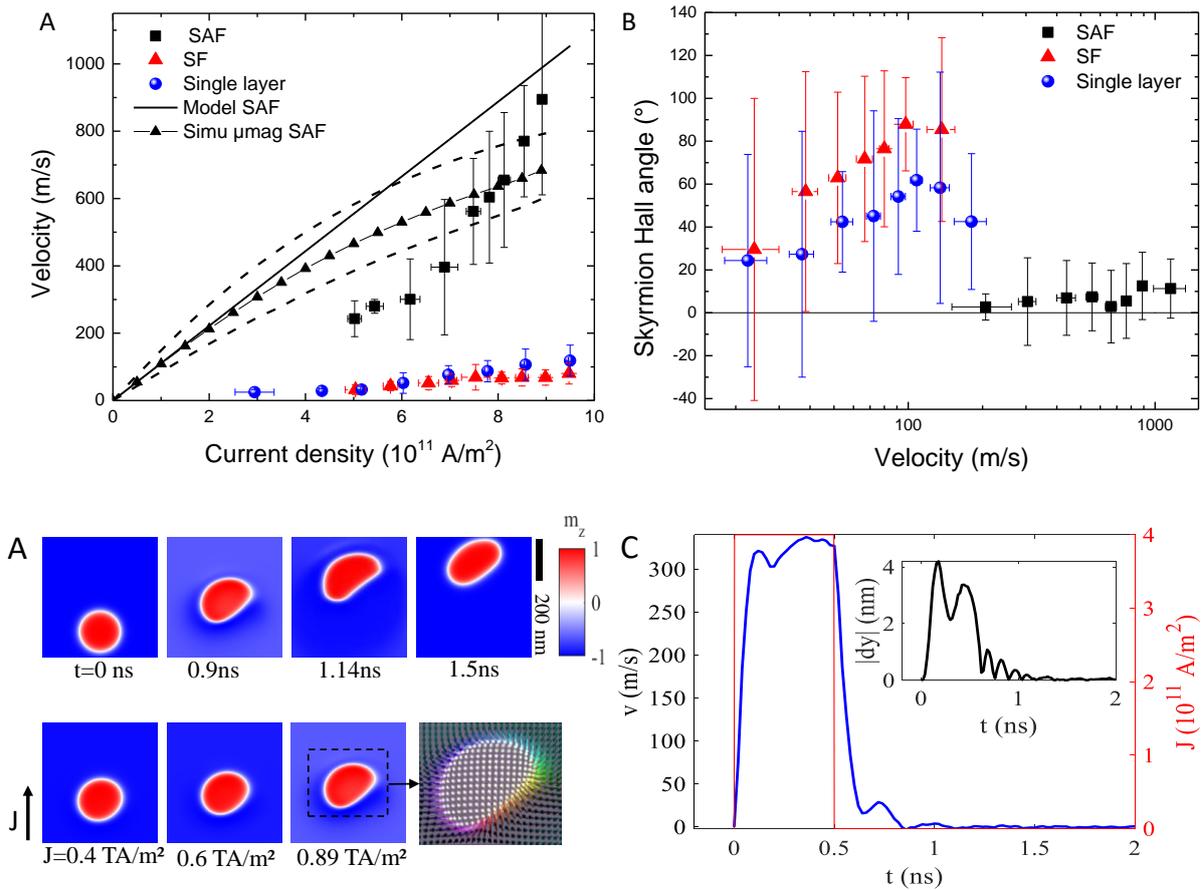

Fig. 4A (black line) (*36*). We observe that the model correctly reproduce the large velocity at large current density. However, it does not reproduce the low mobility regime and non-linear increase observed experimentally. These regimes can be accounted for by the effect of pinning, which is known to play a major role in the skyrmion dynamics in ultrathin films for low driving force and which is not taken into account in the model (*5, 7, 41*).

To shed more light on the skyrmion dynamics in the SAF, we performed micromagnetic simulations using experimental magnetic and transport parameters (*36*). As expected, the simulations show that both AF coupled skyrmions move in the current direction in the steady state, with no skyrmion Hall effect. At low current density, the skyrmion velocity scales linearly with J (

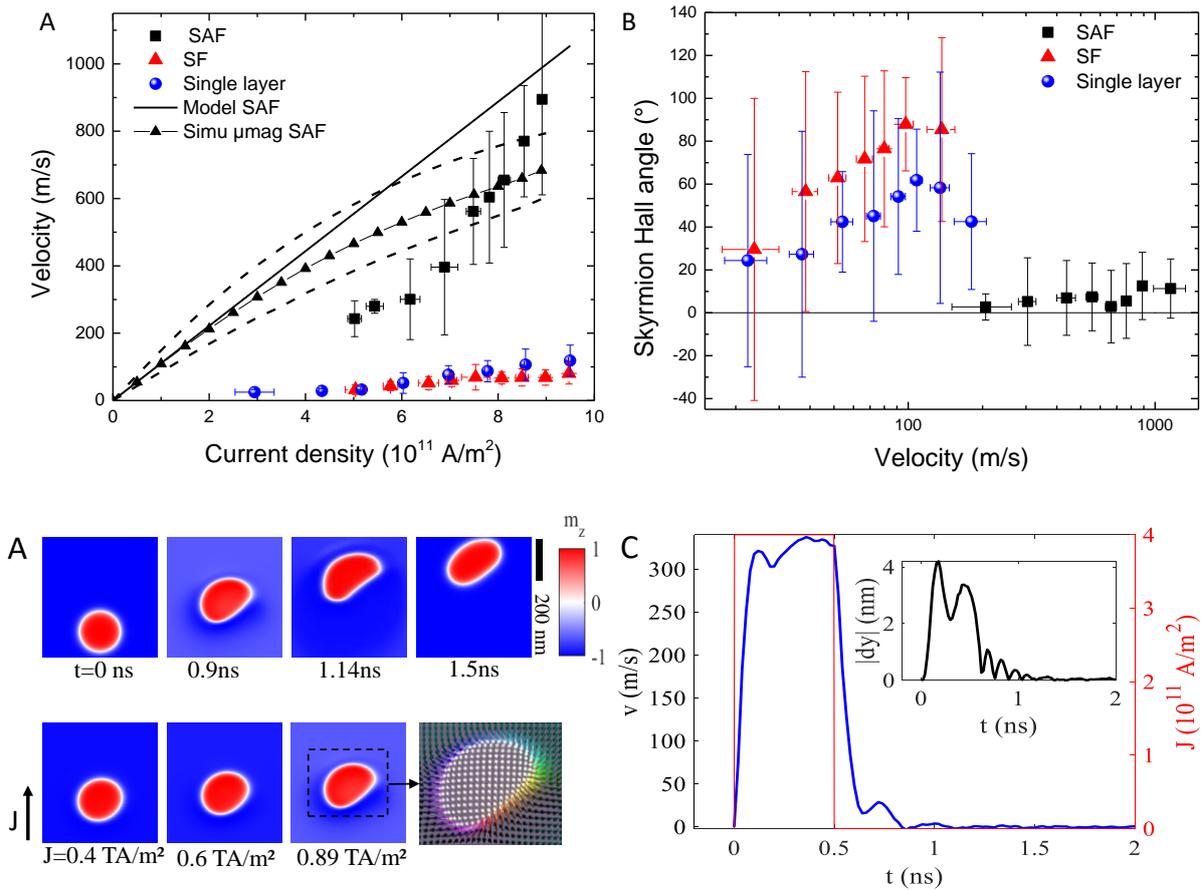

Fig. 4A, black triangles) and is in good agreement with the prediction of the Thiele model (black line). However, above $2 \times 10^{11}$ A/m², a sublinear drop is observed, whose amplitude increases with the current density. The dotted lines in Fig. 4A show the error range resulting from the main parameter uncertainty, namely the damping parameter $\alpha=0.14\pm0.04$ (*36*). One sees that the simulations correctly reproduce the experimental range of velocity measured at large current densities. The simulations also reveal that the non-linear velocity drop at large drive is caused by a dynamical deformation of the skyrmion, which is not taken into account in the Thiele model. Indeed, during the pulse injection, an expansion along with an elliptical deformation of the skyrmion is observed in simulations, and becomes more pronounced with time and increasing current, with the appearance of DW-like sections at large current (see Fig. 5A). This deformation is a pure dynamical effect that was already reported in FM skyrmions (*5, 10, 43, 44*), and is explained by various factors including kinetic effective field (*45, 46*) and inhomogeneous magnetization canting (*47*); for AF skyrmions (*16, 19, 21–24, 48*), the inhomogeneous gyrotropic forces play an additional role (*48*). The deformation is accompanied by an inhomogeneous tilt of the skyrmion magnetization away from the Néel chirality, which increases with current (see Fig. 5B and Fig. S39-S42 (*36*)). This behavior is similar to the one observed in SOT driven FM DW motion, where the SOT-induced spin rotation decreases the resultant driving force on the DW and eventually leads to a saturation of the DW velocity at high current drives (*49*). This picture is confirmed by simulations of the SAF DW dynamics, which show that the skyrmion velocity is close to the DW ones (Fig. S40 (*36*)). Thus, the observed velocity saturation can be explained by a transition to a DW-like dynamics at larger drives induced by the large skyrmion deformation. This behavior was also reported for FM skyrmions (*44*), but is unexpected in AFM, where the AF exchange torque between the sublattices is expected to prevent the current-induced DW spin rotation. This is not the case in our SAF stacks owing to the low RKKY exchange interaction. However, simulations reveal that a larger velocity can be reached by increasing the RKKY interaction (930 m/s for a RKKY field µ0HRKKY=950 mT instead of 680 m/s for µ0HRKKY =205 mT), which can be achieved by stack engineering (Fig. S41, S42, (*36*)). Another way to limit the effects of velocity saturation is to contain the deformation and DW spin rotation using shorter current pulses: for instance, a larger velocity of 760 m/s is predicted using a shorter pulse width of 100 ps (Fig. S38, S42 (*36*)). Reducing the deformation is also crucial for devices with narrow tracks because skyrmions transform into 360 domain walls when touching the track edge (*19*).

Simulations also reveal that SAF skyrmions feature intrinsic inertial dynamics: it takes ~100 ps for the skyrmion to reach a velocity plateau in response to a square current pulse (Fig. 5B). A closer look at the AF coupled skyrmion trajectories shows that this inertia can be explained by the finite AF RKKY interlayer interaction. When injecting a fast rising current pulse, the AF coupled skyrmions move perpendicularly to the current direction but in opposite directions, owing to the opposite gyrotropic forces ±$Gv_x$. The resulting distance dy between the skyrmions leads to a reaction force $F_{RKKY} = \mp k\,dy$ caused by the rise in the RKKY energy (here $k$ is the spring constant). Eventually, the two forces compensate and the skyrmions move along x with a finite distance dy = $\mp Gv_x/k$, which is typically a few nanometers in our simulations (Fig. 5C, inset). The resulting inertia can be described (*36*) by an effective mass $m = \frac{G^2+(\alpha D)^2}{k}$. The corresponding relaxation time $\tau = \frac{m}{2\alpha D}$ (*36*) is expected to scale as $\tau \sim \frac{1}{J_{RKKY}}$, where $J_{RKKY}$ is the amplitude of the RKKY interaction, which is confirmed by micromagnetic simulations (*36*). Note that inertial effects cannot be seen with our quasi-static experiments. The simulations show that the total skyrmion displacement ΔX remains proportional to v0T for a current pulse of length T where v0 is the steady state skyrmion velocity (*36*). Thus, this inertial effect does not affect the measured averaged velocity $v_{ag}$=ΔX /T, nor, for applications, devices based on the position of skyrmions.

**Discussion and outlook**

We now discuss the origin of the large enhancement of the current induced skyrmion velocities in SAF as compared to the FM stacks. Three effects can be invoked. Firstly, the compensation of the gyrotropic force, which leads to an increase of the velocity by a factor $\sqrt{1+\left(\frac{G}{aD}\right)^2} \sim \left(\frac{2\Delta}{\alpha R}\right) \sim 3.8 \pm 1.5$ (Fig.1C). Secondly, the presence of a large SOT effective field (22.1±0.14 mT/(TA/m²), as measured from second harmonic measurements, larger than in a single Pt/Co/Ru trilayer (17±0.14 mT/(TA/m²)), despite a twice larger thickness (Section 1.4, Table S2, Fig. S4, S5 (*36*). This large SOT suggests an efficient spin Hall effect, with an effective spin Hall angle per FM layer ($\theta_{eff}^{SAF} = 0.125$) 1.3 times larger than the one measured in single Pt/Co/Ru trilayer ($\theta_{eff}^{FM} = 0.096$)(*36*). Note that the SOT effective field was measured in the SF stack, where the Ru layer is slightly thicker as compared to the SAF stack (1.4 nm instead of 0.85 nm) (*36*). A similar enhancement of $\theta_{eff}^{FM}$ in multiple repetitions of (Pt/Co/X) trilayers (with X = Pt, Ir, Cu or Ta) as compared to a single trilayer was already reported (*50*, *51*). This may be explained by a SOT that is generated not only from the bottom Pt(3nm) layer but also from the Ru/Pt bilayer (*52*, *53*). Additionally, the bilayer (α=0.14±0.06) features a lower magnetic damping as compared to the single layer (0.33±0.09), which can be accounted for by the increased effective magnetic thickness.

Our results provide an important insight into the dynamics of fast-moving antiferromagnetic skyrmions. They also open promising perspectives for the development of fast-operation skyrmion-based devices, including non-volatile memory, nanoscale oscillators, Boolean and non-conventional logic, in a class of material compatible with the semi-conductor industry and spintronic devices

**References**

1. N. Nagaosa, Y. Tokura, Topological properties and dynamics of magnetic skyrmions. *Nat. Nanotechnol.* **8**, 899–911 (2013).

2. X. Z. Yu, N. Kanazawa, W. Z. Zhang, T. Nagai, T. Hara, K. Kimoto, Y. Matsui, Y. Onose, Y. Tokura, Skyrmion flow near room temperature in an ultralow current density. *Nat. Commun.* **3**, 988 (2012).

3. W. Jiang, X. Zhang, G. Yu, W. Zhang, X. Wang, M. Benjamin Jungfleisch, J. E. Pearson, X. Cheng, O. Heinonen, K. L. Wang, Y. Zhou, A. Hoffmann, S. G. E. te Velthuis, Direct observation of the skyrmion Hall effect. *Nat. Phys.*, doi: 10.1038/nphys3883 (2016).

4. S. Woo, K. Litzius, B. Krüger, M.-Y. Im, L. Caretta, K. Richter, M. Mann, A. Krone, R. M. Reeve, M. Weigand, P. Agrawal, I. Lemesh, M.-A. Mawass, P. Fischer, M. Kläui, G. S. D. Beach, Observation of room-temperature magnetic skyrmions and their current-driven dynamics in ultrathin metallic ferromagnets. *Nat. Mater.* **15**, 501–506 (2016).

5. R. Juge, S.-G. Je, D. de S. Chaves, L. D. Buda-Prejbeanu, J. Peña-Garcia, J. Nath, I. M. Miron, K. G. Rana, L. Aballe, M. Foerster, F. Genuzio, T. O. Mentes, A. Locatelli, F. Maccherozzi, S. S. Dhesi, M. Belmeguenai, Y. Roussigné, S. Auffret, S. Pizzini, G. Gaudin, J. Vogel, O. Boulle, Current-Driven Skyrmion Dynamics and Drive-Dependent Skyrmion Hall Effect in an Ultrathin Film. *Phys. Rev. Appl.* **12**, 044007 (2019).

6. K. Litzius, I. Lemesh, B. Krüger, P. Bassirian, L. Caretta, K. Richter, F. Büttner, K. Sato, O. A. Tretiakov, J. Förster, R. M. Reeve, M. Weigand, I. Bykova, H. Stoll, G. Schütz, G. S. D. Beach, M. Kläui, Skyrmion Hall effect revealed by direct time-resolved X-ray microscopy. *Nat. Phys.* **13**, 170–175 (2017).

7. W. Jiang, P. Upadhyaya, W. Zhang, G. Yu, M. B. Jungfleisch, F. Y. Fradin, J. E. Pearson, Y. Tserkovnyak, K. L. Wang, O. Heinonen, S. G. E. te Velthuis, A. Hoffmann, Blowing magnetic skyrmion bubbles. *Science* **349**, 283–286 (2015).

8. O. Boulle, J. Vogel, H. Yang, S. Pizzini, D. de Souza Chaves, A. Locatelli, T. O. Menteş, A. Sala, L. D. Buda-Prejbeanu, O. Klein, M. Belmeguenai, Y. Roussigné, A. Stashkevich, S. M. Chérif, L. Aballe, M. Foerster, M. Chshiev, S. Auffret, I. M. Miron, G. Gaudin, Room-temperature chiral magnetic skyrmions in ultrathin magnetic nanostructures. *Nat. Nanotechnol.* **11**, 449–454 (2016).

9. C. Moreau-Luchaire, C. Moutafis, N. Reyren, J. Sampaio, C. a. F. Vaz, N. V. Horne, K. Bouzehouane, K. Garcia, C. Deranlot, P. Warnicke, P. Wohlhüter, J.-M. George, M. Weigand, J. Raabe, V. Cros, A. Fert, Additive interfacial chiral interaction in multilayers for stabilization of small individual skyrmions at room temperature. *Nat. Nanotechnol.* **11**, 444–448 (2016).

10. K. Litzius, J. Leliaert, P. Bassirian, D. Rodrigues, S. Kromin, I. Lemesh, J. Zazvorka, K.-J. Lee, J. Mulkers, N. Kerber, D. Heinze, N. Keil, R. M. Reeve, M. Weigand, B. Van Waeyenberge, G. Schütz, K. Everschor-Sitte, G. S. D. Beach, M. Kläui, The role of temperature and drive current in skyrmion dynamics. *Nat. Electron.* **3**, 30–36 (2020).

11. K. Zeissler, S. Finizio, C. Barton, A. J. Huxtable, J. Massey, J. Raabe, A. V. Sadovnikov, S. A. Nikitov, R. Brearton, T. Hesjedal, G. van der Laan, M. C. Rosamond, E. H. Linfield,

G. Burnell, C. H. Marrows, Diameter-independent skyrmion Hall angle observed in chiral magnetic multilayers. *Nat. Commun.* **11**, 428 (2020).

12. A. A. Thiele, Theory of the Static Stability of Cylindrical Domains in Uniaxial Platelets. *J. Appl. Phys.* **41**, 1139–1145 (1970).

13. A. Hrabec, J. Sampaio, M. Belmeguenai, I. Gross, R. Weil, S. M. Chérif, A. Stashkevich, V. Jacques, A. Thiaville, S. Rohart, Current-induced skyrmion generation and dynamics in symmetric bilayers. *Nat. Commun.* **8**, ncomms15765 (2017).

14. F. Büttner, I. Lemesh, G. S. D. Beach, Theory of isolated magnetic skyrmions: From fundamentals to room temperature applications. *Sci. Rep.* **8**, 4464 (2018).

15. J. Barker, O. A. Tretiakov, Static and Dynamical Properties of Antiferromagnetic Skyrmions in the Presence of Applied Current and Temperature. *Phys. Rev. Lett.* **116**, 147203 (2016).

16. C. Jin, C. Song, J. Wang, Q. Liu, Dynamics of antiferromagnetic skyrmion driven by the spin Hall effect. *Appl. Phys. Lett.* **109**, 182404 (2016).

17. R. Tomasello, V. Puliafito, E. Martinez, A. Manchon, M. Ricci, M. Carpentieri, G. Finocchio, Performance of synthetic antiferromagnetic racetrack memory: domain wall versus skyrmion. *J. Phys. Appl. Phys.* **50**, 325302 (2017).

18. X. Zhang, Y. Zhou, M. Ezawa, Antiferromagnetic Skyrmion: Stability, Creation and Manipulation. *Sci. Rep.* **6**, 24795 (2016).

19. P. E. Roy, Method to suppress antiferromagnetic skyrmion deformation in high speed racetrack devices. *J. Appl. Phys.* **129**, 193902 (2021).

20. X. Zhang, Y. Zhou, M. Ezawa, Magnetic bilayer-skyrmions without skyrmion Hall effect. *Nat. Commun.* **7**, 10293 (2016).

21. H. Velkov, O. Gomonay, M. Beens, G. Schwiete, A. Brataas, J. Sinova, R. A. Duine, Phenomenology of current-induced skyrmion motion in antiferromagnets. *New J. Phys.* **18**, 075016 (2016).

22. A. Salimath, F. Zhuo, R. Tomasello, G. Finocchio, A. Manchon, Controlling the deformation of antiferromagnetic skyrmions in the high-velocity regime. *Phys. Rev. B* **101**, 024429 (2020).

23. J. Xia, X. Zhang, K.-Y. Mak, M. Ezawa, O. A. Tretiakov, Y. Zhou, G. Zhao, X. Liu, Current-induced dynamics of skyrmion tubes in synthetic antiferromagnetic multilayers. *Phys. Rev. B* **103**, 174408 (2021).

24. S. Komineas, N. Papanicolaou, Traveling skyrmions in chiral antiferromagnets. *SciPost Phys.* **8**, 086 (2020).

25. S. Gao, H. D. Rosales, F. A. G. Albarracín, V. Tsurkan, G. Kaur, T. Fennell, P. Steffens, M. Boehm, P. Čermák, A. Schneidewind, E. Ressouche, D. C. Cabra, C. Rüegg, O. Zaharko, Fractional antiferromagnetic skyrmion lattice induced by anisotropic couplings. *Nature*, doi: 10.1038/s41586-020-2716-8 (2020).

26. H. Jani, J.-C. Lin, J. Chen, J. Harrison, F. Maccherozzi, J. Schad, S. Prakash, C.-B. Eom, A. Ariando, T. Venkatesan, P. G. Radaelli, Antiferromagnetic half-skyrmions and bimerons at room temperature. *Nature* **590**, 74–79 (2021).

27. K. G. Rana, R. Lopes Seeger, S. Ruiz-Gómez, R. Juge, Q. Zhang, K. Bairagi, V. T. Pham, M. Belmeguenai, S. Auffret, M. Foerster, L. Aballe, G. Gaudin, V. Baltz, O. Boulle, Imprint from ferromagnetic skyrmions in an antiferromagnet via exchange bias. *Appl. Phys. Lett.* **119**, 192407 (2021).

28. W. Legrand, D. Maccariello, F. Ajejas, S. Collin, A. Vecchiola, K. Bouzehouane, N. Reyren, V. Cros, A. Fert, Room-temperature stabilization of antiferromagnetic skyrmions in synthetic antiferromagnets. *Nat. Mater.* **19**, 34–42 (2020).

29. R. Juge, N. Sisodia, J. U. Larrañaga, Q. Zhang, V. T. Pham, K. G. Rana, B. Sarpi, N. Mille, S. Stanescu, R. Belkhou, M.-A. Mawass, N. Novakovic-Marinkovic, F. Kronast, M. Weigand, J. Gräfe, S. Wintz, S. Finizio, J. Raabe, L. Aballe, M. Foerster, M. Belmeguenai, L. D. Buda-Prejbeanu, J. Pelloux-Prayer, J. M. Shaw, H. T. Nembach, L. Ranno, G. Gaudin, O. Boulle, Skyrmions in synthetic antiferromagnets and their nucleation via electrical current and ultra-fast laser illumination. *Nat. Commun.* **13**, 4807 (2022).

30. Y. Hirata, D.-H. Kim, S. K. Kim, D.-K. Lee, S.-H. Oh, D.-Y. Kim, T. Nishimura, T. Okuno, Y. Futakawa, H. Yoshikawa, A. Tsukamoto, Y. Tserkovnyak, Y. Shiota, T. Moriyama, S.-B. Choe, K.-J. Lee, T. Ono, Vanishing skyrmion Hall effect at the angular momentum compensation temperature of a ferrimagnet. *Nat. Nanotechnol.*, 1 (2019).

31. S. Woo, K. M. Song, X. Zhang, Y. Zhou, M. Ezawa, X. Liu, S. Finizio, J. Raabe, N. J. Lee, S.-I. Kim, S.-Y. Park, Y. Kim, J.-Y. Kim, D. Lee, O. Lee, J. W. Choi, B.-C. Min, H. C. Koo, J. Chang, Current-driven dynamics and inhibition of the skyrmion Hall effect of ferrimagnetic skyrmions in GdFeCo films. *Nat. Commun.* **9**, 959 (2018).

32. T. Dohi, S. DuttaGupta, S. Fukami, H. Ohno, Formation and current-induced motion of synthetic antiferromagnetic skyrmion bubbles. *Nat. Commun.* **10**, 1–6 (2019).

33. Y. Quessab, J.-W. Xu, E. Cogulu, S. Finizio, J. Raabe, A. D. Kent, Zero-Field Nucleation and Fast Motion of Skyrmions Induced by Nanosecond Current Pulses in a Ferrimagnetic Thin Film. *Nano Lett.* **22**, 6091 (2022).

34. S. S. P. Parkin, N. More, K. P. Roche, Oscillations in exchange coupling and magnetoresistance in metallic superlattice structures: Co/Ru, Co/Cr, and Fe/Cr. *Phys. Rev. Lett.* **64**, 2304–2307 (1990).

35. J. U. Larrañaga, N. Sisodia, V. T. Pham, I. Di Manici, A. Masseboeuf, K. Garello, F. Disdier, B. Fernandez, S. Wintz, M. Weigand, M. Belmeguenai, S. Pizzini, R. Sousa, L. Buda-Prejbeanu, G. Gaudin, O. Boulle, Electrical detection and nucleation of a magnetic skyrmion in a magnetic tunnel junction observed via operando magnetic microscopy. arXiv arXiv:2308.00445 [Preprint] (2023). https://doi.org/10.48550/arXiv.2308.00445.

36. See supplementary materials.

37. I. Gross, W. Akhtar, A. Hrabec, J. Sampaio, L. J. Martínez, S. Chouaieb, B. J. Shields, P. Maletinsky, A. Thiaville, S. Rohart, V. Jacques, Skyrmion morphology in ultrathin magnetic films. *Phys. Rev. Mater.* **2**, 024406 (2018).

38. R. Juge, S.-G. Je, D. de Souza Chaves, S. Pizzini, L. D. Buda-Prejbeanu, L. Aballe, M. Foerster, A. Locatelli, T. O. Menteş, A. Sala, F. Maccherozzi, S. S. Dhesi, S. Auffret, E. Gautier, G. Gaudin, J. Vogel, O. Boulle, Magnetic skyrmions in confined geometries: Effect of the magnetic field and the disorder. *J. Magn. Magn. Mater.* **455**, 3–8 (2018).

39. L. Ranno, M. A. Moro, Design Rules for DMI-Stabilised Skyrmions. *ArXiv210700767 Cond-Mat* (2021).

40. J.-V. Kim, M.-W. Yoo, Current-driven skyrmion dynamics in disordered films. *Appl. Phys. Lett.* **110**, 132404 (2017).

41. C. Reichhardt, C. J. O. Reichhardt, M. V. Milosevic, Statics and Dynamics of Skyrmions Interacting with Pinning: A Review. *Rev. Mod. Phys.* **94**, 035005 (2022).

42. A. A. Thiele, Steady-State Motion of Magnetic Domains. *Phys. Rev. Lett.* **30**, 230 (1973).

43. J. Masell, D. R. Rodrigues, B. F. McKeever, K. Everschor-Sitte, Spin-transfer torque driven motion, deformation, and instabilities of magnetic skyrmions at high currents. *Phys. Rev. B* **101**, 214428 (2020).

44. H. Vakili, J.-W. Xu, W. Zhou, M. N. Sakib, M. G. Morshed, T. Hartnett, Y. Quessab, K. Litzius, C. T. Ma, S. Ganguly, M. R. Stan, P. V. Balachandran, G. S. D. Beach, S. J. Poon, A. D. Kent, A. W. Ghosh, Skyrmionics—Computing and memory technologies based on topological excitations in magnets. *J. Appl. Phys.* **130**, 070908 (2021).

45. R. E. Troncoso, Á. S. Núñez, Brownian motion of massive skyrmions in magnetic thin films. *Ann. Phys.* **351**, 850–856 (2014).

46. K. Yu. Guslienko, K.-S. Lee, S.-K. Kim, Dynamic Origin of Vortex Core Switching in Soft Magnetic Nanodots. *Phys. Rev. Lett.* **100**, 027203 (2008).

47. Z. Chen, X. Zhang, Y. Zhou, Q. Shao, Skyrmion Dynamics in the Presence of Deformation. *Phys. Rev. Appl.* **17**, L011002 (2022).

48. E. A. Tremsina, G. S. D. Beach, Atomistic simulations of distortion-limited high-speed dynamics of antiferromagnetic skyrmions. *Phys. Rev. B* **106**, L220402 (2022).

49. A. Thiaville, S. Rohart, É. Jué, V. Cros, A. Fert, Dynamics of Dzyaloshinskii domain walls in ultrathin magnetic films. *EPL Europhys. Lett.* **100**, 57002 (2012).

50. B. Jinnai, C. Zhang, A. Kurenkov, M. Bersweiler, H. Sato, S. Fukami, H. Ohno, Spin-orbit torque induced magnetization switching in Co/Pt multilayers. *Appl. Phys. Lett.* **111**, 102402 (2017).

51. K.-F. Huang, D.-S. Wang, H.-H. Lin, C.-H. Lai, Engineering spin-orbit torque in Co/Pt multilayers with perpendicular magnetic anisotropy. *Appl. Phys. Lett.* **107**, 232407 (2015).

52. J. Kim, Y. Otani, Orbital angular momentum for spintronics. *J. Magn. Magn. Mater.* **563**, 169974 (2022).

53. R. Xu, H. Zhang, Y. Jiang, H. Cheng, Y. Xie, Y. Yao, D. Xiong, Z. Zhu, X. Ning, R. Chen, Y. Huang, S. Xu, J. Cai, Y. Xu, T. Liu, W. Zhao, Giant orbit-to-charge conversion induced via the inverse orbital Hall effect. arXiv arXiv:2308.13144 [Preprint] (2023). https://doi.org/10.48550/arXiv.2308.13144.

54. S. Panigrahy, S. Mallick, J. Sampaio, S. Rohart, Skyrmion inertia in synthetic antiferromagnets. *Phys. Rev. B* **106**, 144405 (2022).

55. O. Boulle, Data of "Fast current induced skyrmion motion in synthetic antiferromagnets." doi: 10.17605/OSF.IO/8QHWR (2024).

56. S. Bandiera, R. C. Sousa, S. Auffret, B. Rodmacq, B. Dieny, Enhancement of perpendicular magnetic anisotropy thanks to Pt insertions in synthetic antiferromagnets. *Appl. Phys. Lett.* **101**, 072410 (2012).

57. P. J. Metaxas, J. P. Jamet, A. Mougin, M. Cormier, J. Ferré, V. Baltz, B. Rodmacq, B. Dieny, R. L. Stamps, Creep and Flow Regimes of Magnetic Domain-Wall Motion in Ultrathin $\mathrm{Pt}/\mathrm{Co}/\mathrm{Pt}$ Films with Perpendicular Anisotropy. *Phys. Rev. Lett.* **99**, 217208 (2007).

58. S.-G. Je, D.-H. Kim, S.-C. Yoo, B.-C. Min, K.-J. Lee, S.-B. Choe, Asymmetric Magnetic Domain-Wall Motion by the Dzyaloshinskii-Moriya Interaction. *Phys. Rev. B* **88** (2013).

59. T. H. Pham, J. Vogel, J. Sampaio, M. Vaňatka, J.-C. Rojas-Sánchez, M. Bonfim, D. S. Chaves, F. Choueikani, P. Ohresser, E. Otero, A. Thiaville, S. Pizzini, Very large domain wall velocities in Pt/Co/GdOx and Pt/Co/Gd trilayers with Dzyaloshinskii-Moriya interaction. *EPL* **113**, 67001 (2016).

60. C. O. Avci, K. Garello, C. Nistor, S. Godey, B. Ballesteros, A. Mugarza, A. Barla, M. Valvidares, E. Pellegrin, A. Ghosh, I. M. Miron, O. Boulle, S. Auffret, G. Gaudin, P. Gambardella, Fieldlike and antidamping spin-orbit torques in as-grown and annealed Ta/CoFeB/MgO layers. *Phys. Rev. B* **89**, 214419 (2014).

61. H.-B. Braun, Fluctuations and instabilities of ferromagnetic domain-wall pairs in an external magnetic field. *Phys. Rev. B* **50**, 16485–16500 (1994).

62. N. Romming, A. Kubetzka, C. Hanneken, K. von Bergmann, R. Wiesendanger, Field-Dependent Size and Shape of Single Magnetic Skyrmions. *Phys. Rev. Lett.* **114**, 177203 (2015).

63. U. Ritzmann, S. von Malottki, J.-V. Kim, S. Heinze, J. Sinova, B. Dupé, Trochoidal motion and pair generation in skyrmion and antiskyrmion dynamics under spin–orbit torques. *Nat. Electron.* **1**, 451–457 (2018).

64. A. Vansteenkiste, J. Leliaert, M. Dvornik, M. Helsen, F. Garcia-Sanchez, B. V. Waeyenberge, The design and verification of MuMax3. *AIP Adv.* **4**, 107133 (2014).

65. L. Belliard, A. Thiaville, S. Lemerle, A. Lagrange, J. Ferré, J. Miltat, Investigation of the domain contrast in magnetic force microscopy. *J. Appl. Phys.* **81**, 3849–3851 (1997).

66. D. W. Abraham, F. A. McDonald, Theory of magnetic force microscope images. *Appl. Phys. Lett.* **56**, 1181–1183 (1990).

67. A. Hubert, W. Rave, S. L. Tomlinson, Imaging Magnetic Charges with Magnetic Force Microscopy. *Phys. Status Solidi B* **204**, 817–828 (1997).

68. O. Hellwig, A. Berger, J. B. Kortright, E. E. Fullerton, Domain structure and magnetization reversal of antiferromagnetically coupled perpendicular anisotropy films. *J. Magn. Magn. Mater.* **319**, 13–55 (2007).

69. S. Hamada, K. Himi, T. Okuno, K. Takanashi, MFM observation of perpendicular magnetization and antiferromagnetically coupled domains in Co/Ru superlattices. *J. Magn. Magn. Mater.* **240**, 539–542 (2002).

70. O. Hellwig, A. Berger, E. E. Fullerton, Domain Walls in Antiferromagnetically Coupled Multilayer Films. *Phys. Rev. Lett.* **91**, 197203 (2003).

71. A. Baruth, L. Yuan, J. D. Burton, K. Janicka, E. Y. Tsymbal, S. H. Liou, S. Adenwalla, Domain overlap in antiferromagnetically coupled [Co/Pt]/NiO/[Co/Pt] multilayers. *Appl. Phys. Lett.* **89**, 202505 (2006).

72. L. Berges, Size-dependent mobility of skyrmions beyond pinning in ferrimagnetic GdCo thin films. *Phys. Rev. B* **106** (2022).

73. C. A. Akosa, O. A. Tretiakov, G. Tatara, A. Manchon, Theory of the Topological Spin Hall Effect in Antiferromagnetic Skyrmions: Impact on Current-Induced Motion. *Phys. Rev. Lett.* **121**, 097204 (2018).

74. P. M. Buhl, F. Freimuth, S. Blügel, Y. Mokrousov, Topological spin Hall effect in antiferromagnetic skyrmions. *Phys. Status Solidi RRL – Rapid Res. Lett.* **11**, 1700007 (2017).

75. B. Göbel, A. Mook, J. Henk, I. Mertig, Antiferromagnetic skyrmion crystals: Generation, topological Hall, and topological spin Hall effect. *Phys. Rev. B* **96**, 060406 (2017).

Notes During the review process, Panigrahy et al. reported similar results on the skyrmion inertia in synthetic antiferromagnets (*54*).

**Acknowledgement** We would like to thank Robert Morel for his technical support in the MFM experiments, Jan Vogel for his help in VSM measurement. We acknowledge National Supercomputing Mission (NSM) for providing computing resources of 'PARAM Ananta' at IIT Gandhinagar, which is implemented by C-DAC and supported by the Ministry of Electronics and Information Technology (MeitY) and Department of Science and Technology (DST), Government of India. This work was partly supported by the French RENATECH network, implemented at the Upstream Technological Platform in Grenoble PTA (ANR-22-PEEL-0015).

**Funding** We acknowledge the support of the Agence Nationale de la Recherche, project ANR-17-CE24-0045 (SKYLOGIC), France 2030 plan project Chirex PEPR SPIN ANR-22-EXSP-0002, the DARPA TEE program through grant MIPR# HR0011831554 from the DOI. VJ acknowledges the support of the European Union's Horizon 2020 research and innovation programme, Grant Agreement 866267 (ERC CoG - EXAFONIS). IDM acknowledges support from the Focus Numérique Frugal program from CEA.

**Authors Contributions** OB conceived and supervised the experiments.
VTP fabricated the SAF sample, performed the MFM SAF experiments. VTP, IDM and OB analyzed the SAF data.
IDM, RG, KB fabricated the SF and single trilayer samples, performed the MFM experiments and analyzed the data.
JUL and VTP performed the characterization of the magnetic properties.
JUL performed the spin orbit torque sample preparation and measurements.
VTP performed the magnetic damping measurements with the help of S.P.
OB, KB, JUL, AL, OTM performed the XMCD-PEEM measurements.
NS, IDM, OB performed the micromagnetic simulations.
JPP and OB performed the analytical modelling.
JPP performed the numerical modelling.
PK, AF performed the NV center magnetometry measurements.
SA carried out the sputtering deposition of the thin films.
GG participated in the nanofabrication.
OB and VTP wrote the manuscript.
All co-authors discussed the results and commented on the manuscript.
**Competing interests:** Authors declare they have no competing interests
**Data and materials availability:** Data presented in this paper are open access and canbe found on an open science framework server(*55*)

**Supplementary Materials**

Materials and Methods

Supplementary Text

Figs. S1 to S44

Table S1-S2

References (57)-(81)

**Supplementary Materials for**

Fast current induced skyrmion motion in synthetic antiferromagnets

Van Tuong Pham, Naveen Sisodia, Ilaria Di Manici, Joseba Urrestarazu-Larrañaga, Kaushik Bairagi, Johan Pelloux-Prayer, Rodrigo Guedas, Liliana D. Buda-Prejbeanu, Stéphane Auffret, Andrea Locatelli, Tevfik Onur Mentes, Stefania Pizzini, Pawan Kumar, Aurore Finco, Vincent Jacques, Gilles Gaudin and Olivier Boulle

Correspondence to: olivier.boulle@cea.fr

**This PDF file includes:**

Materials and Methods
Supplementary Text
Figs. S1 to S44
 Tables S1 to S2

# Materials and Methods
1. **Characterization of the magnetic and transport properties of the synthetic antiferromagnetic stack**

   ### 1.1 Growth optimization and characterization

We describe in this section the material optimization of the synthetic antiferromagnetic (SAF) stack. The stack was deposited by magnetron sputtering on a highly resistive Si substrate with native $SiO_2$. It is composed of two antiferromagnetically coupled Pt/Co layers through a thin Ru layer to achieve RKKY AF coupling:
Ta(3)/Pt(3)/Co(t)/Ru(0.85)/Pt(0.45)/Co(t)/Ru(0.6)/Ta(1.5), nominal thicknesses in nm. The Co(t) layers in the stacks were deposited as a wedge on a 100 mm wafer using off-axis deposition, such that the Co thickness varies linearly along the wafer between 1.4 and 1.9 nm. The Ru spacer is inserted to obtain the RKKY coupling between both ferromagnetic layers with $t_{Ru}$ = 0.85 nm to reach the maximum AF coupling (*34*, *56*). The thickness of the interlayer Pt layer (0.45nm) in contact with Ru was optimized in order to achieve large RKKY interaction as well as to have similar anisotropies in both layers, which favors the stabilization of SAF skyrmions. The Pt/Co interface was used to achieve PMA and DMI. The Pt layers allow to generate the spin-orbit torques to move the skyrmions.

To characterize the magnetic properties of the bottom and top Co layers constituting the SAF, two stacks were deposited, where the thickness of one of the Co layer is 0.3nm, such that it is dead magnetically. These stacks are represented in Figure S6A and Figure S6C respectively. The range of the Co thickness was optimized to be close to the spin reorientation transition. Figure S6B and Figure S6D show the out of plane hysteresis loops measured by MOKE in these two samples. The thickness *t* is varied between 1.42nm and 1.62 nm using a wedge deposition and the loops display the gradual transition from perpendicular to in-plane anisotropy.

For the current induced skyrmion nucleation and motion experiments, the full SAF stack was deposited on a highly resistive Si wafer with native $SiO_2$ and 3 µm wide tracks contacted to gold metallic electrodes were patterned using standard nanofabrication processes (see Figure S11). The thickness of the Co layer was chosen to be around 1.58 nm, which corresponds to a position of +14 mm on the wafer (see Figure S6B and Figure S6D), such that it is close to the in-plane to out-of-plane transition.

The stacks of Figure S6A and Figure S6C were used to measure the perpendicular effective magnetic anisotropy of the top and bottom Co(1.58 nm) layers respectively. It was estimated by measuring the saturation field from in-plane hysteresis loop using vibrating sample magnetometry. This leads to an effective anisotropy field $\mu_0 H_K$=12.4±16 mT for the bottom layer (Figure S6A) and 35.4 ± 14 mT for the top layer (Figure S6C).

### 1.2 Measurement of the magnetic damping

The measurement of the magnetic damping and spin orbit torque usually requires the manipulation of the magnetization using an external magnetic field as well as the detection of its orientation. However, this cannot be achieved in the compensated SAF owing to its vanishing magnetic moment. To solve this issue, we considered a bilayer stack, named synthetic ferromagnet (SF), where the Ru interlayer is thicker (1.4nm). This allows us to achieve a FM coupling of both Co layers instead of an AF one (*34*). Under the assumption that the SOT sand magnetic damping are little affected by the change in the Ru thickness, this enables the measurement of the average effective magnetic damping and SOTs constituting the SAF.

We describe in this section the measurement of the magnetic damping in the SF sample. To measure the magnetic damping, we consider magnetic field-induced domain wall (DW) dynamics experiments using polar magneto-optical Kerr effect (pMOKE) microscopy (*57*). At remanence, a dense multidomain state was observed, which prevents the study the domain wall dynamics. To obtain larger domain, a slightly thinner Co layers $t_{Co}$ = 1.3 nm (instead of 1.58 nm) was used, which allows an increase of the magnetic anisotropy and thus the stabilization of larger magnetic domains. Eventually, the sample we studied had the following composition: *substrate/Ta(3 nm)/Pt(3 nm)/Co(1.3 nm)/Ru(1.4 nm)/Pt(0.45nm)/Co(1.3 nm)/Ru(0.6 nm)/Ta(1.5 nm)* deposited on a Si/SiO2(500nm) substrate.

The reversed domain is generated by applying a local out-of-plane magnetic field pulse using a micro coil. The differential pMOKE image shows the displacement of domain walls induced by the application of a local ouf-of-plane magnetic pulses (see Figure S7A, inset). The velocity of the domain wall is estimated from the average value of the domain wall displacement in all directions in the sample plane.

The velocity of the domain wall as a function of the applied field $\mu_0 H$ is shown in Figure S7A, circle. In the steady-state regime, the domain wall velocity writes $v_{DW} = \mu_{DW}\, \mu_0 H$ where $\mu_{DW}$ is the domain wall mobility. The mobility is extracted from the linear fit of the experimental results in the flow regime (see Figure S7A, black line). In the flow regime (magnetic field lower than the Walker field), $\mu_{DW} = \gamma \Delta / \alpha$ where $\gamma$ is the gyromagnetic ratio and $\Delta$ the domain wall width. The domain wall width is estimated from the relation $\Delta = \sqrt{A/K_{eff}}$, with A the magnetic stiffness of Co and $K_{eff}$ the effective anisotropy. Here we used A = 16 pJ/m (*5*) and $K_{eff}=\mu_0 H_{k,eff} M_s/2 = 0.31 \pm 0.05$ MJ/m$^3$ using an effective anisotropy field $\mu_0 H_{K,eff}=0.43\pm0.05$ T obtained from the in-plane saturation hysteresis loop (Figure S7B) and a saturation magnetization $M_s = 1.43 \pm 0.05$ MA/m (*29, 56*). Using these parameters and assuming the domain wall dynamics is in the flow regime, one gets $\alpha_F = 0.18 \pm 0.03$. Since the damping is obtained for a thickness $t_{Co} = 1.3$ nm instead of $t_{Co}=1.6$ nm in the SAF stack, the final damping of the SAF $\alpha_{SAF}$, was estimated assuming the damping scales as $1/t_{Co}$, which leads to $\alpha_{SAF} = 0.14 \pm 0.04$. Table S1 summarizes the magnetic parameters as well as the damping and the resulting damping parameters.

| Ms (MA.m$^{-1}$) | V$_W$ (ms$^{-1}$) | $\mu_0 H_k$ (T) | $K_{eff}$ (MJ.m$^{-3}$) | $\alpha_F$ | $\alpha_{SAF}$ |
|---|---|---|---|---|---|
| 1.43 ± 0.05 | 173 ± 11 | 0.43 ± 0.05 | 0.307± 0.047 | 0.18 ±0.03 | 0.14 ±0.04 |

**Table S1 Magnetic parameters used to estimate the magnetic damping constant α$_{SAF}$ of the SAF stack.**

Similar measurements were performed in a stack with only one ferromagnetically active layer, namely Si/Ta(3)/Pt(3)/Co(t)/Ru(0.85)/Pt(0.45)/Co(0.3)/Ru(0.6)/Ta(1.5) (thickness in nm). A magnetic damping of 0.33±0.09 is found.

### 1.3 Measurement of the Dzyaloshinskii-Moriya interaction (DMI)

The DMI was measured in the same SF sample as the one used in section 1.2 using domain wall dynamics experiments in the presence of an in-plane magnetic field (*58, 59*). In these experiments, circular magnetic domains are expanded by out-of-plane magnetic field pulses in the presence of a static in-plane field Bx (Figure S8A). The DMI can be seen as a longitudinal field (H$_{DMI}$) localized on the Néel DW, with opposite directions for up/down and down up DWs (*49*). The velocity of the DW is minimal when the in-plane field compensates the DMI field, i.e when the DW gets a Bloch DW configuration (*58*). The velocity of the DW vs B$_x$ is shown in Figure S8B. The minimum of the curve allows the determination of $\mu_0 H_{DMI} = 78.0 \pm 7.7$ mT. Since $\mu_0 H_{DMI} = \frac{D_s}{M_s t \Delta}$ (t is the total Co thickness), we extract $D_s = 2.02 \pm 0.84$ pJ/m and a value $D = 0.62 \pm 0.24$ mJ/m² for the thickness of the AF sample. This value is in agreement with Brillouin Light Scattering experiments performed in Pt/Co/Ru stacks (*29*).

### 1.4 Spin Orbit Torque (SOT) measurements

The spin orbit torques were measured in (i) the SF stack of composition *substrate/Ta(3 nm)/Pt(3 nm)/Co(1.58 nm)/Ru(1.4 nm)/Pt(0.45nm)/Co(1.58 nm)/Ru(0.6 nm)/Ta(1.5 nm)* where both Pt/Co layers are ferromagnetically coupled and (ii) a stack that would mimic the bottom layer of the SAF *(cf Figure S6C)*. The SOT were measured using the second harmonic method described in Ref. (*60*). The method is based on the measurement of the first and second harmonic response of the anomalous Hall voltage and planar Hall voltage to a small alternating current (10 Hz). To this end, the sample was patterned into 30 µm wide Hall crosses. The setup allows us to rotate the sample 360° around the normal to the sample plane while an external field is applied. Two different measurements were performed: in-plane (IP) and out-of-plane (OOP) angle scans. The OOP angle scans are used to extract the anisotropy field and the IP scans allow us to extract the torques. By systematic angle scan measurements at different external magnetic fields, we are able to extract the effective anisotropy field parameters $\mu_0 H_K = 56 \pm 19$ mT for the SF stack and $\mu_0 H_K = 35 \pm 14$ mT for the bottom layer of the SAF stack.

The second harmonic planar Hall resistance $R_H^{2f}$ includes the contribution of the damping like (DL) $R_{DL}^{2f}$, field like (FL) $R_{FL}^{2f}$, and thermal components $R_{\nabla T}^{2f}$ resistance: $R_H^{2f} = R_{DL}^{2f} + R_{FL}^{2f} + R_{\nabla T}^{2f}$. The different contributions can be separated from their distinct angular symmetries. The damping-like and thermal components have the same dependence on the in-plane angle ϕ, while the FL has a different one:

$$R_{DL}^{2f} + R_{\nabla T}^{2f} \sim \cos \phi_M$$
$$R_{FL}^{2f} \sim 2 \cos^3 \phi_M - \cos \phi_M$$

Figure S9A and B are showing an example of an in-plane angle scan measurement of the first $R_H^f$ and second harmonic $R_H^{2f}$ planar Hall resistances as a function of the IP angle field orientation ($\phi$). The component $R_{DL}^{2f} + R_{\nabla T}^{2f}$ and $R_{FL}^{2f}$ can then be extracted from their in-plane field angular dependence (Figure S9C and D). One can show (*60*) that $R_{FL}^{2f} \propto 1/H_{ext}$ and $R_{DL}^{2f} + R_{\nabla T}^{2f} \propto 1/(H_{ext} - H_K)$ where $H_{ext}$ is the applied magnetic field.

By systematic measurements of these contributions for different external IP magnetic field amplitudes, we obtained the dependence of $R_{FL}^{2f}$ on $1/H_{ext}$ and $R_{DL}^{2f} + R_{\nabla T}^{2f}$ on $1/(H_{ext} - H_K)$ (Figure S10A and Figure S10B, dots). The DL and FL components of the torques are extracted from the slope in the linear regime (red lines) (Figure S10A and Figure S10B). The model assumes a macrospin model, *i.e* the magnetization is saturated. The deviation from the linearity observed at large $1/H_{ext}$ in Figure S10 can be explained by the formation of domains at low field. The data in Figure S10 show the results for the torques for the SF system. The same procedure was followed for the bottom layer.

From these experiments, the damping like and field like spin orbit torques for the ferromagnetically coupled bilayer SF are:

| FL Torque (T/(A/m²)) | (0,53 ±0,06) ×10⁻¹⁴ |
|---|---|
| DL Torque (T/(A/m²)) | (2,21 ±0,14) ×10⁻¹⁴ |

For the bottom FM layer, the torques are

| FL Torque (T/(A/m$^2$)) | $(0.86\pm0.07) \times 10^{-14}$ |
|---|---|
| DL Torque (T/(A/m$^2$)) | $(1.70\pm0.14) \times 10^{-14}$ |

### 1.5 Summary of magnetic and transport parameters

| Parameter | Symbol | Value | Unit |
|---|---|---|---|
| **Saturation Magnetization** | $M_S$ | 1.43± 0.05 | MA m$^{-1}$ |
| **Exchange Constant** | $A_{ex}$ | 16 ±6 | pJ m$^{-1}$ |
| **DMI constant** | $D$ | 0.62± 0.24 | mJ m$^{-2}$ |
| **Anisotropy field (top)** | $H_{k,top}$ | 12.4 ±16 | mT |
| **Anisotropy field (bottom)** | $H_{k,bottom}$ | 35.4 ±13 | mT |
| **Anisotropy constant (top)** | $K_{top}$ | 1.294±0.09 | MJ/m$^3$ |
| **Anisotropy constant (bottom)** | $K_{bottom}$ | 1.31±0.09 | MJ/m$^3$ |
| **Damping like spin orbit torque field** | DL-SOT | 2.21 ±0.14 | 10$^{-14}$ TA$^{-1}$m$^2$ |
| **Field like spin orbit torque** | FL-SOT | 0.53±0,06 | 10$^{-14}$ TA$^{-1}$m$^2$ |
| **Gilbert Damping Constant** | $\alpha$ | 0.14±0.04 | |
| **Effective magnetic thickness of the Co layer** | $t_{Co}$ | 1.3±0.15 | nm |
| **RKKY Field** | $H_{RKKY}$ | 205 ±5 | mT |
| **Domain wall width** | $\Delta$ | 27±7.6 | nm |
| **Gyromagnetic ratio** | $\gamma$ | 194.8±6.3 | GHz/T |

**Table S2 Magnetic and transport parameters.**
Table S2 summarizes the magnetic and transport parameters of the SAF sample. The saturation magnetization was measured from the dependence of the magnetic moment on the Co magnetic film thickness in Pt/Co/Ru (*56*). The exchange constant was taken from Ref. (*5*). The effective magnetic thickness t of the Co layer was measured from the magnetic moment per unit area $M_s \times t$ measured by SQUID magnetometry. The domain wall width Δ is obtained from micromagnetic simulations (see below) by fitting a line scan of the z component of the magnetization along the diameter of the two skyrmions using the function $m_z(r) = 1 - \frac{4\cosh^2 c}{\cosh 2c + \cosh\left(\frac{2r}{\Delta}\right)}$ where c is the skyrmion radius (*61–63*) and averaging over the two skyrmions. The gyromagnetic ratio was measured by ferromagnetic resonance in Pt/Co/Pt and Pt/Co/Ru ultrathin films. To extract the current density, we assume that the current flows homogeneously with the SAF stack, neglecting the 1.5 nm oxidized Ta capping layer and the 3 nm Ta seed layer with large sheet resistance.

## 2  Magnetic force microscopy experiments

The MFM tip is a low moment commercial tip from Nanosensor with reference PPP-LM-MFMR. According to the manufacturer specifications, the hard magnetic coating on the AFM tip is characterized by a coercivity of approximatively 250 Oe and a remanence magnetization of app. 150 emu/cm3 (these values were determined on a flat surface) with an effective magnetic moment 0.5x of standard AFM probes, guaranteed AFM tip radius of curvature < 30 nm magnetic resolution better than 35 nm, AFM tip height 10 - 15 µm
For the current injection, the sample was patterned into 3 µm wide tracks (10 in parallel) contacted to gold electrodes using a standard nanofabrication process. For these experiments, a bipolar voltage pulse generator from Kentech with a rise time of approximatively 300 ps and a maximum voltage pulse of 80 V was used. The amplitude and full width at half maximum (FWHM) of the transmitted voltage pulse was measured for each voltage pulse application using a sampling oscilloscope with 50 Ω input impedance, which allows us to extract the current pulse amplitude. Figure S28 shows an example the (normalized) transmitted voltage pulse  and a Gaussian fit (red curve) leading to a FWHM of 0.53 ns. The average velocity $v_{av}$ of the skyrmions was calculated from the displacement ΔX of the skyrmion induced by the current pulse as $v_{av}$ =ΔX / FWHM. Note that for a Gaussian pulse $v_{av}$ relates to the maximum instantaneous velocity $v_{max}$ as $\frac{v_{av}}{v_{max}} = \frac{\sqrt{\pi}}{2\sqrt{\ln 2}} \approx 1.0645$. To measure ΔX, MFM images were corrected for drifts using a defect on the topography images.

**Current induced skyrmion motion in synthetic ferromagnetic (SF) stacks**
To evaluate the impact of the magnetic compensation on the current induced skyrmion dynamics, synthetic ferromagnetic (SF) multilayer stacks were grown. The composition of the stack was the same as the SAF stack except for the Ru thickness, which was set to 1.4 nm instead of 0.85 nm to obtain a ferromagnetic coupling:
Ta(3)/Pt(3)/Co(1.58)/Ru(1.4)/Pt(0.5)/Co(1.58)/Ru(0.6)/Ta(1.5) (thickness in nm). Hysteresis cycles display linear and reversible hysteresis showing that the sample is close to the perpendicular to in-plane reorientation transition (see Figure S12). Magnetic force microscopy images show stripe domains at remanence (Figure S23) and skyrmions can be nucleated by applying an out-of-plane magnetic field (see Figure S13A for a magnetic field $B_z$=15 mT) with an average diameter of 215 nm (standard deviation of 102 nm). To study the current induced dynamics, the SF stack was patterned into 3 µm wide track using nanofabrication tools. The track design was the same as the SAF one (see Figure S11). To study the current induced skyrmion dynamics, MFM images were acquired before and after the injection of a short (3-5 ns) current pulse. An example is shown Figure S13 where the injection of the current pulse leads to the displacement of several skyrmions in the track (see white arrows**).** The average velocity for each skyrmion is then obtained by dividing the displacement by the pulse width. These experiments were repeated for different current densities. The velocity and skyrmion Hall angle of all events are shown in Figure S14. The resulting average velocity vs current density and skyrmion Hall angle curves are shown in Figure S15 and Figure 4 of the main text.

**Current induced skyrmion motion in single ferromagnetic layer stack**

To evaluate the impact of the magnetic compensation on the current induced skyrmion dynamics, single ferromagnetic layer stacks were also grown. The composition of the stack was: Ta(3)/Pt(3)/Co(1.58)/Ru(0.85)/Ta(1.5) (thickness in nm).

Skyrmions can be nucleated by applying an out-of-plane magnetic field (see Figure S13A for a magnetic field of 15 mT) with an average diameter of 107 nm (standard deviation of 17 nm). To study the current induced dynamics, the stack was patterned into 3 µm wide track and MFM images were acquired before and after the injection of a short (3-5 ns) current pulse. An example is shown in Figure S16 where the injection of the current pulse leads to the displacement of several skyrmions in the track (see white arrows). The average velocity for each skyrmion is then obtained by dividing the displacement by the pulse width. These experiments were repeated for different current densities. The velocity and skyrmion Hall angle of all events are shown in Figure S17. The resulting average velocity vs current density and skyrmion Hall angle curves are shown in Figure S18 and Figure 4 of the main text.

**Micromagnetic simulation and analytical modelling**
### 1.1 Magnetic parameters and static micromagnetic simulations

We performed micromagnetic simulations for a ferromagnet (FM)/spacer(S)/ferromagnet (FM) trilayer structure where the FM layers are anti-ferromagnetically coupled via RKKY interaction. The magnetic parameters used for the micromagnetic simulations were measured experimentally and are given in Table S2. We assume that the value are the same for both Co layers. However, the magnetic anisotropy of the top layer was slightly increased (36 mT instead of 12 mT) within error bar of the effective anisotropy of the SAF, such that the magnetization of the domain is not reversed by SOT for the maximal current density. In addition, the DMI was slightly adjusted within error bars (D=0.76 mJ/m² instead 0.62 mJ/m²) such that the skyrmion diameter is similar to the experimental ones, ~215 nm (see section 3). Simulations are performed using the open-source finite-difference micromagnetic solver Mumax3 (*64*).

To obtain the static configuration, we initialize a Néel skyrmion in both FM layers and minimize the total energy using the in-built steepest conjugate gradient method in Mumax3. In Figure S39(a) and Figure S39 (b), we show the static magnetization configuration of a stabilized skyrmion in both FM layers ($m_z$ in color code). Figure S39 (c) shows a 1D linescan of the z component of the magnetization of the top and bottom layers along the skyrmion diameter and Figure S39(d) shows the corresponding magnetization profile in the form of vector arrows. A left-handed chiral Néel-type skyrmion is obtained in both FM layers in agreement with the sign of the DMI at the Pt/Co interface and the XMCD-PEEM magnetic microscopy experiments. Using these parameters, an average domain wall width of $\Delta=22.7\pm6.4$ nm is obtained.

### 1.2 Micromagnetic simulations of the current induced skyrmion dynamics

For the micromagnetic simulations of the current induced skyrmion dynamics, the experimental values of the SOT shown in Table S2 were used. The SOT is applied on both Co layers. To mimic experimental results, the current induced average skyrmion velocities $v_{av}$ (show in Fig4A of the main text) were deduced from the net displacement of the skyrmions $\Delta X$ using their positions before and after the application of a current pulse with FWHM of 500 ps (see Figure S28) with $v_{av}=\Delta X/\text{FWHM}$.

**Supplementary Text**
**Observation of SAF skyrmions and chiral Domain Walls using X-PEEM**

X-ray magnetic circular dichroism combined with photoemission electron microscopy was performed at the Nanospectroscopy beamline at the Elettra synchrotron, Trieste (Italy) in order to determine the chirality of the skyrmion and domain walls in the SAF sample(*8*). Figure S20A shows a XMCD-PEEM image acquired at the Co L3 absorption edge and at zero external magnetic field. The image displays a multi-domain state including an isolated SAF skyrmion with a diameter approximately of 250 nm. Although the SAF is fully compensated, a magnetic contrast is observed. This is explained by the surface sensitivity of the PEEM such that most of the photo-emitted electrons come from the top Co layer. The XMCD contrast is proportional to the projection of the magnetization on the beam direction. Since the X-ray beam impinges on the sample surface plane at a grazing angle of 16°, the contrast is approximately three times larger for the in-plane component of the magnetization than for the out-of-plane one. This feature allows the observation of the internal in-plane spin structure of DWs or skyrmions(*8*). In Figure S19A, the white and black contrast in the domains corresponds to the out-of-plane component of the magnetization, respectively down and up. However, the stronger black/white contrast observed for domain walls perpendicular to the X-ray beam is explained by the in-plane component of the DW magnetization with a Néel configuration. In Figure S19B, we plot the XMCD contrast measured along the white line in Figure S20A, which is along the X-ray beam and perpendicular to the domain walls. A peak in the XMCD contrast is observed at the DW position separating white and black domains, and a dip at the DW position separating the black and white domains. The peak means that the DW magnetization is oriented *opposite* to the in-plane component of the X-ray beam for an *up–down* DW, while the dip means that the in-plane component of the magnetization is oriented *along* the in-plane component of the beam direction for *down-up* DW. This demonstrates that the magnetization rotates as ↑←↓→↑ corresponding to a left handed Néel chirality of the DW, which is in agreement with the sign of the DMI expected in Pt/Co/Ru interfaces(*29*). Another example is shown in Figure S20. In Figure S20B, a peak in the dichroic contrast is observed for black/white DWs along the red line in Figure S20B, and a peak for white/black domain walls. We also performed linescans along DWs which are parallel to the beam direction. We show for example in Figure S20C the linescan along the blue line in Figure S20A, crossing a domain wall parallel to the X-ray beam. No peak or dip are observed, showing that the domain wall magnetization is perpendicular to the X-ray beam direction, and thus to the domain wall surface, in agreement with its Néel configuration. We note that we only observe the top Co layer. Nevertheless, the AF coupling and Pt/Co interface of the bottom layer promotes DWs of the same chirality in both FM layers so a left-handed Néel spin texture is also expected in the bottom Co layer.

**Magnetic contrast in the magnetic force magnetic microscopy experiments**

We discuss in this section the magnetic contrast observed in the magnetic force microscopy experiments. Figure S21A reproduces Fig. 2C of the main paper, which shows an MFM image of domains and skyrmions measured at zero field. We observe a domain contrast, which can be accounted for the perturbation of the domain magnetization by the stray field of the magnetic tip (*65–67*). This effect is expected to scale as $1/K_{eff}^2$ where $K_{eff}=K-\mu_0 M_s^2/2$ is the effective perpendicular magnetic anisotropy (*65, 66*). Hence, the small $K_{eff}$ in our sample favors this domain contrast despite the relatively low magnetic moment of the tip.

In addition, a larger dark contrast is observed at the domain wall positions. Figure S21B shows the MFM contrast along a linescan crossing HH/TT and TT/HH domain walls (black line in Figure S21A) where HH and TT stands for head-to-head and tail-to-tail SAF configuration. A dip in the MFM contrast is observed at both domain wall positions. Similarly, two successive dips are observed when crossing magnetic skyrmions (see Figure S21C).

To better understand these results, we performed micromagnetic simulations of the magnetic contrast using the MFM module of the Mumax3 code (*64*) and the parameters of Table S2. Figure S22 (A-B) shows the MFM contrast of a domain wall assuming the tip behaves as a point monopole (Figure S22A), with a tip distance of 20 nm. A black/white contrast is observed around the domain wall. The corresponding contrasts along a line scan crossing the domain walls is shown in Figure S22 B. The anti-Lorentzian shape (succession of a peak and a dip) was already reported in the literature for perpendicularly magnetized SAF domain walls (*68*, *69*) and can be explained by the different distance of the two AF coupled domain walls from the tip. Corresponding simulations of skyrmion profiles are also shown in Figure S22(C-D).

Dips in the MFM contrast for DWs were already reported in perpendicularly magnetized SAF (*68*, *70*, *71*). They were accounted for a shift of the DWs in the AF coupled layers leading to the creation of a small perpendicularly magnetized domain where all layers are coupled ferromagnetically. The creation of this ferromagnetic domain is explained by a gain in the stray field energy at the cost of an increased RKKY energy. The ferromagnetic domain size $\delta$ results from a balance between these two energies and scale as (*71*) $\delta \sim M_s^2 t^2 / J_{RKKY}$, with t the layer thickness, and $J_{RKKY}$ the interlayer coupling.

However, our micromagnetic simulations (see Figure S39C) do not predict a ferromagnetic domain at the domain wall position: only a tiny displacement (~ nm) of the domain wall. Thus this cannot explain the systematic dips observed experimentally in the MFM contrast.

A possible explanation for the observed domain wall contrast is that the formation of a small ferromagnetic domain at the DW position is favored by the magnetic tip stray field. This would explained that the domain wall always appear as a dip, namely that the DW magnetization is aligned along the tip magnetization. A similar feature was reported in Ref. (*68*) (see Figure 15 of Ref. (*68*)). This scenario is further supported by the comparison of the MFM contrast amplitude of domain walls in synthetic antiferromagnetic and ferromagnetic stacks. Figure S23(A,C) shows a strip domain texture measured at remanence by MFM in the SF stack described in section 0. Linescans of the phase contrast show that worm domains lead to a peak in the MFM phase with an amplitude of around 0.07-0.08°, which is similar to the ones measured for domain walls in SAF stack, which are around 0.05° (Figure S21). We note however that the amplitude and shape of the stray field measured by scanning NV magnetometry imaging experiments (see next section) are in agreement with the skyrmion profile obtained from the micromagnetic simulations, which is nearly fully compensated. Since NV magnetometry is non-perturbative, this further supports that the formation of this ferromagnetic domain at the domain wall position is favored by the tip stray field.

**Scanning NV magnetometry imaging experiments**

In order to evaluate the effect of the magnetic moment of the MFM tip on the magnetic texture, we performed additional scanning nitrogen vacancy (NV) magnetometry imaging for comparison. Scanning NV magnetometry is a non-perturbative technique which measures quantitatively the magnetic stray field produced by the magnetic texture by monitoring the Zeeman shift of the magnetic resonance of a single NV center in diamond. The comparison of the MFM and NV images shown in Figure S24A and B demonstrates that, although the MFM tip may lead to the formation of a small ferromagnetic domain wall, it has a negligible effect on the observed magnetic bubble shape, which is similar in both experiments.

In addition, since the NV magnetometry measurement is quantitative, we can compare the stray field map with calculations of the stray field produced by a skyrmion of diameter 200 nm whose profile can be determined using the micromagnetic parameters detailed in Table S2. Figure S24C presents the simulated stray field map produced by such a skyrmion at 36 nm from the sample surface, which is the height of the NV center in the experiment from Figure S24B. The magnetic field is also projected along the direction of the NV center quantification axis in order to allow a direction comparison. We find a good agreement between the simulation and the data on the small skyrmion at the top of the image, indicating that the SAF is compensated and that the computed skyrmion profile corresponds to the actual skyrmion profile.

**Additional data from magnetic force microscopy experiment**

### 1.3 Skyrmion nucleation induced by external magnetic fields

In this section, we present some additional examples of skyrmion nucleation using the external magnetic field. Figure S25 shows MFM images measured at zero applied magnetic field obtained after the nucleation procedure. A large negative magnetic field of -0.7 T was first applied perpendicularly to the sample plane. Then, positive out-of-plane magnetic fields of +220 mT and +175 mT were applied sequentially. This leads to the formation of isolated domains as well as of skyrmions (Figure S25A). Further application of an out-of-plane magnetic fields of 185 mT leads to the formation of small isolated skyrmions (Figure S25B) measured at zero applied magnetic field). The sensitivity of SAF skyrmions to external fields can be explained by the uncompensated moments in the AF coupled DWs (*68*, *71*) that act as nucleation centers for magnetization reversal in SAFs. Note that the nucleation procedure using magnetic field is stochastic and varies from sample to samples, which may be explained by a thermally activated process sensitive to defects. More deterministic nucleation of SAF skyrmions can be achieved using local current injection as well ultrafast laser excitation as show in Ref. (*29*).

### 1.1 Impact of the skyrmion size on the skyrmion dynamics

The diameter of the skyrmion is extracted from the area of the MFM contrast as seen in the inset of Figure S26A. The distribution of the diameter of the skyrmions that were moved by current pulses is shown in Figure S26A. The average value of skyrmion diameter is 215± 102 nm. Larger skyrmion diameter, above 300 nm, may be explained by a predominant effect of pinning on the skyrmion size (*37*, *38*). The variation of the skyrmion size before and after the current pulse is shown in Figure S26A. Excluding the few events with large deformation (>100%), the average deformation is small (5%), meaning the current pulse leads nearly equally to a decrease or an increase of the skyrmion size. Since it is expected that the current induced torques leads to an enlargement of the skyrmion diameter, this points to skyrmion pinning as the main origin of the observed deformation. The standard deviation of the skyrmion diameter deformation is found to be 35%. Thus, although large current induced skyrmion deformations are observed, these are uncommon events

We plot in Figure S27A the velocity vs the skyrmion diameter for current densities between $7.5\times10^{11}$ A/m² and $8.9\times10^{11}$ A/m². No dependence of the average velocity is observed for diameters above 300 nm, while a small decrease is noticed for skyrmion diameters below 300 nm. These measurements are compared to the prediction of the Thiele model using experimental parameters,: $v_m \approx \frac{\pi\mu_0 H_{DL}\gamma R}{2\alpha(\frac{R}{\Delta}+2\exp(-\frac{R}{2\Delta}))}$ (see section 1.7), shown in red line in Figure S27A. As pointed out in the main text, the model overestimates the velocity. However, it also shows that in our material stack, the velocity depends little on the diameter for diameters above 250 nm, with $v \approx \frac{\pi\mu_0 H_{DL}\gamma\Delta}{2\alpha}$ for $R \gg \Delta$, in agreement with the experimental data. The small decrease of the skyrmion velocity observed experimentally for diameters below 300 nm is also accounted for by the model. Figure S27B compares the velocity for skyrmion diameters below 400 nm (red points) and for all skyrmion diameters (black points). No significant difference is observed.

### 1.2 Additional experimental data on current-induced skyrmion motion

We present in this section additional experimental results on current-induced skyrmion motion. In Figure S29, we show examples of current induced skyrmion motion events of small isolated skyrmions at zero external magnetic field. In Figure S29A, a 0.8 ns current pulse with density J=7.1x10$^{11}$ A/m² leads to a displacement of around 310 nm of an isolated skyrmion with a diameter of around 90 nm in the current direction. This corresponds to an average velocity of around 380 m/s. In Figure S29B, the back and forth motions of a 125 nm diameter skyrmion induced by 0.8 ns current pulses with successive positive and negative current polarities are shown (J=7x10$^{11}$ A/m²). A minimal skyrmion Hall deviation is observed. An average velocity of respectively 370 m/s and 560 m/s is deduced from the skyrmion displacement.

In Figure S30, we show another example of a current induced skyrmion motion event where a skyrmion moves around 750 nm after the injection of a 1.1 ns current pulse of density 8.65×10$^{11}$ Am$^{-2}$. The skyrmion is moving along the current direction.

Figure S31 and Figure S32 show other examples of current induced skyrmion motion. Figure S33 shows an example of the current induced motion of larger elongated skyrmions. Such events were not included in the skyrmion velocity statistics due to the significant error in the measurement of the center of mass of the skyrmion compared to the skyrmion displacement. Figure S34 shows other examples of current induced skyrmion motion events with very high velocities. Here 4 skyrmions are moved back and forth by injecting successively positive and negative current pulses. A significant deformation of the skyrmion is observed after being moved by the current pulse, which is explained by pinning on inhomogeneities and/or dynamical deformation of the skyrmions [12]. The injection of the first current pulse of width 0.53 ns and density 8.85×10$^{11}$ A/m² (Figure S34 top-middle) leads to a displacement of all the skyrmions in the current direction, respectively by 795 nm (dotted red line), 639 nm (dotted blue line), 537 nm (dotted white line), 353 nm (dotted black line), that corresponds to velocities of 1505, 1210, 1017, and 668 m/s, respectively. The domain wall on the right is also moving by 760 nm corresponding to a velocity of 1440 m/s. Note that the injection of higher current density/longer pulses leads to the nucleation of multiple domains, which can be explained by Joule heating and/or spin orbit torques.

Figure S35 shows the same images as Figure 3 in the main text but with a larger field of view such that topological defects can be seen on the images.

Figure S36 shows the velocities extracted from all the current induced skyrmion motion events for both positive and negative current densities. The experimental velocity vs current density curves of Figure 4 of the main text was obtained by averaging the velocity over events with the same current velocity with: 3 events averaged for J = 5.0 × 10$^{11}$ A/m²; 5 events averaged for J = 5.4 × 10$^{11}$ A/m², 14 events averaged for J = 6.2 × 10$^{11}$ A/m²; 6 events averaged for J = 6.9 × 10$^{11}$ A/m²; 6 events averaged for J=7.5 × 10$^{11}$ A/m², 8 events averaged for J = 7.8× 10$^{11}$ A/m², 47 events averaged for J = 8.1 × 10$^{11}$ A/m², 28 events averaged for J = 8.5 × 10$^{11}$ A/m², 12 events averaged, for J = 8.9 × 10$^{11}$ A/m².

Figure S37 shows the skyrmion Hall angle of all events of current induced skyrmion motion and their average vs current density and velocity. Figure S38 compares the velocities for positive and negative current pulses, where no difference is observed as expected.

**Additional micromagnetic simulations and analytical modelling data**
### 1.3 Static and dynamical topological charge distribution

In Figure S40, we show the topological charge density maps of the top and bottom FM layers for static as well as moving skyrmion configurations. Mathematically, the topological charge density is defined as $\boldsymbol{\eta} = \boldsymbol{m} \cdot (\frac{\partial \boldsymbol{m}}{\partial x} \times \frac{\partial \boldsymbol{m}}{\partial y})$. The topological charge Q can then be calculated as the spatial integral of η, $\boldsymbol{Q} = \frac{1}{4\pi} \iint \boldsymbol{\eta} d\boldsymbol{x} d\boldsymbol{y}$. Figure S40(a) and (b) show the topological charge density for static skyrmion configuration at t=0 ns. It can be seen that the topological charge density has opposite signs for the top and bottom layers, as expected, with highest amplitude in the skyrmion domain wall region. However, the maps are not completely anti-symmetric due to different anisotropy values in the top and bottom layers that modifies the domain wall width of the skyrmion. The numerically calculated topological charges for top and bottom layers are -0.998 and 0.997, respectively. Figure S40(c) and (d) show the topological charge density map for the top and bottom FM layers, of the moving skyrmion for a DC current pulse of $J=5.6\times10^{11}$A/m$^2$ at t=0.53ns. Due to the skyrmion deformation, the topological charge density no longer contained symmetrically in the domain wall region but is concentrated at some specific points. However, the cancellation of the topological charge is still maintained with a charge Q for the top and bottom layers of -0.9998 and 0.9996, respectively.

### 1.4 Comparison of the instantaneous and average velocities

Since the average skyrmion velocity $v_{av}$ is measured experimentally from the net-displacement as $v_{av}=\Delta X/t_{pulse}$, one may wonder how $v_{av}$ compares to the instantaneous velocity. We plot in Figure S41 micromagnetic simulations of the instantaneous velocity for $J=1\times10^{11}$ A/m$^2$ as a function of time for a DC square pulse of $t_{pulse}=2$ ns. The instantaneous velocity reaches a plateau after around 200 ps indicating inertial effects. We also show the average velocity calculated from the net-displacement as $\Delta X/t_{pulse}$ with dashed black line. We find that the average velocity is the same as the steady-state velocity.

We also studied the impact of the current density amplitude. Figure S42a shows the instantaneous velocity of the skyrmion for Gaussian current pulses with different densities. The shape of the applied current pulse is also shown alongside using dashed pink line (right y-axis). Due to the inertial effects, the maximum instantaneous velocity for each case does not coincide with the maximum of the Gaussian pulse. The maximum value of the instantaneous velocity as a function of current density is shown in Figure S42B together with the corresponding average velocity calculated from the skyrmion displacement. The lines refer to the analytically calculated velocities. While the average and maximum velocities coincide at lower current density, the maximum velocity is significantly smaller at the larger current density which may be attributed to the deformation of the skyrmion. The deformation also explains the sublinear drop in the simulations compared to the model, as outlined in the main text.

We also studied the impact of inertial dynamics on the skyrmion dynamics for current pulses with a width similar to the relaxation time. Figure S43a shows the average skyrmion velocity as a function of the pulse width (FWHM) ranging from 100 ps to 530 ps for $J=4\times10^{11}$ A/m² and $J=8.9\times10^{11}$ A/m². For $J=4\times10^{11}$ A/m², little dependence of the velocity on the pulse width is observed. Thus, the inertial dynamics affect little the average velocities deduced from the skyrmion displacement. On the other hand, for $J=8.9\times10^{11}$ A/m², the velocity increases as the pulse width is decreased. As discussed below, this can be attributed to the lower skyrmion deformation at shorter pulse.

In Figure S43b, we show the effect of the skyrmion size on the velocities by performing simulations using two different DMI values. A smaller DMI leads to a smaller skyrmion diameter. We find that the velocity is larger for the bigger skyrmions consistent with the 1D-model presented in the main paper.

### 1.5 Micromagnetic simulations of the current induced skyrmion deformation

This section discusses the current induced dynamical deformation and its impact on the skyrmion dynamics. As underlined in the main text, the simulations reveal that during the pulse injection, an expansion along with an elliptical deformation perpendicular to the skyrmion motion occur, that becomes more pronounced with time and increasing current, with the appearance of DW like sections at large current (see figure 5 of the main text). Figure

S44A shows the time dependence of the average skyrmion diameter (black line) during the injection of a 0.5 ns current pulse for a current density of $8.9\times10^{11}$ A/m² (show as dotted pink line). The diameter increases during the pulse injection and features inertial dynamics, the maximum in diameter being observed around 250 ps after the current pulse maximum. The elliptical deformation also increases with the current density (see Figure S44B, blue symbols for the dimension along the major and minor axes). Interestingly, the major axis of the deformation is not perpendicular to the current, as observed in previous works on AF skyrmion dynamics (*16*, *19*, *21–24*, *48*) but slightly tilted. The elliptical deformation is concomitant with a sublinear drop of the skyrmion velocity (around $4\times10^{11}$ A/m², Figure S44B). The deformation is also accompanied by a rotation of the DW magnetization away from its Néel orientation in both layers, which increases with the current density (see Figure S44C).

This behavior is similar to the one observed in SOT driven FM DW motion, where the SOT induced spin rotation decreases the resultant driving force on the DW and eventually leads to a saturation of the DW velocity at high current drives (*49*). This picture is confirmed by simulations of the DW dynamics in the SAF stacks, which show that the skyrmion velocity is close to the DW one (see Figure S45A) as well as the DW magnetization angle (see Figure S45B). Thus, the observed velocity saturation can be explained by a transition to a DW like dynamics at larger drives induced by the large skyrmion deformation.

This behavior was also reported for FM skyrmions(*44*), but is unexpected in AFM, where the AF exchange torque between the sublattices is expected to prevent the current induced DW spin rotation. This is not the case in our SAF stacks due to the low RKKY exchange interaction (RKKY field of around 205 mT). However, a larger velocity can be achieved by increasing the RKKY interaction. Figure S46A shows that the velocity monotonously increases with the RKKY interaction (930 m/s for $H_k$=950 mT instead of 680 m/s for $H_k$=200 mT), which can be achieved by stack engineering. Another way to limit the velocity saturation is to contain the deformation using shorter current pulses: Figure S46B shows that the velocity increases and the skyrmion diameter decreases when decreasing the pulse width from 500 ps to 100 ps.

Figure S47 compares the velocity (A) domain wall angle (B), and skyrmion dimensions along the major (C) and minor axes (D) vs current density for (i) a 0.5 ns current pulse and a RKKY field $\mu_0H_{RKKY}$=205 mT (red square), (ii) a larger RKKY interaction of $\mu_0H_{RKKY}$=410 mT (blue circles) and (iii) a shorter current pulse of 0.1 ns (black triangles). Using shorter pulse width allows a significant reduction of the elliptical deformation of the skyrmion (see snapshot in Figure S47F) as well as the rotation of the DW magnetization angle. A larger RKKY interaction affects little the lateral deformation of the skyrmion (see snapshot Figure S47G), but significantly decreases the DW magnetization rotation. Thus a significant enhancement of the skyrmion velocity can be achieved by increasing the RKKY interaction as well as using a shorter pulse width thanks to the reduced DW magnetization rotation as well as reduced skyrmion deformation.

### 1.6 Micromagnetic simulations of the impact of the RKKY interaction on the inertial dynamics in SAF skyrmions

To investigate the impact of the RKKY interaction on the inertial dynamics of the SAF skyrmion observed in Fig5B of the main text, we performed simulation of the SAF skyrmion velocity in response to a square pulse for different RKKY field ($H_{RKKY}$) (see Figure S48A and B for a zoom on the rise/fall of the velocity). One observes that the larger the RKKY interaction, the smaller the rise/fall time of velocity. The rise and fall time constant $\tau$ for each $J_{RKKY}$ and for each of the skyrmions composing the SAF can be extracted by assuming an exponential fall/rise time dependence. Figure S48C shows that the average $1/\tau$ increases approximately linearly with $H_{RKKY}$ in agreement with the expectation of the Thiele model (see below). This demonstrates the clear link between the inertial behavior seen in the rise and fall of the SAF skyrmion speed and the RKKY exchange coupling. Phenomenologically, when the SOT pulse triggers the SAF skyrmion motion, the Magnus effect splits the AF coupled FM skyrmions apart since they have opposite core polarity. The RKKY exchange coupling is a recovering force that compensates for this splitting and stores energy, until a steady state is reached which explains the inertia seen in our simulations. We show in the next section that this inertia can be described from the Thiele equations of two skyrmions coupled by the RKKY exchange interaction. Note that a similar analysis was recently published in Ref.(*54*).

### 1.7 Thiele model of the SAF skyrmions dynamics

To model the dynamics of the SAF skyrmion, we consider the Thiele model, which assumes that the skyrmion spin texture is rigid during its motion (*42*). We describe the dynamics of the SAF skyrmions as two AF skyrmions (Sk1, Sk2) coupled by the RKKY exchange interactions:

$$F_{SOT}^1 + G_1 \times v_1 - k(R_1 - R_2) - \alpha D_1 v_1 = 0$$
$$F_{SOT}^2 + G_2 \times v_2 - k(R_2 - R_1) - \alpha D_2 v_2 = 0$$

Here $R_i$ the position of the skyrmion i, $v_i$ the velocity, $D_i$ the dissipative tensor, $G_i$ the gyrovector, $F_{SOT}^i$ the force due to the DL-SOT and k is a spring constant that describes the restoring force between the two skyrmions due to the RKKY exchange interaction $J_{RKKY}$. We write $d = (R_1 - R_2)$, $F_{SOT} = F_{SOT}^1 + F_{SOT}^2$, $\Delta F_{SOT} = F_{SOT}^1 - F_{SOT}^2$, $v_m = \frac{(v_1+v_2)}{2}$, $\Delta v = v_1 - v_2$. We assume $D_1 = D_2 = D$ and since the two skyrmions have opposite core polarities, $G_1 = -G_2 = G u_z$. Then

$$F_{SOT} + G \times \Delta v - 2\alpha D v_m = 0 \quad (1)$$
$$\Delta F_{SOT} + 2G \times v_m - 2kd - \alpha D \Delta v = 0 \quad (2)$$

For a Néel skyrmion and a current flowing along x, $F_{SOT} = F_{SOT} u_x$ and assuming $D$ is diagonal with $D_{xx}=D_{yy}=D$, (3) and (4) lead to

$F_{SOT} - G\Delta v_y - 2\alpha D v_{m,x} = 0$ (3)
$G\Delta v_x - 2\alpha D v_{m,y} = 0$ (4)
$2G v_{m,x} - 2k d_y - \alpha D \Delta v_y = 0$ (5)
$\Delta F_{SOT,x} - 2G v_{m,y} - 2k d_x - \alpha D \Delta v_x = 0$ (6)

*Steady state*
In the steady state, the skyrmions have the same speed thus $\Delta v = 0$. This leads to the solution:

$$v_m = \frac{F_{SOT}}{2\alpha D}$$

Note that this result is equivalent to considering a single SAF skyrmion with G=0 and a Thiele equation: $F_{SOT} - 2\alpha D v_m = 0$ (Note here that D is the dissipative tensor for 1 skyrmion in the SAF).

$F_{SOT}$ and D write (*13*) $\mathbf{F_{SOT}} \approx \mu_0 M_s t \pi^2 H_{DL} R u_x$ and $D \approx \frac{2\pi M_s t'}{\gamma}\left(\frac{R}{\Delta} + \frac{\Delta}{R}\right)$ assuming R>>Δ. Here t is the film thickness, t'=t/2, R the skyrmion radius, $\mu_0 H_{DL}$ the damping like effective field. Thus, the steady state velocity writes for R>>Δ

$$v_m = v_{m,x} \approx \frac{\pi B_{SOT} J R \gamma}{2\alpha \left(\frac{R}{\Delta} + \frac{\Delta}{R}\right)}$$

This expression was used in Figure S4, red line, of the main paper using the parameters of Table S2.

An expression valid for smaller R can also be obtained using (72) $D \approx \frac{M_s t'}{\gamma}(\frac{2\pi R}{\Delta} + 4\pi \exp\left(-\frac{R}{2\Delta}\right))$, which leads to

$$v_m \approx \frac{\pi \mu_0 H_{DL} \gamma R}{2\alpha(\frac{R}{\Delta} + 2\exp\left(-\frac{R}{2\Delta}\right))}$$

*Inertial dynamics*

Combining (3) and (5) and writing $\Delta v_y = \frac{d\, d_y}{dt}$

One gets

$$\left(F_{SOT} + \frac{\alpha D}{2k}\frac{dF_{SOT}}{dt}\right) - 2\alpha D v_{m,x} = m \frac{dv_{m,x}}{dt} \quad (7)$$

Where $m = \frac{G^2}{k}\left(1 + \left(\frac{\alpha D}{G}\right)^2\right)$

Thus, the RKKY exchange interaction leads to an additional inertial term in the Thiele equation with an inertial mass m and characteristic time constant $\tau = \frac{m}{2\alpha D} = \frac{G^2}{2k\alpha D}\left(1 + \left(\frac{\alpha D}{G}\right)^2\right)$. For circular skyrmions with no deformation and small lateral shift d, one can show that $k \approx 2\pi J_{RKKY}\left(\frac{R}{\Delta} + \frac{\Delta}{R}\right)$, with R the skyrmion radius and $\Delta$ the domain wall width. Thus, the inverse of the relaxation time scales as $\frac{1}{\tau} \sim k \sim J_{RKKY}$, a result in agreement with the micromagnetic simulations. Note that during the review process, Panigrahy et al. reported similar results on the skyrmion inertia in synthetic antiferromagnets (*54*).

## 1.8 Topological spin Hall torque

We explore numerically the effect of a topological spin Hall torque in an effort to explain the high speed measured for AF skyrmions (*22, 73–75*). The resulting torque writes (*22*) $\frac{d\boldsymbol{m}}{dt} = -b_j \lambda^2 N_{xy}\partial_y \boldsymbol{m}$ with $b_j = \frac{\mu_B JP}{M_s q}$ where P is the spin polarization, q the electron charge, $\mu_B$ the Bohr magneton, J the applied current density and $\lambda$ a parameter describing the amplitude of the topological spin Hall torque (*22, 73*).

The resulting Thiele force can be expressed as:

$$F_{tst,x} = -\frac{M_s t}{\gamma} b_j \lambda^2 \int\int ((\partial_x \boldsymbol{m} \times \partial_y \boldsymbol{m}) \cdot \boldsymbol{m})^2 \, dx\, dy$$

The steady state motion of AF skyrmion writes:

$$v_{sts,x} = \frac{F_{tst,x}}{2\alpha D}$$

We perform a numerical integration of the torque to estimate its effect on an AF skyrmion speed. We vary the core radius and domain wall width of the skyrmion and estimate the speed for a range $\lambda^2$ values from 3 to 50 nm² (*22, 73*) assuming J=8×10¹¹ A/m² and P=0.5. The skyrmion spin texture is expressed as:

$$\theta(r) = 2\operatorname{atan}\left(exp\left(\frac{r-R}{\Delta}\right)\right)$$

and the magnetization is expressed in polar coordinates $(r, \theta, \varphi)$:

$$\boldsymbol{m} = \begin{pmatrix} \sin(\theta(r))\cos(\varphi) \\ \sin(\theta(r))\sin(\varphi) \\ \cos(\theta(r)) \end{pmatrix}$$

This leads to a Néel skyrmion spin texture with left handed domain wall. We perform a numerical integration of the force resulting from the topological spin Hall torque for $\Delta$=24.5 nm with a range of R/$\Delta$ ratio from 1 to 5. We compute the speed of the skyrmion and compare it with the DL-SOT speed of an AF skyrmion according to Thiele equation. The results are plotted in Figure S49. We see that the effect of the topological torque is negligible for large skyrmions, in agreement with literature. However, we also see that there is a strong increase in speed as R/$\Delta$ gets close to 1. Moreover, this increase in speed due to the topological torque is proportional to $\lambda^2$ which shows that a large $\lambda^2$ can lead to very high speed even for small skyrmions.

**Supplementary Figures**

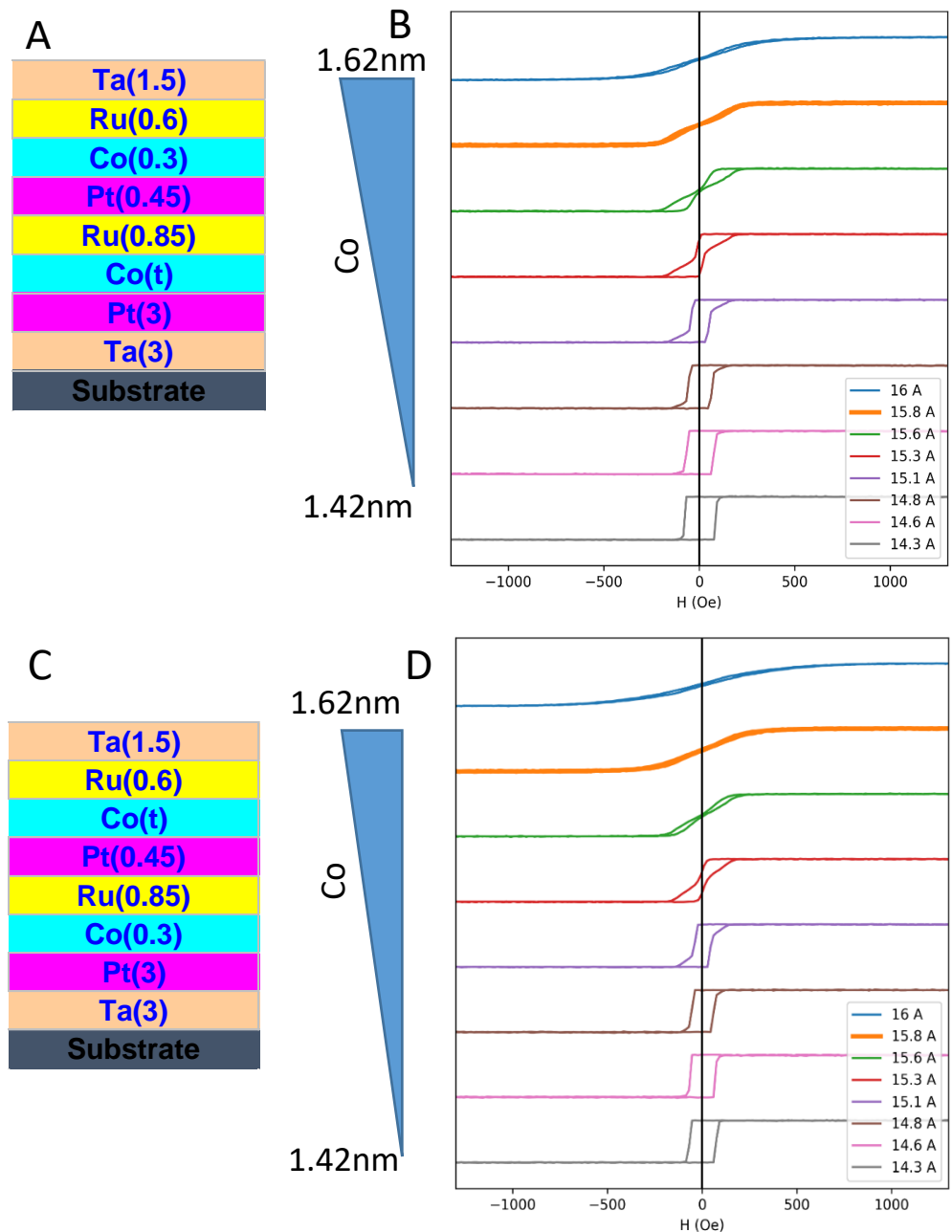

**Figure S6 Thickness dependence of the magnetic properties of the top and bottom Co layer of the SAF stacks** (A, C): Schematic representations of the stacks used to characterize the magnetic properties of the bottom (A) and top layer (C) constituting the SAF stacks. The thicknesses are in nanometer. The Co(t) layer is deposited as a wedge, such that its thickness varies along the 100 mm Si wafer. (B) Magneto-optical Kerr effect (MOKE) hysteresis loops (magnetic field applied perpendicularly to the sample plane) of the stack A (bottom FM layer) measured at different positions along the 100 mm Si wafer. The different positions correspond to different thicknesses of the Co layer. The curves are offset for clarity. (D) Same for the stack in C (top FM layer). The chosen thickness for experiments is highlighted with the thicker line in orange (15.8A).

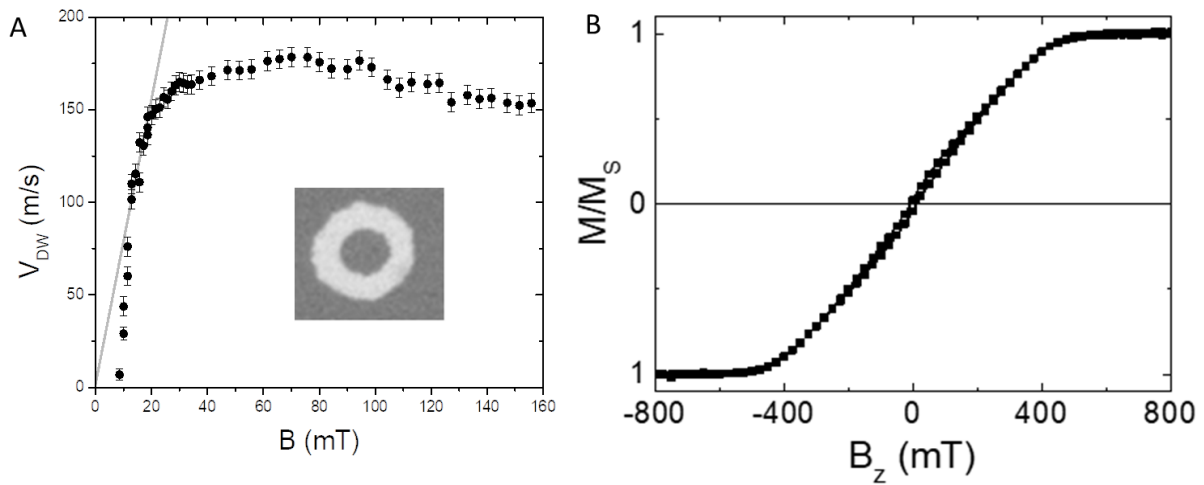

**Figure S7 Field-induced DW dynamics experiments for magnetic damping measurements** (A) DW velocity as a function of the applied out-of-plane magnetic field measured at room-temperature. The blue dots denote the experimental data and the solid line is a linear fit in the flow regime. The error bars denote the standard deviation. (Inset) Differential pMOKE image shows the expansion of the magnetic domain after applying the applied out-of-plane magnetic field pulse. (B) Magnetic moment vs. the external in-plane field measured by vibrating sample magnetometry. The extracted anisotropy field is $\mu_0 H_K = 0.43 \pm 0.05$ T.

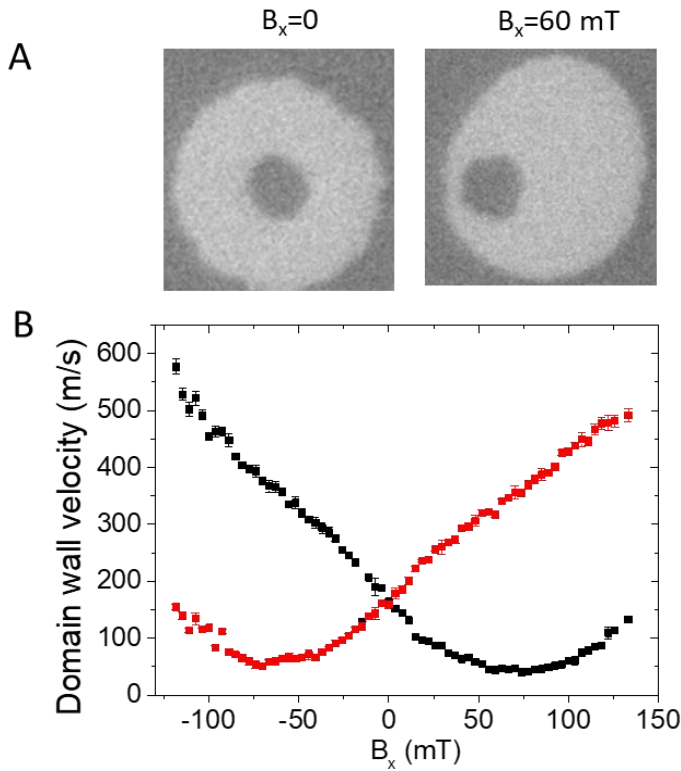

**Figure S8 Field-induced domain wall dynamics experiments to determine $H_{DMI}$.** (A) Differential pMOKE images showing the expansion of the circular domain (white contrast) induced by a magnetic field pulse of 57 mT for a constant in-plane field Bx = 0 (left) and Bx = 60 mT (right) (B) Domain wall velocity vs. in plane field $B_x$ for a perpendicular magnetic

field pulse $B_z= 57$ mT and a pulse width of 30 ns for up/down and down/up domain wall (black and red curves).

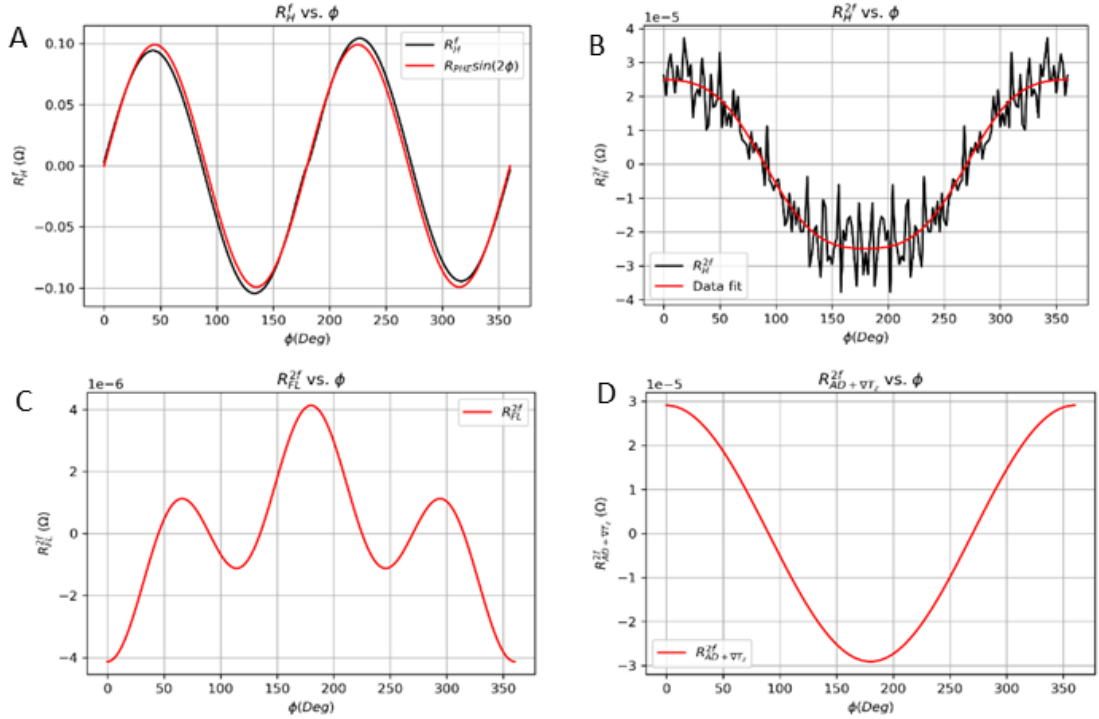

**Figure S9: Second harmonic planar Hall effect measurements of the spin orbit torque** (A-B) First (A) and second (B) harmonic Hall resistance vs the in-plane field angle orientation $\phi$ (B) Second harmonic planar Hall resistance (black) and the fit (red) vs the field angle. (C) Field like and (D) Damping like and thermal contributions of the second harmonic resistance as a function of $\phi$. These measurements are performed for an external field of 1.3T.

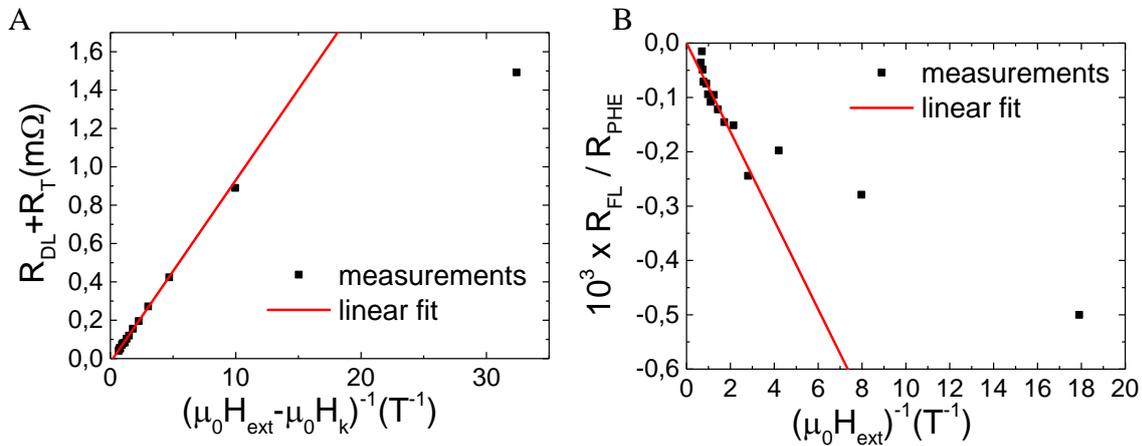

**Figure S10: Fit to extract the SOT** (A) $R_{DL} + R_T$ vs $1/(\mu_0 H_{ext} - \mu_0 H_k)$ and (B) $10^3 R_{FL}/R_{PHE}$ vs $1/(\mu_0 H_{ext})$ for the ferromagnetically coupled bilayer system (note the presence of $\mu_0 H_k$, the magnetic anisotropy field).

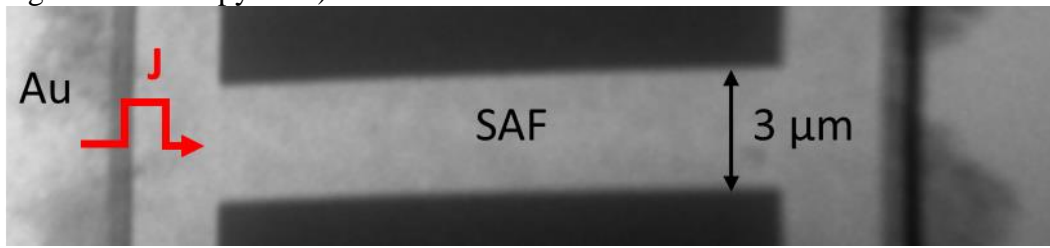

**Figure S11 X-PEEM image of a track for the current induced skyrmion motion experiments**

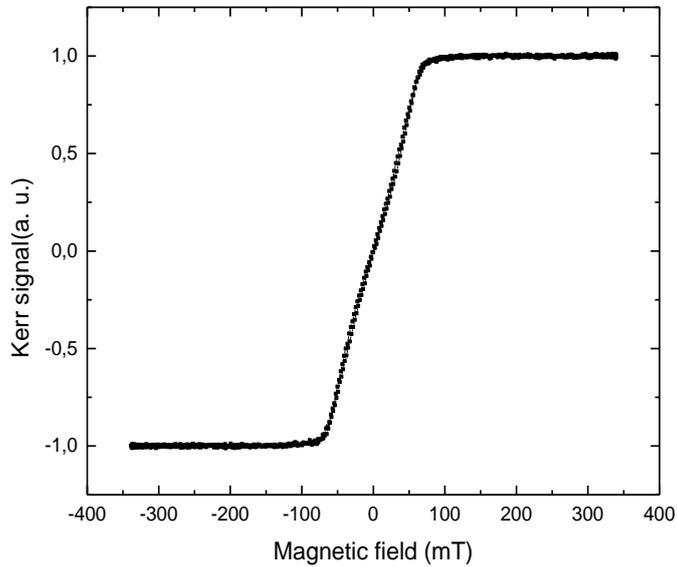

**Figure S12 Hysteresis cycle of the synthetic ferromagnetic stack.** Kerr signal as a function of the perpendicular magnetic field.

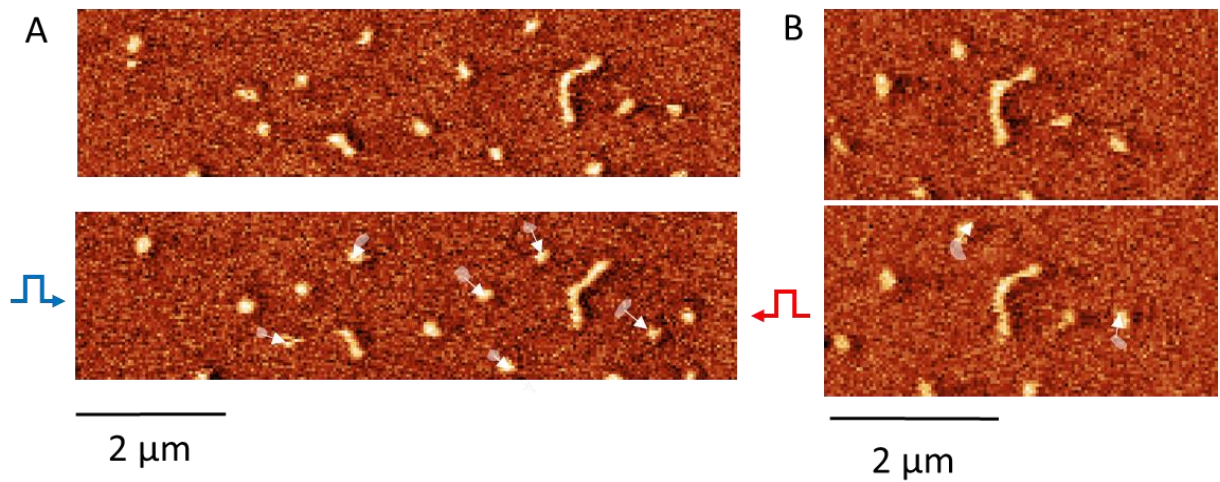

**Figure S13 Current induced skyrmion motion in synthetic ferromagnet observed by magnetic force microscopy A** Sequence of MFM images acquired before and after the injection of a current pulse of 4.95 ns with density $J = 6{,}95 \times 10^{11}\ A/m^2$ . B Sequence of images acquired before and after the injection of a pulse of 5.01 ns with current density of $J = 6{,}70 \times 10^{11}\ A/m^2$. The white arrows in the bottom images show the displacement of the skyrmions induced by the current pulse.

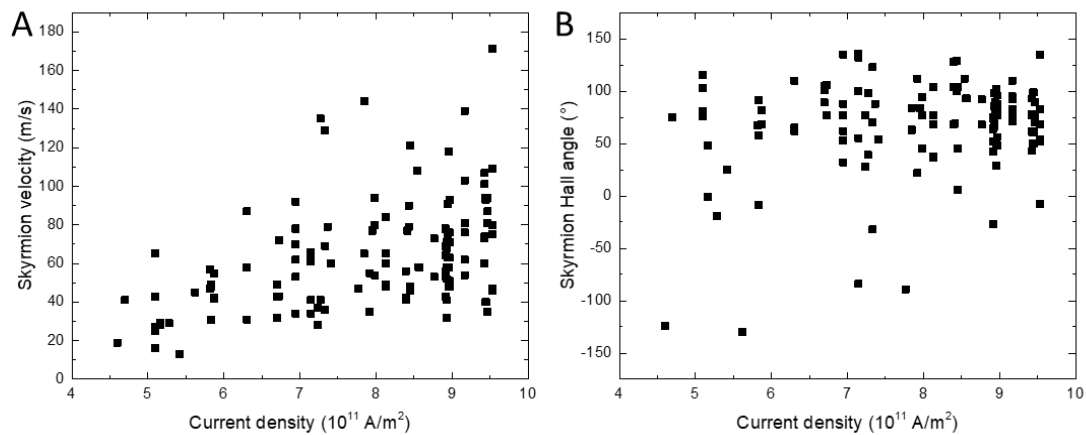

Figure S14 **Current induced skyrmion velocity and skyrmion Hall angle in synthetic ferromagnetic stack** Skyrmion velocity A and skyrmion Hall angle B as a function of the current density. One square dot corresponds to one event. An external out-of-plane magnetic field of 15 mT is applied.

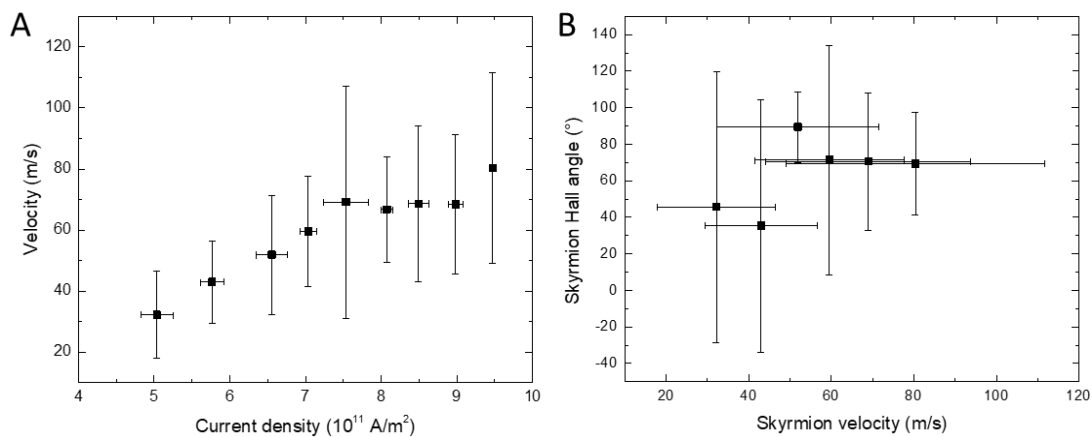

Figure S15 **Current induced skyrmion velocity and skyrmion Hall angle in synthetic ferromagnetic stack** A Average velocity vs current density of moving skyrmions. B Average skyrmion Hall angle vs skyrmion velocity. The error bars stand for the standard deviation. An external out-of-plane magnetic field of 15 mT is applied.

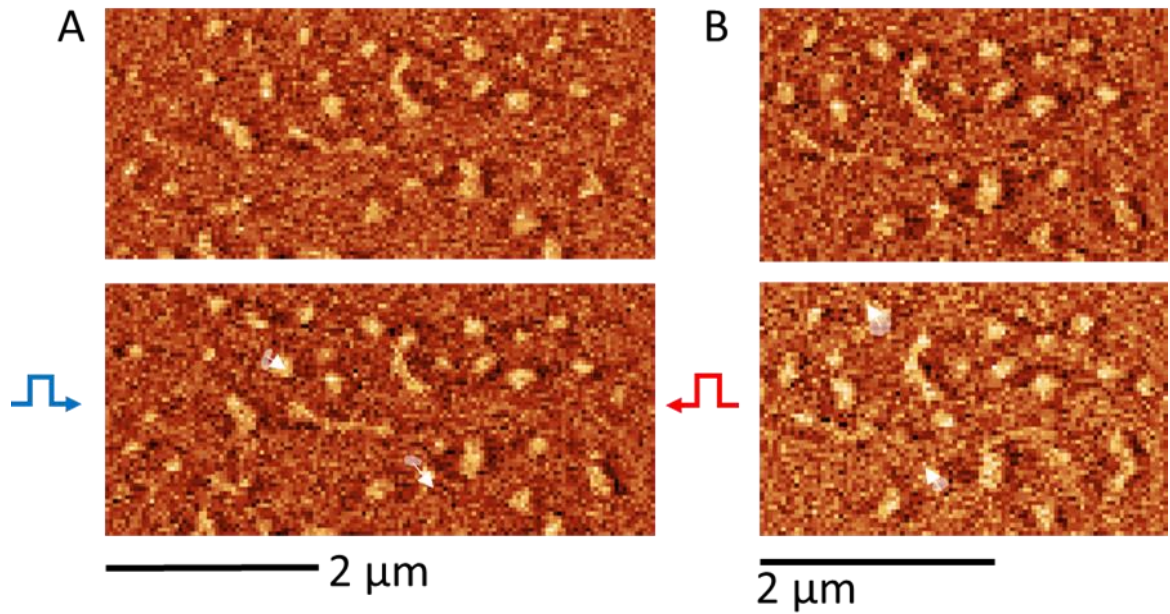

Figure S16 **Current induced skyrmion motion in single ferromagnetic layer stack observed by magnetic force microscopy A** Sequence of MFM images acquired before and after the injection of a pulse of 5.05 ns with current density of $J = 4{,}36 \times 10^{11}\ A/m^2$. **B** Sequence of images acquired before and after the injection of a pulse of 4.99 ns with current density of $J = 5{,}14 \times 10^{11}\ A/m^2$. An external out-of-plane magnetic field of 15 mT is applied. The white arrows in the bottom images show the displacement of the skyrmions induced by the current pulse.

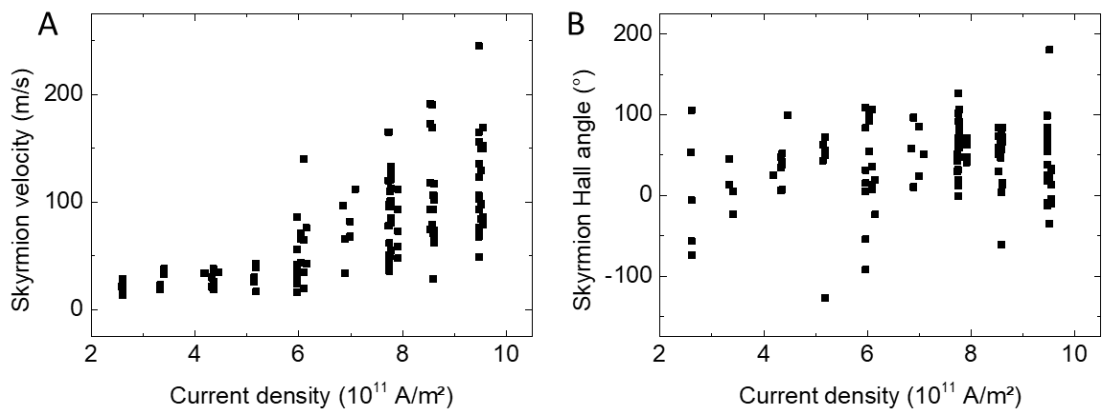

Figure S17 **Current induced skyrmion velocity and skyrmion Hall angle in single ferromagnetic layer stack** Skyrmion velocity A and skyrmion Hall angle B as a function of the current density. One square dot corresponds to one event.

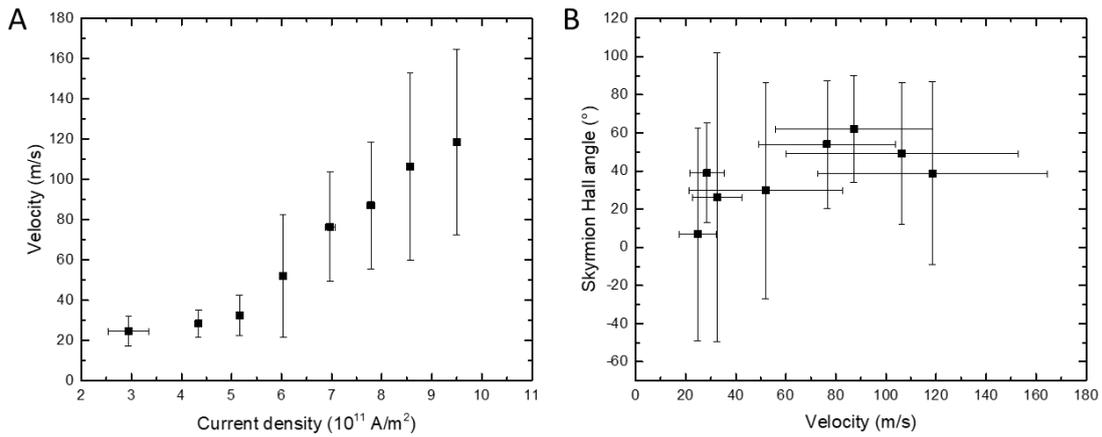

Figure S18 **Current induced skyrmion velocity and skyrmion Hall angle in single ferromagnetic layer stack** A Average velocity vs current density of moving skyrmions. B Average skyrmion Hall angle vs skyrmion velocity. The error bars stand for the standard deviation.

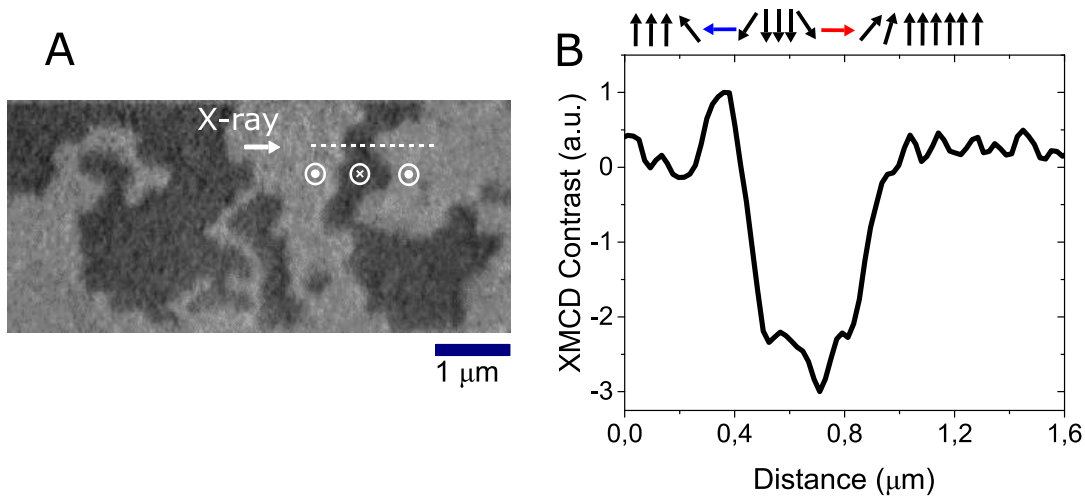

**Figure S19: Observation of the chiral structure of domain walls using XMCD-PEEM magnetic microscopy.** (A) XMCD-PEEM image of a multidomain state of the SAF stack (B) Line profile of the magnetic contrast along the white dotted line in (A). The line scans were performed with a certain width in order to get an average signal.

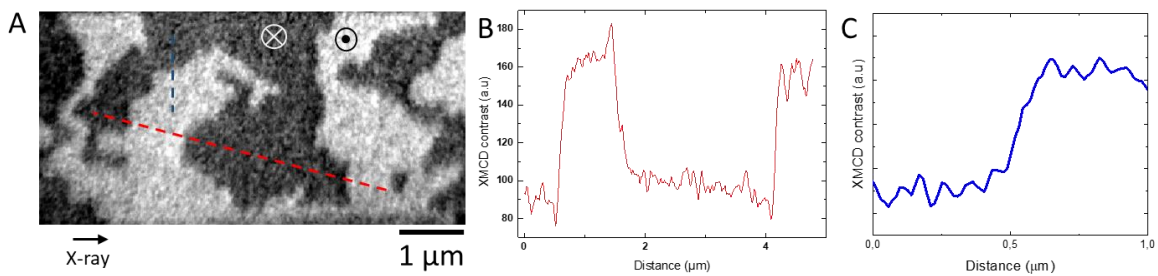

**Figure S20: Observation of the chiral structure of domain walls using XMCD-PEEM magnetic microscopy.** (A) XMCD-PEEM image of a multidomain state of the SAF stack. (B) Line profile of the magnetic contrast along the red dotted line in (A) (Domain wall

perpendicular to the X-ray beam). (C) Line profile of the magnetic contrast along the blue dotted line in (A) (domain wall along the X-ray beam). The line scans were performed with a certain width in order to get an average signal.

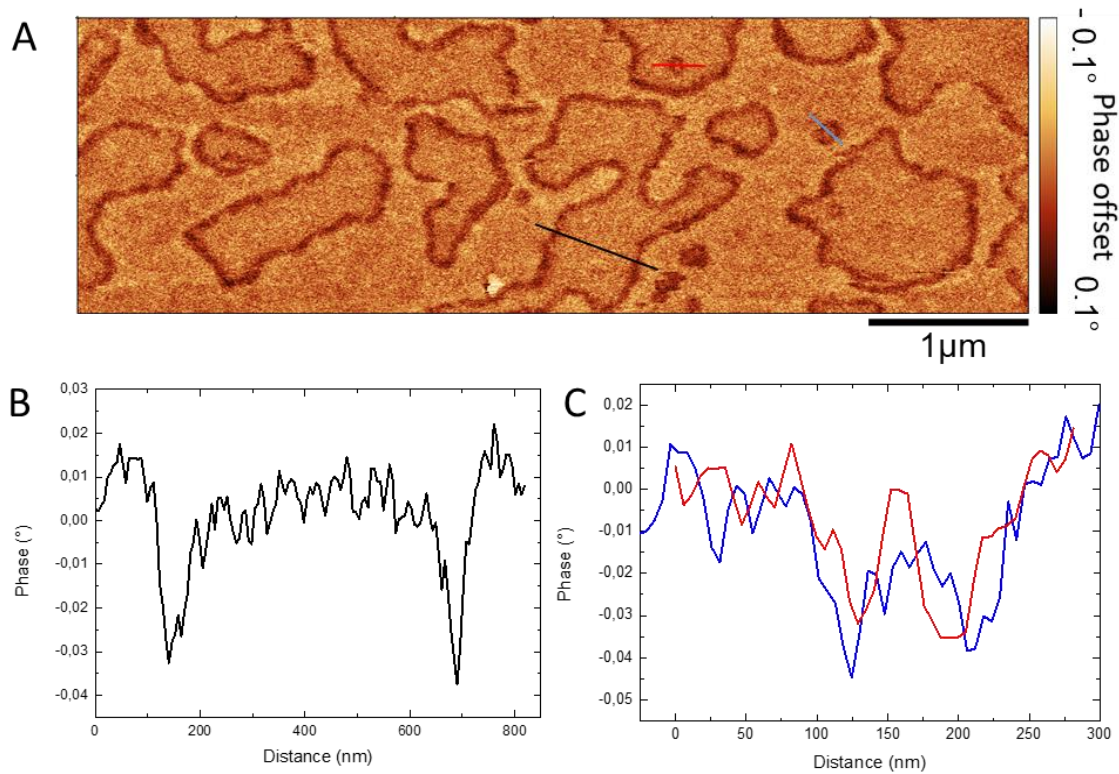

**Figure S21 MFM contrast of domain wall and skyrmions.** A MFM image measured at zero applied magnetic field showing domain walls and skyrmions (same as Figure 2C of the main paper). B MFM phase along the black line in A crossing HH/TT and TT/HH domain walls. C MFM phase along the red and blue lines in A crossing two skyrmions with opposite core magnetization.

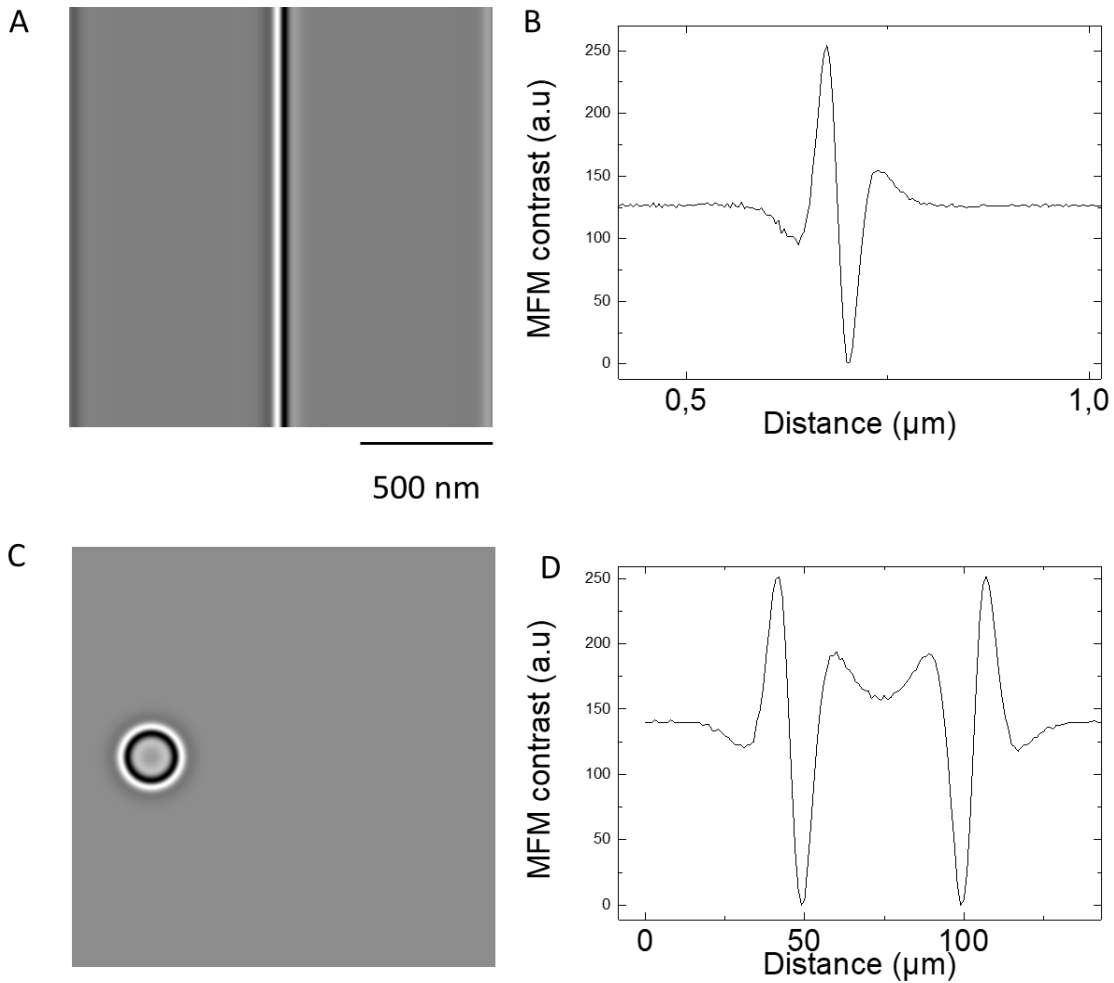

**Figure S22 Micromagnetic simulation of the MFM contrast** A Simulation of a MFM contrast of a magnetic domain wall. B MFM contrast along a line scan perpendicular to the domain wall. C-D. Same for a skyrmion.

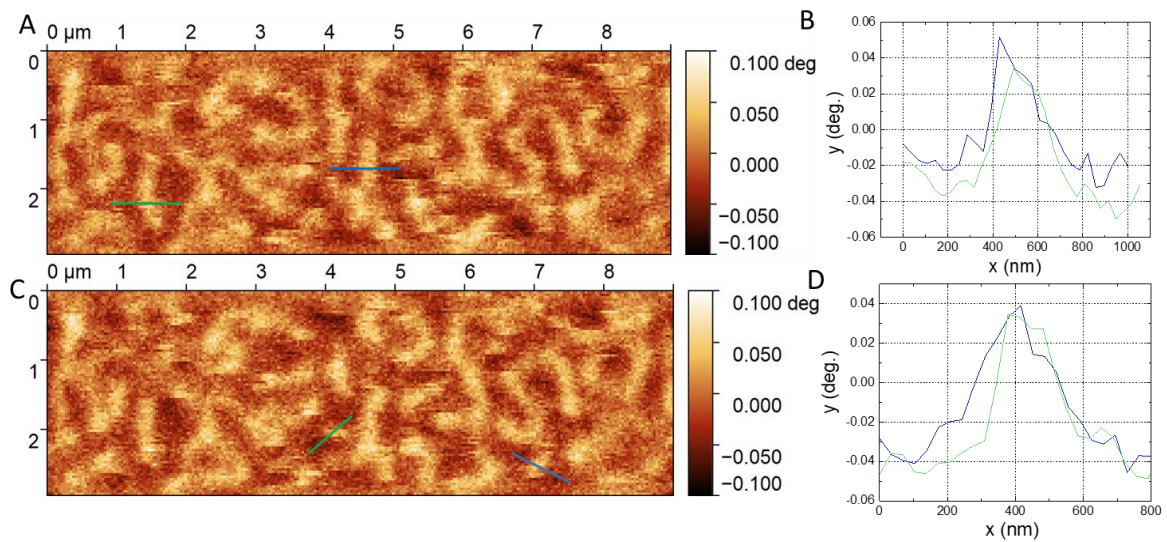

**Figure S23 Magnetic force microscopy experiments of the synthetic ferromagnetic stack.** (A,C) MFM images at zero field. (B,D) Magnetic contrast along the linescan in A and C. The magnetic tip is the same as the one used for the SAF stack.

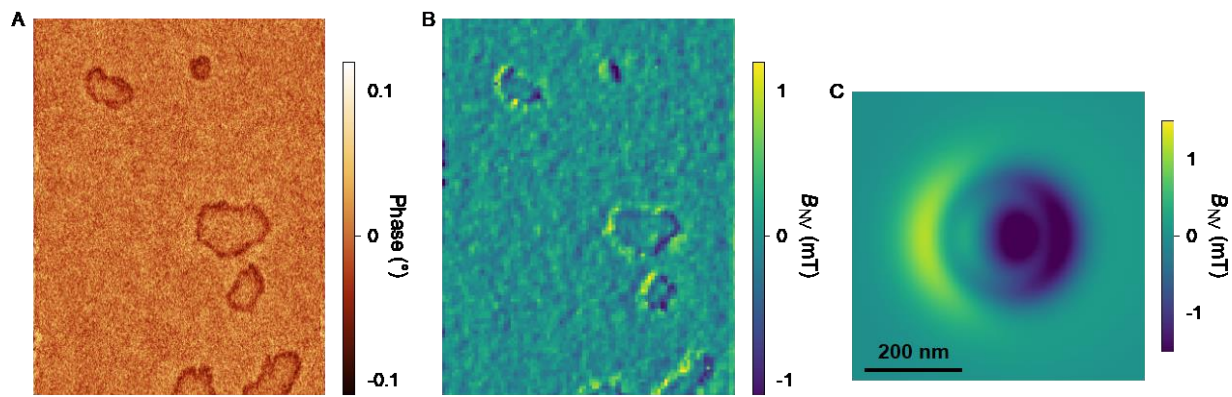

**Figure S24 Comparison of MFM and NV center magnetometry data** A) MFM image measured at zero external magnetic field, on the same track geometry as the one used for the current induced motion after applying an out-of-plane magnetic field sequence of -1 T then 175 mT. B) Stray field map of the same region as in A measured with scanning NV center magnetometry, at a distance of 36 nm from the sample surface. The field component shown here is the projection onto the NV center quantification axis (polar angle 61°, azimuthal angle 0°). The measurements were performed at zero external magnetic field. C) Simulation of the expected stray field map above a skyrmion of diameter 200 nm, whose profile is calculated using the micromagnetic parameters shown in Table S2.

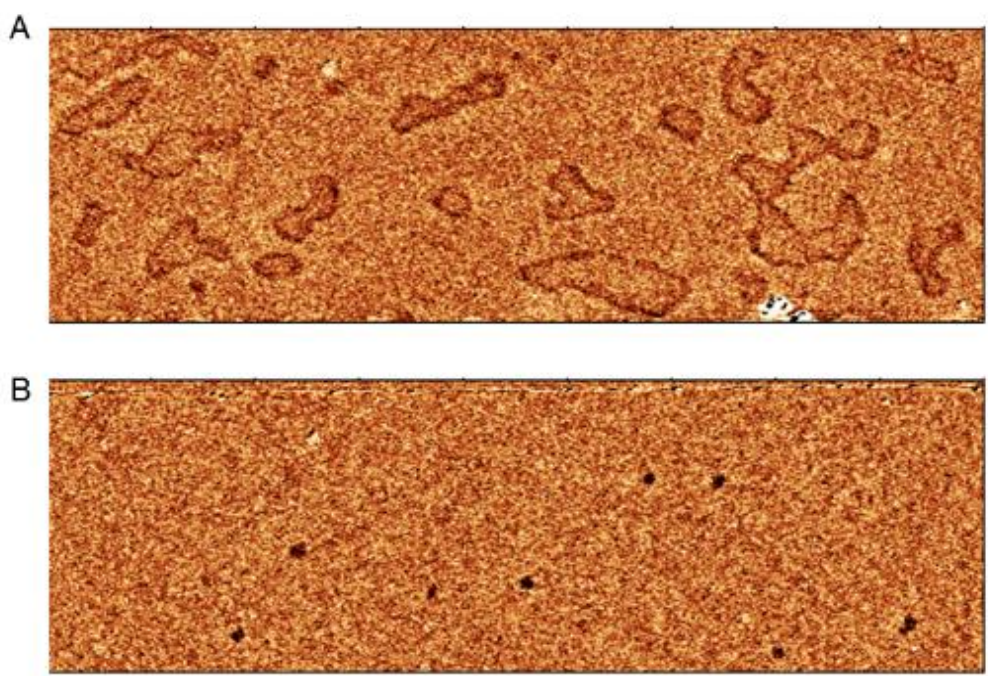

**Figure S25 Nucleation of SAF skyrmions using external magnetic fields** (A) MFM images of skyrmions in a magnetic stripe at B = 0 at room temperature acquired after applying sequentially out-of-plane magnetic fields of 220 mT and 175 mT. (B) MFM image measured at zero applied field obtained from A after an additional application of an out-of- plane field of 185 mT. The scale bar is 1 µm.

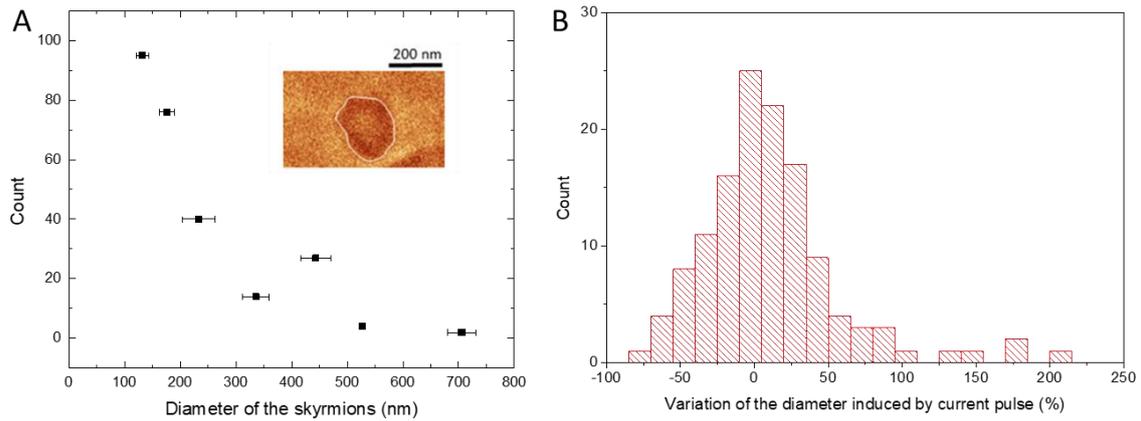

**Figure S26 Distribution of the diameters of the skyrmions** The inset is a MFM image of a typical skyrmion. The diameter d of a skyrmion is calculated from the area A inside the border (drawn in white) with $d = \sqrt{\frac{4A}{\pi}}$. The error bars are the sample standard deviation. We only considered here the diameter of skyrmions for which a motion was observed during the current induced motion experiments B Distribution of the variation of the skyrmion diameter before and after the pulse (average 5%, standard deviation of 35%, excluding events with deformation >100%).

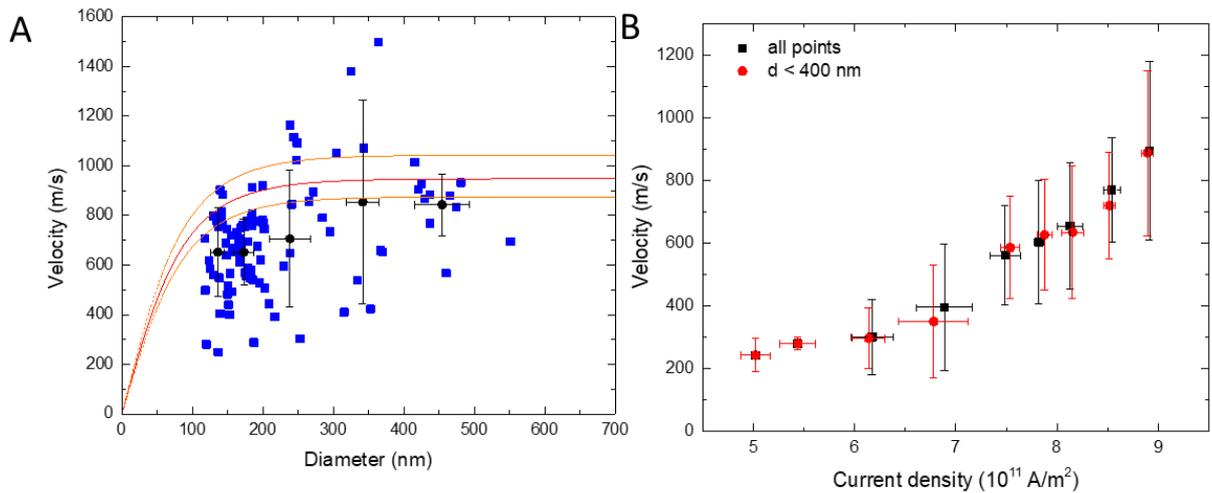

**Figure S27 Velocity vs skyrmion diameter** for current densities between $7.5\times10^{11}$ A/m² and $8.9\times10^{11}$ A/m². Blue dots are single events, black dots are averaged values for different skyrmion diameters (error bars stand for standard deviation), the red line is the prediction of the model for J=$8.12\times10^{11}$ A/m², the dotted orange lines corresponds to the prediction of the model for J=$7.5\times10^{11}$ A/m² and $8.9\times10^{11}$ A/m². The skyrmion diameter corresponds to the one measured before the pulse injection. B Velocity vs current density for skyrmion diameter smaller than 400 nm (red points) and skyrmion with all diameters (black points).

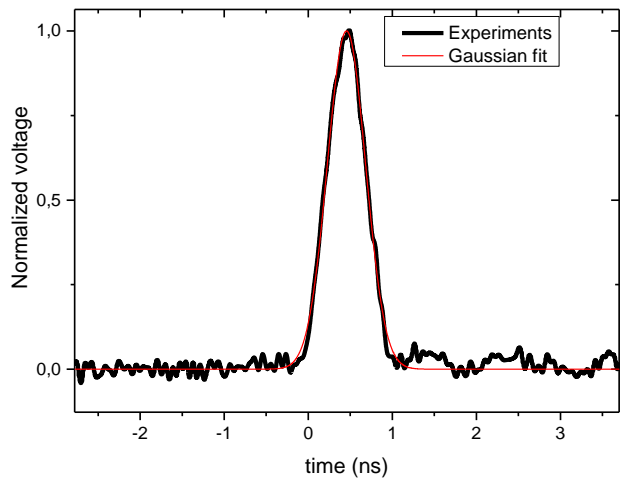

**Figure S28 Example of current pulse used for the current induced skyrmion experiments** Experimental normalized voltage pulse as a function of time (black line) and its Gaussian fit (red line).

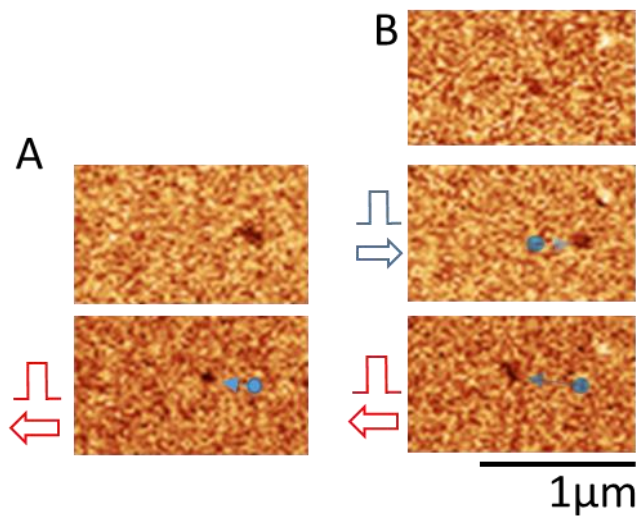

**Figure S29 Current induced skyrmion motion of isolated skyrmions at zero external magnetic field** A MFM image of an isolated skyrmion before (top) and after (bottom) the injection of a 0.81 ns current pulse with density J=7.1x10$^{11}$ A/m². The skyrmion move by around 310 nm. B MFM image of an isolated skyrmions skyrmion before (top), after (middle) the injection of a positive current pulse of 0.84 ns (skyrmion displacement of 310 nm) and density J= 6.9x10$^{11}$ A/m² and after (bottom) the injection of a 0.81 ns current pulse of density J=7.1x10$^{11}$ A/m², skyrmion displacement of 450 nm.

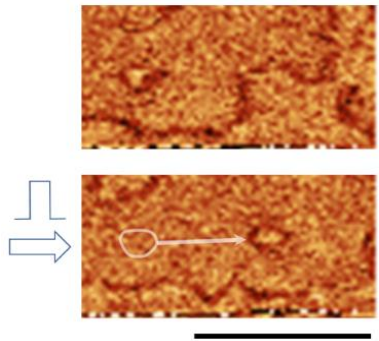

**Figure S30 Current induced motion of a SAF skyrmion observed by MFM** Magnetic force microscopy image before (top) and after (bottom) the injection of a 1.1 ns current pulse of density 8×10¹¹ A/m². The scale bas is 1µm.

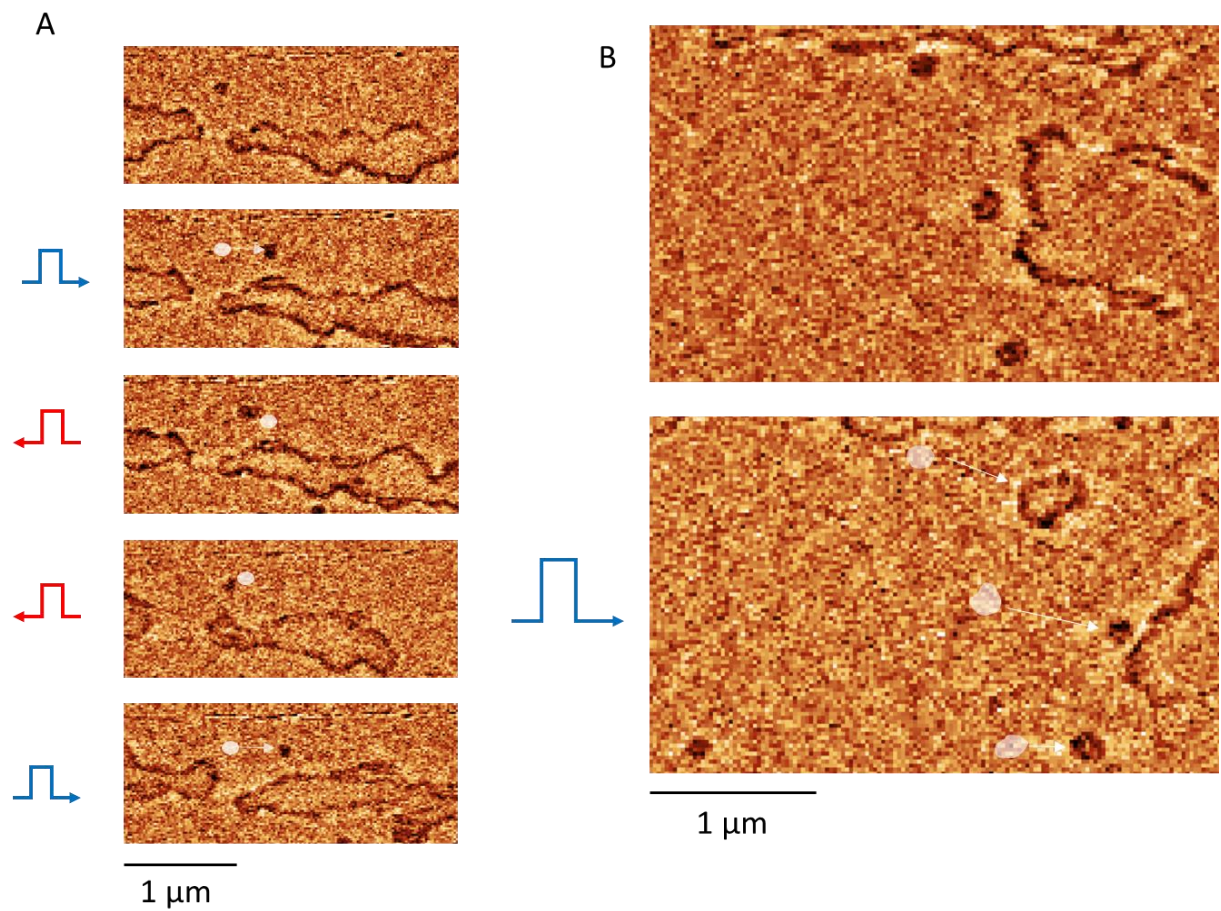

**Figure S31 Current induced motion of SAF skyrmions observed by magnetic force microscopy (MFM).** A Sequence of images acquired after injection of subsequent pulses of 1.2 ns, 1.3 ns, 1.2 ns 1.2 ns with respectively current density $J = 6.4 \times 10^{11}\ A/m^2$, $-5.8 \times 10^{11}\ A/m^2$, $-7.2 \times 10^{11}\ A/m^2$, $7.2 \times 10^{11}\ A/m^2$ The corresponding velocities are 430 m/s, 160 m/s, 340 m/s, 440 m/s respectively. B Sequence of images acquired before and after the injection of a pulse of 0.9 ns with current density of $J = 8,1 \times 10^{11}\ A/m^2$. The corresponding velocities are 780, 800, 440 m/s respectively.

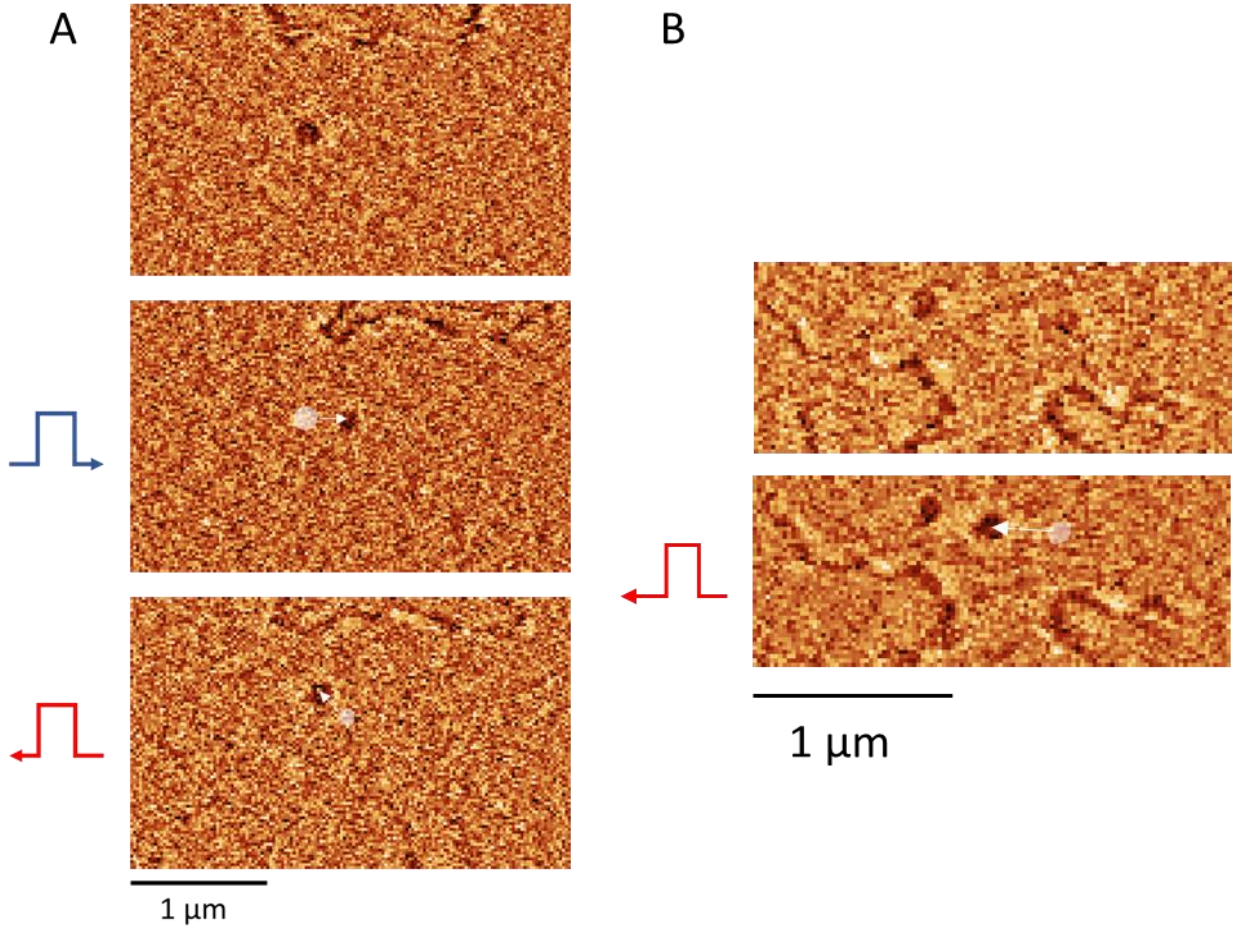

**Figure S32 Current induced motion of SAF skyrmions observed by magnetic force microscopy (MFM).** A Sequence of images acquired before and after the injection of two subsequent pulses of 0.54 ns and 0.53 ns with current density respectively of $J=7.7 \times 10^{11}$ $A/m^2$ and $-8.0 \times 10^{11} A/m^2$. The corresponding velocities are 780 m/s, 800 m/s and 440 m/s. B Sequence of images acquired before and after the injection of a pulse of 1,3 ns with current density of $J=5,19 \times 10^{11}$ $A/m^2$. The corresponding velocity is 280 m/s.

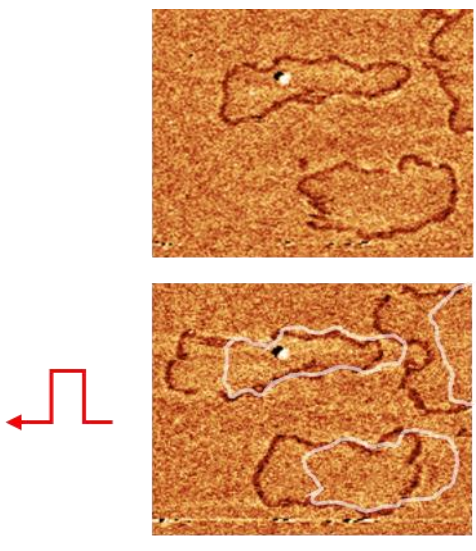

**Figure S33 Current induced motion of domain walls and SAF bubbles** Magnetic force microscopy image before (top) and after (bottom) the injection of a 0.62 ns current pulse of density $8.0\times10^{11}$ A/m². The white lines in the bottom image corresponds to the position of the elongated skyrmion before the pulse injection. The scale bar is 1 µm.

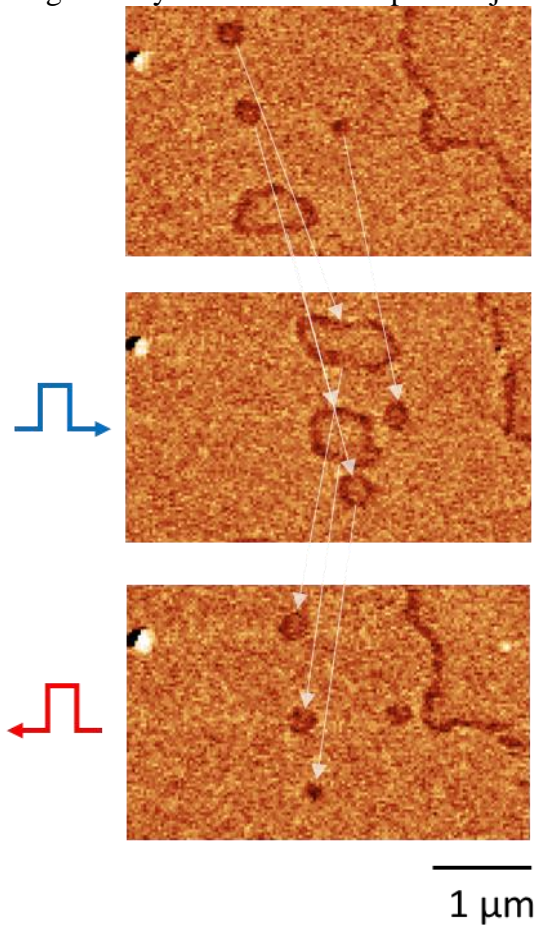

**Figure S34 Back and forth motion of SAF skyrmions induced by current pulses** Magnetic force microscopy image before (top) and after the injection of a positive (middle) and negative (bottom) 0.53 ns current pulse of density $8.85\times10^{11}$ A/m². Scale bar is 1 µm.

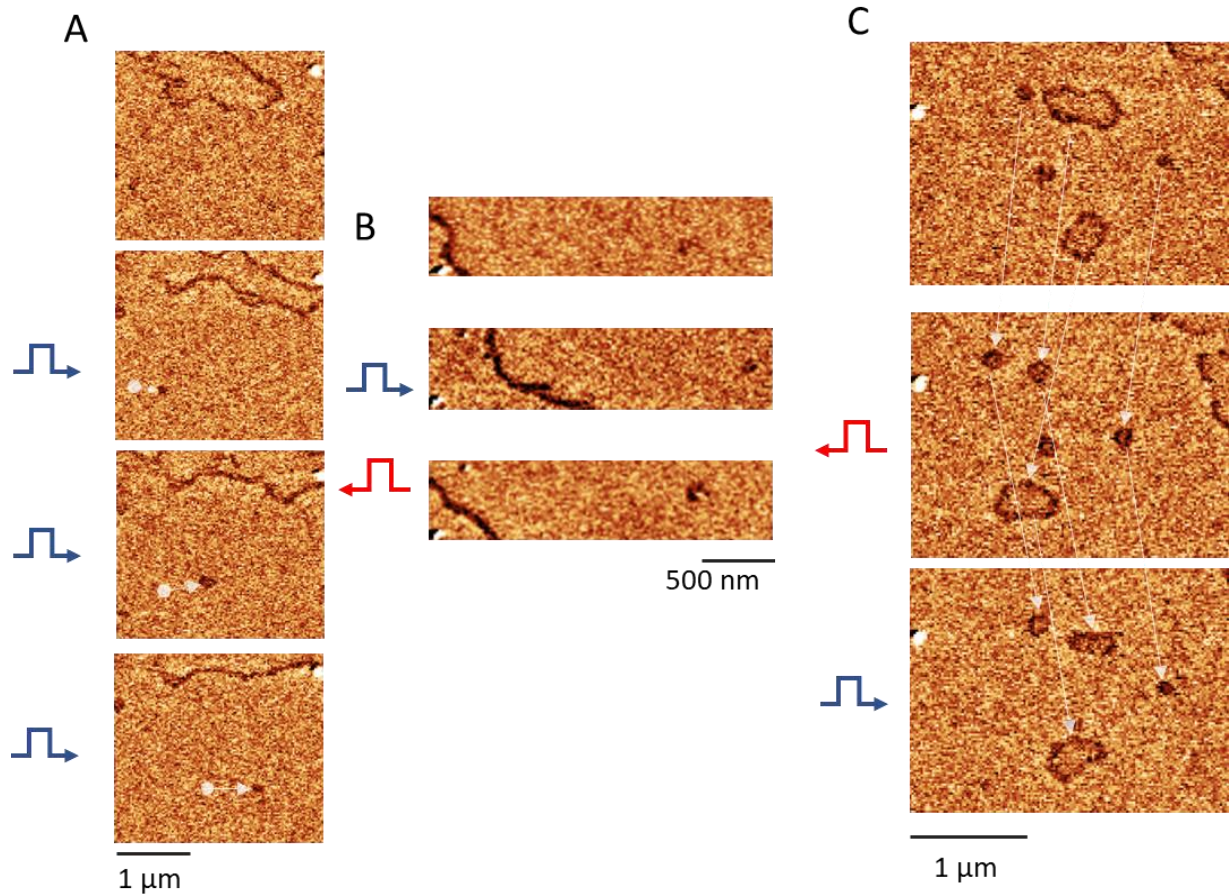

**Figure S35** MFM images showing current induced motion experiments. The images are the same as the one of Figure 3 of the main text but zoomed out such that a topological defect can be seen on the image. The scale bars are 1 µm.

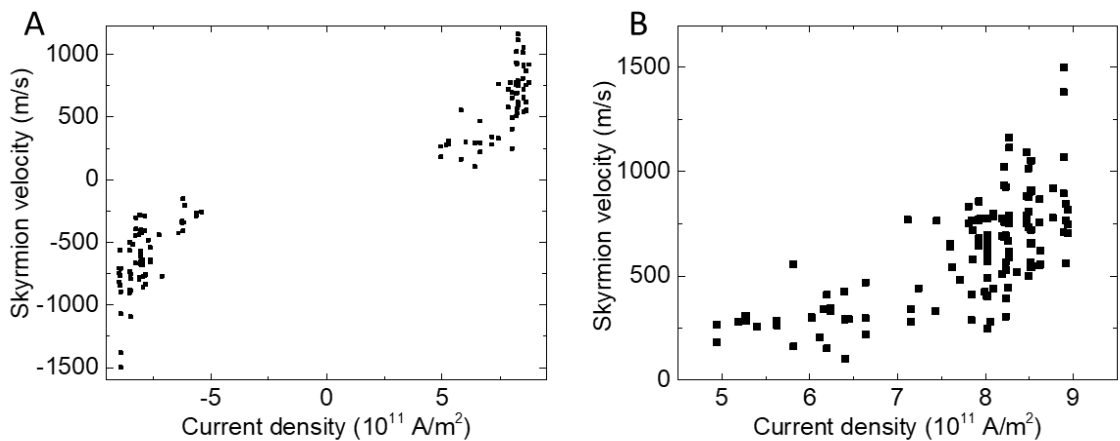

**Figure S36 A Skyrmion velocities vs current density of all the current induced skyrmion motion events.** B same as A but plotted in absolute value.

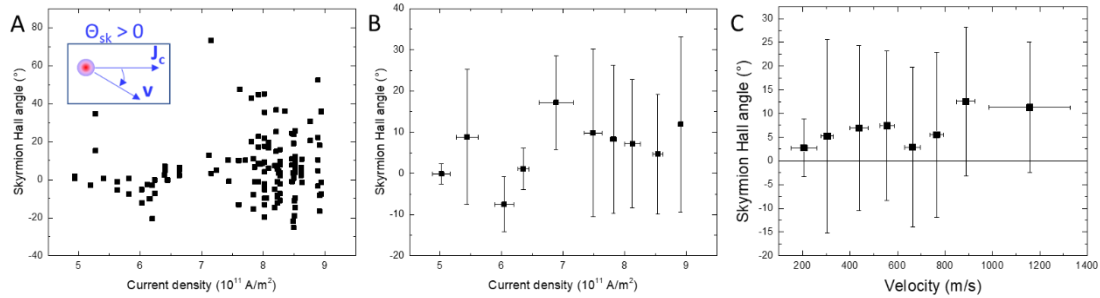

**Figure S37 Skyrmion Hall effect in the SAF stack** A Skyrmion Hall angle vs current density of all the current induced motion events. B. Average skyrmion Hall angle vs current density. C. Average skyrmion Hall angle vs velocity.

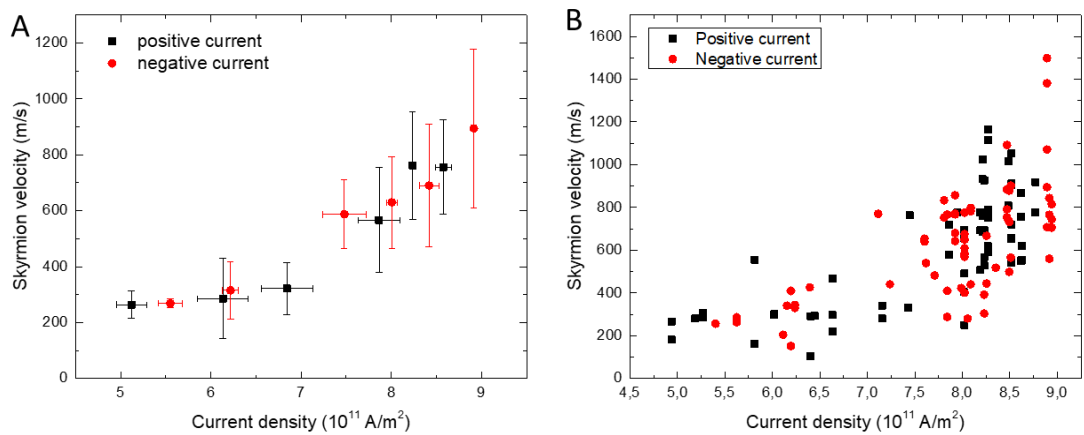

**Figure S38 Comparison between the velocities measured for positive and negative current pulses.** A shows the values averaged over several events (the error bars stand for the standard deviation) and B shows the corresponding velocities distribution.

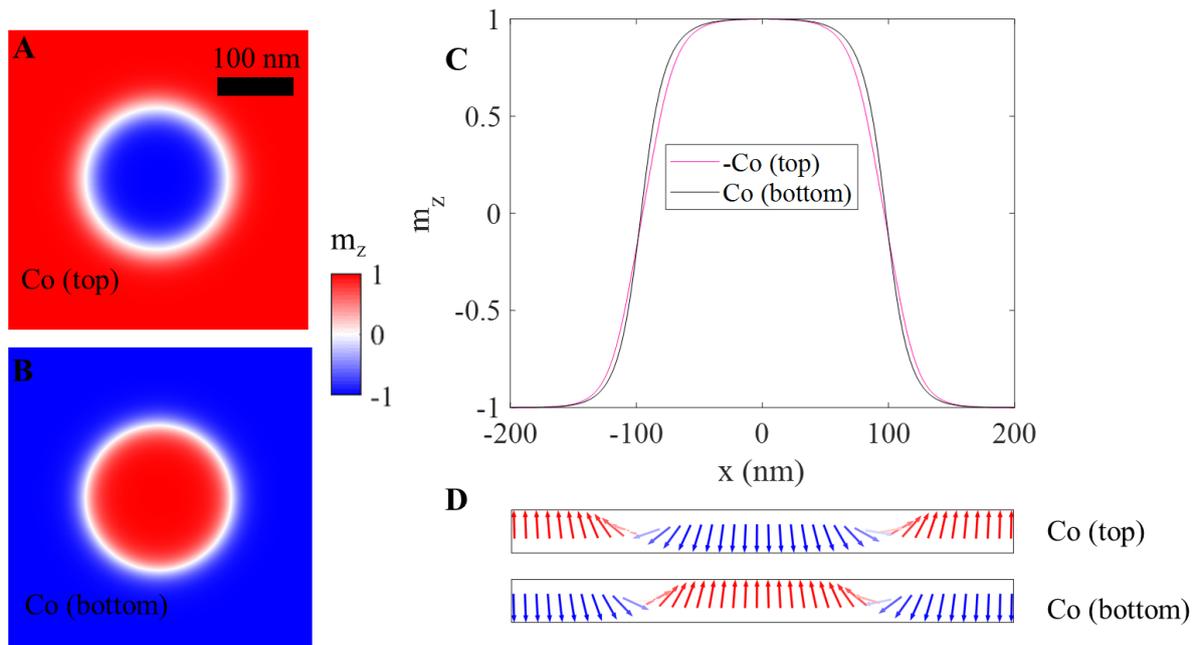

**Figure S39: Micromagnetic simulation of the SAF skyrmions spin texture** (a) and (b) show the z-component of magnetization in top and bottom FM layers, respectively. (c) shows

a 1D linescan through (a) and (b). (d) shows the linescan using the vector arrows representing the magnetic moments.

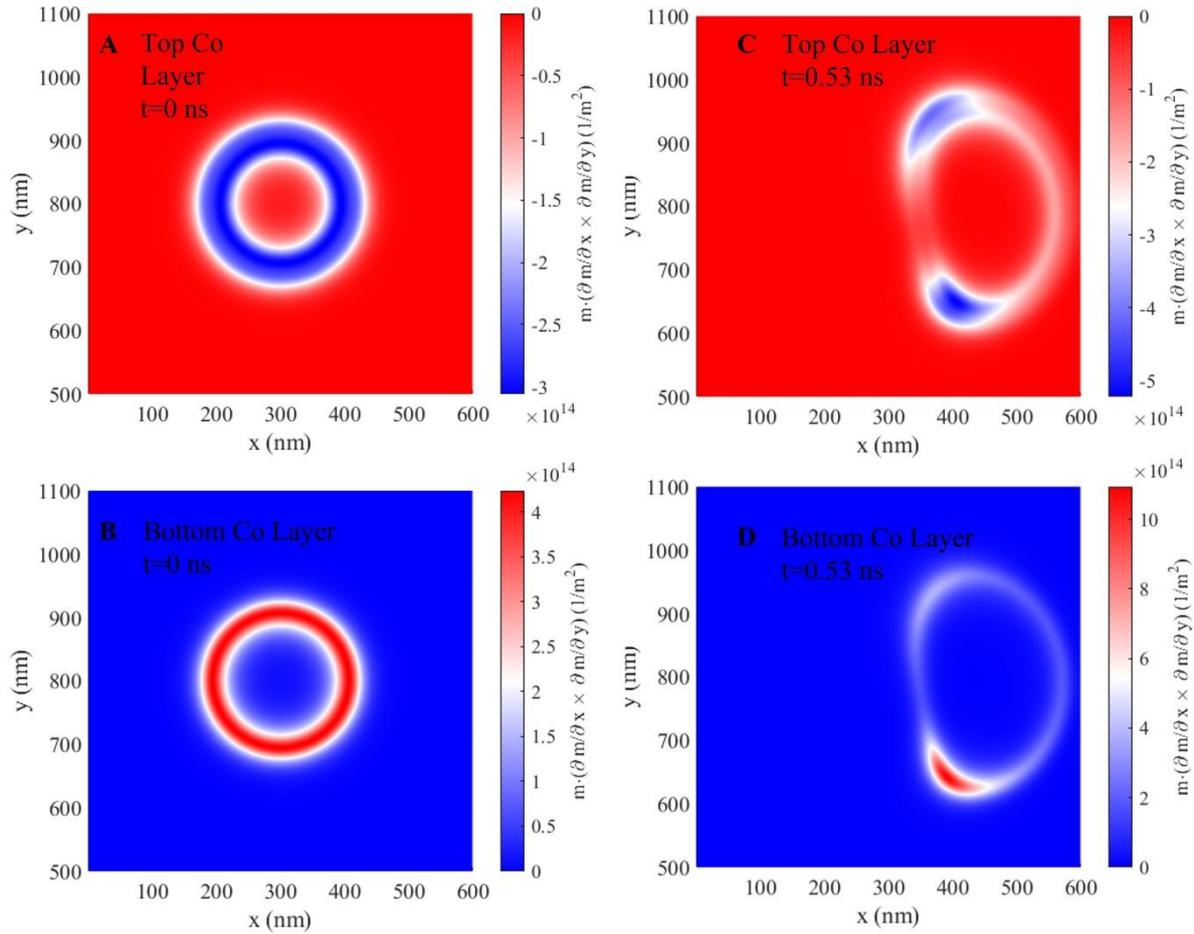

**Figure S40 Maps of topological charge density for different cases**. (a) and (c) represent the topological charge density map for the top and bottom FM layer at t=0ns (static), respectively. Similarly, (c) and (d) represent the topological charge density map for the top and bottom FM layers, of the moving skyrmion for a DC pulse with J=5.6×10$^{11}$A/m2 at t=0.53ns. Note that slightly different parameters were used for these micromagnetic simulations with α=0.216, H$_{k,top}$=12.4 mT, H$_{k,bottom}$=35.4 mT, γ=175.9 GHz/T, D=0.62 mJ/m².

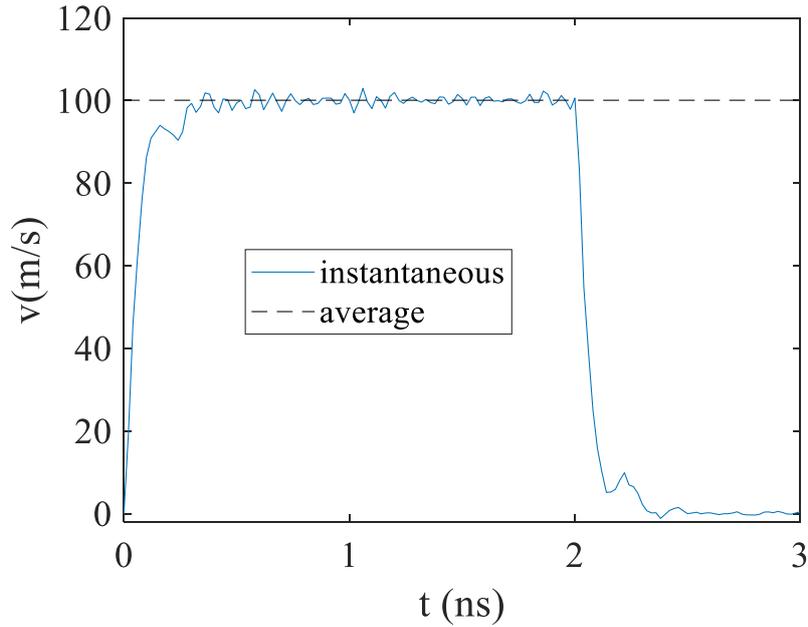

**Figure S41: Instantaneous velocity vs time for J=1×10¹¹A/m².** Black dashed line shows the average velocity calculated from the displacement of skyrmion.

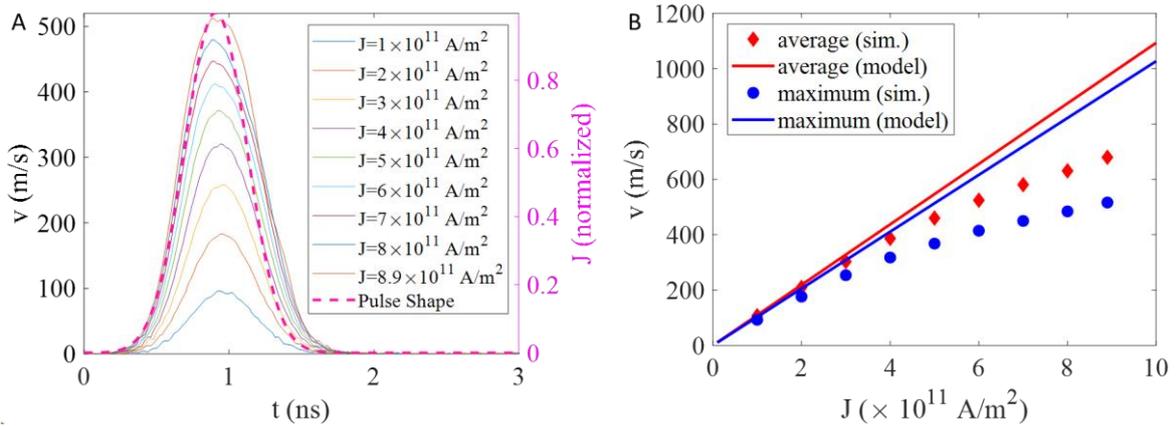

**Figure S42 Instantaneous velocity and average velocities for varying current densities** (a) Instantaneous velocity for a Gaussian current pulse of varying amplitudes. The right y-axis shows the shape of the injected pulse (dashed pink line). (b) Maximum instantaneous velocity and average velocity as a function of current pulse amplitude. The solid lines show the analytically calculated velocity.

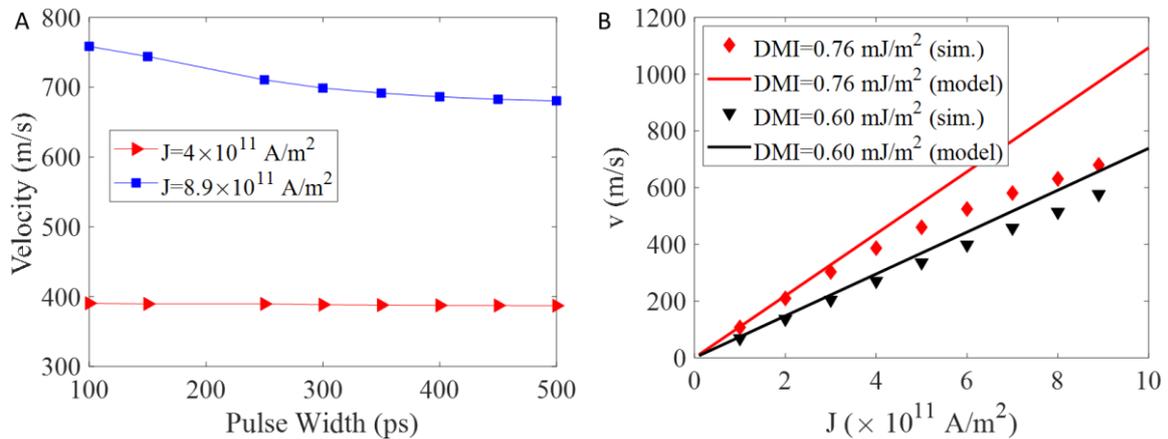

**Figure S43 Average velocities for short current pulses** (a) Average velocity as a function of the Gaussian pulse width (FWHM) for J=4×10$^{11}$ A/m² and J=8.9×10$^{11}$ A/m². (b) Average velocity vs J for two different skyrmion diameters obtained for two different sets of parameters.

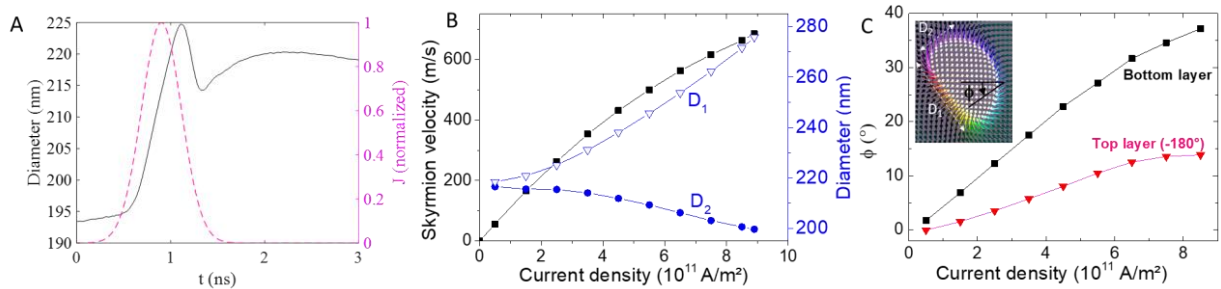

**Figure S44 Current induced skyrmion deformation** A Time dependence of the average skyrmion diameter during the application of a 0.5 ns current pulse with density J=8.9×10$^{11}$ A/m². The diameter is calculated from the area of the skyrmion. B skyrmion velocity vs current density for a 0.5 ns current pulse (black squares). Skyrmion dimension along the major ("D$_1$", open blue triangles) and minor ("D$_2$", blue circles) axes of the skyrmion elliptical shape at the maximum of the pulse. C Angle ϕ of the DW magnetization along the x axis (see inset, right DW) for the bottom (black squares) and top (pink triangles) layers. The angle is calculated with respect to the –x axis. Inset: skyrmion magnetization at the maximum of the current pulse during the injection of a 0.5 ns pules with density 8.9×10$^{11}$ A/m².

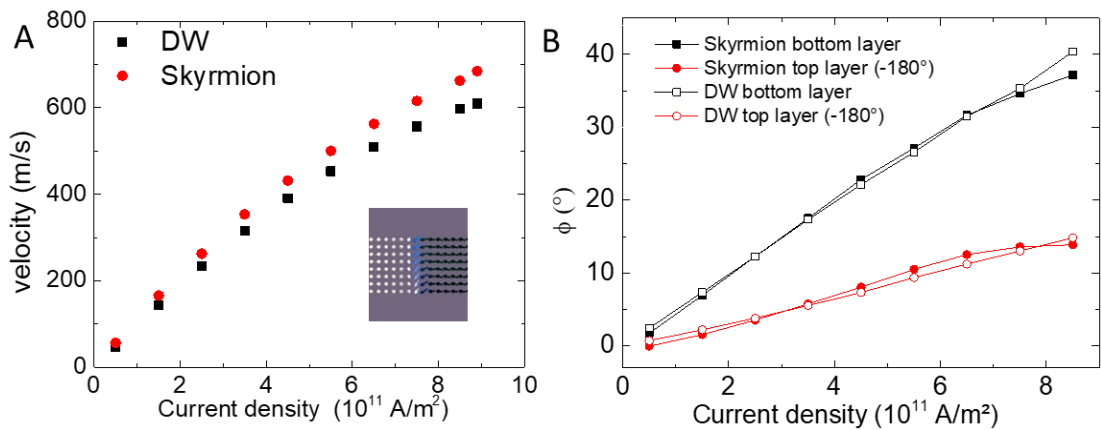

**Figure S45 Comparison of current induced DW and skyrmion dynamics** A Current induced DW (black square) and skyrmion (red circles) velocity vs current density. Inset: example of the magnetization pattern of one layer in the SAF during the current pulse injection. B DW angle ϕ vs current density of the top (red dots) and bottom (black dots) layers of a DW in a track (open symbols) and a skyrmion (DW position same as Figure S44C, inset).

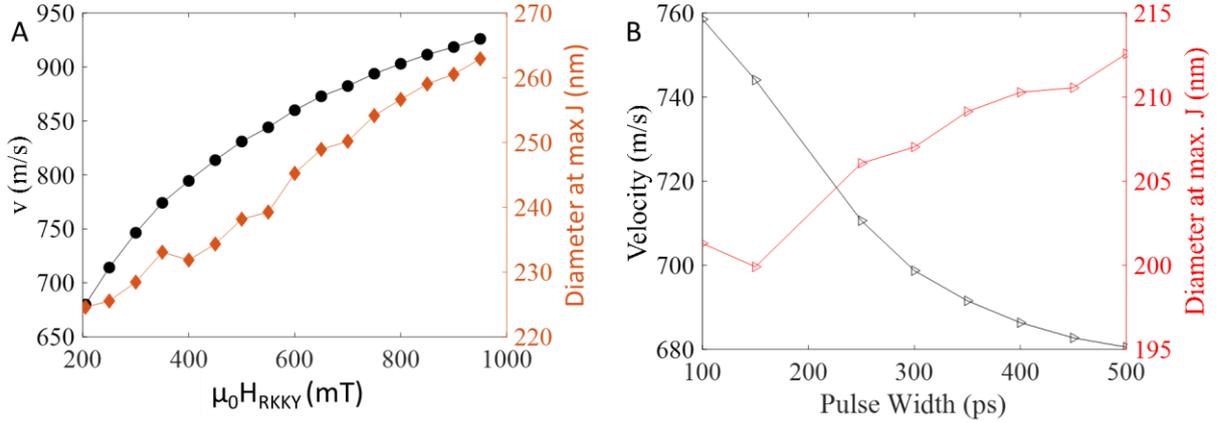

**Figure S46 Impact of the RKKY interaction and pulse width on the skyrmion velocity** A Skyrmion velocity and diameter at max current vs A. RKKY field for a pulse width of 0.5 ns and B. pulse width for an RKKY field of $\mu_0 H_{RKKY}$=205 mT for J=8.9×$10^{11}$ A/m².

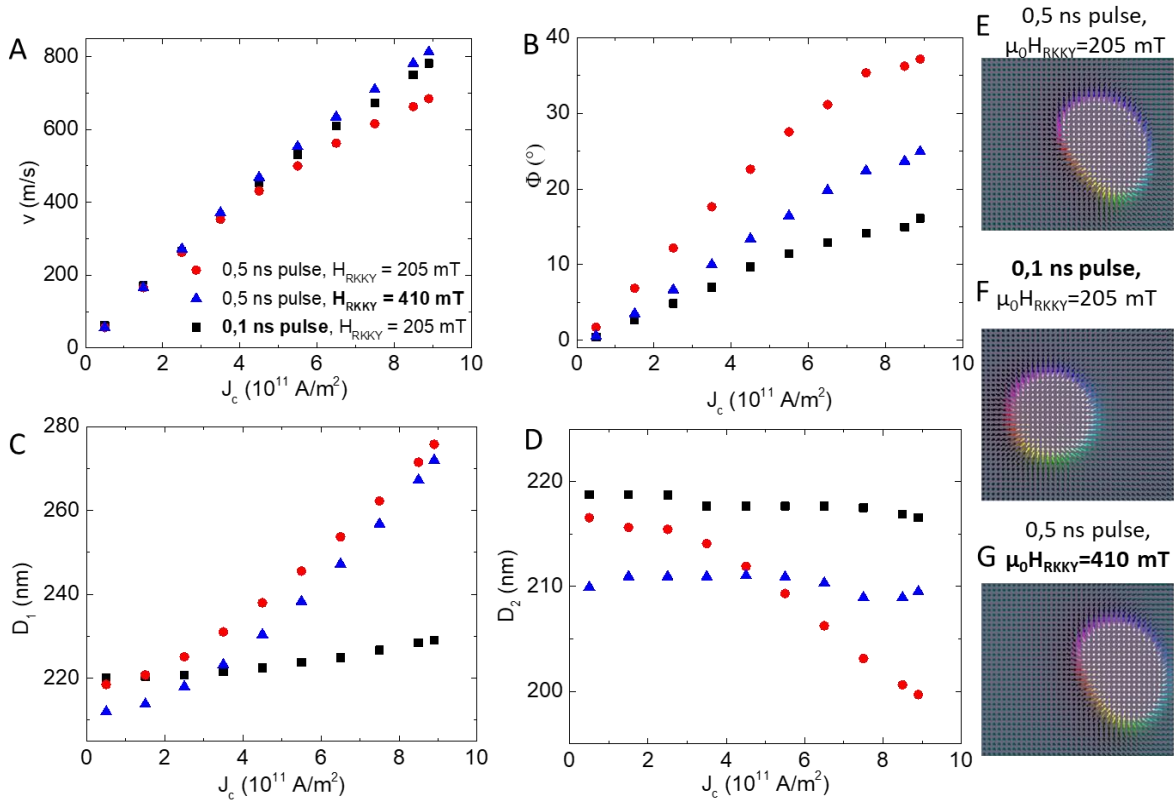

**Figure S47 Impact of the RKKY interaction and pulse width on the skyrmion dynamics** (A-D) A Velocity v, B domain wall magnetization angle ϕ (bottom layer), dimension of the skyrmion C along the major (D1) and D minor (D2) axis for (i) a 0.5 ns current pulse and a RKKY field $\mu_0 H_{RKKY}$=205 mT (red square), (ii) a 0.5 ns current pulse and $\mu_0 H_{RKKY}$=410 mT (blue circles) (iii) a 0.1 ns current pulse and $\mu_0 H_{RKKY}$=205 mT (black triangles) (E-G) Magnetization pattern at the maximum of the current pulse corresponding to E (i), F (ii) and G (iii). The angle ϕ, and dimensions D1 and D2 are measured at the current pulse maximum.

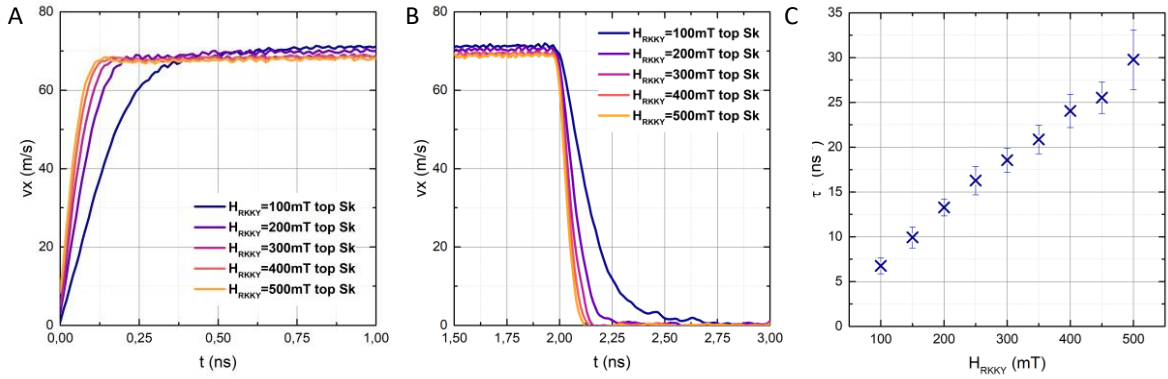

**Figure S48 Impact of the RRKY exchange interaction on the inertial dynamics** (a-b) Time dependence of the skyrmion velocity in response to a 2 ns square pulse (J=1011 A/m²) for different HRKKY interaction fields ((a) and (b) are zoom on the rise and fall of the velocity). (c) inverse of the time constant of the SAF skyrmion displacement computed by fitting the rising and falling edge of the skyrmion speed. The mean and the error bar are plotted, where the error bar is the standard deviation over the data. The mean value takes into account the top and bottom skyrmions and the rise and fall time constants (i.e. 4 data points each). Note that slightly different parameters were used for these micromagnetic simulations with α=0.216, $H_{k,top}$=12.4 mT, $H_{k,bottom}$=35.4 mT, γ=175.9 GHz/T, D=0.62 mJ/m².

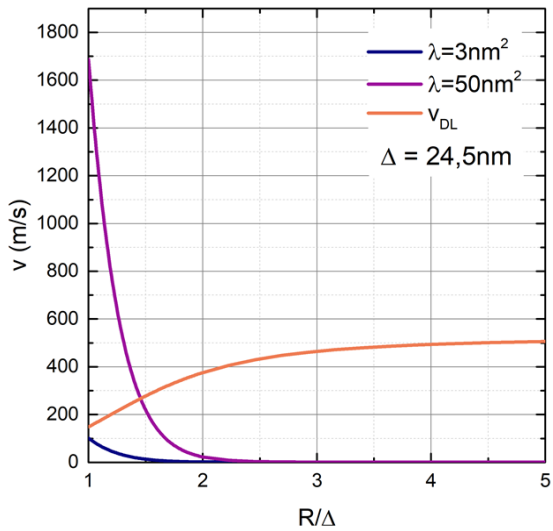

**Figure S49: Calculation of the topological spin Hall torque** Comparison of the speed of AF skyrmion due to SOT driving (in orange) and the equivalent speed due to the topological spin torque (in blue and violet). We compare the results for two values of the topological spin torque range: $\lambda^2$ = 3 nm and 50 nm². The applied current density is $8\times10^{11}$ A/m² along the x direction, the domain wall width is fixed at 24.5 nm and we vary the core radius of the skyrmion from 24.5nm (R/Δ =1) to 122.5nm (R/Δ = 5). Note that slightly different parameters were used for these micromagnetic simulations with α=0.216, $H_{k,top}$=12.4 mT, $H_{k,bottom}$=35.4 mT, γ=175.9 GHz/T, D=0.62 mJ/m².

**References**

1. N. Nagaosa, Y. Tokura, Topological properties and dynamics of magnetic skyrmions. *Nat. Nanotechnol.* **8**, 899–911 (2013).

2. X. Z. Yu, N. Kanazawa, W. Z. Zhang, T. Nagai, T. Hara, K. Kimoto, Y. Matsui, Y. Onose, Y. Tokura, Skyrmion flow near room temperature in an ultralow current density. *Nat. Commun.* **3**, 988 (2012).

3. W. Jiang, X. Zhang, G. Yu, W. Zhang, X. Wang, M. Benjamin Jungfleisch, J. E. Pearson, X. Cheng, O. Heinonen, K. L. Wang, Y. Zhou, A. Hoffmann, S. G. E. te Velthuis, Direct observation of the skyrmion Hall effect. *Nat. Phys.*, doi: 10.1038/nphys3883 (2016).

4. S. Woo, K. Litzius, B. Krüger, M.-Y. Im, L. Caretta, K. Richter, M. Mann, A. Krone, R. M. Reeve, M. Weigand, P. Agrawal, I. Lemesh, M.-A. Mawass, P. Fischer, M. Kläui, G. S. D. Beach, Observation of room-temperature magnetic skyrmions and their current-driven dynamics in ultrathin metallic ferromagnets. *Nat. Mater.* **15**, 501–506 (2016).

5. R. Juge, S.-G. Je, D. de S. Chaves, L. D. Buda-Prejbeanu, J. Peña-Garcia, J. Nath, I. M. Miron, K. G. Rana, L. Aballe, M. Foerster, F. Genuzio, T. O. Mentes, A. Locatelli, F. Maccherozzi, S. S. Dhesi, M. Belmeguenai, Y. Roussigné, S. Auffret, S. Pizzini, G. Gaudin, J. Vogel, O. Boulle, Current-Driven Skyrmion Dynamics and Drive-Dependent Skyrmion Hall Effect in an Ultrathin Film. *Phys. Rev. Appl.* **12**, 044007 (2019).

6. K. Litzius, I. Lemesh, B. Krüger, P. Bassirian, L. Caretta, K. Richter, F. Büttner, K. Sato, O. A. Tretiakov, J. Förster, R. M. Reeve, M. Weigand, I. Bykova, H. Stoll, G. Schütz, G. S. D. Beach, M. Kläui, Skyrmion Hall effect revealed by direct time-resolved X-ray microscopy. *Nat. Phys.* **13**, 170–175 (2017).

7. W. Jiang, P. Upadhyaya, W. Zhang, G. Yu, M. B. Jungfleisch, F. Y. Fradin, J. E. Pearson, Y. Tserkovnyak, K. L. Wang, O. Heinonen, S. G. E. te Velthuis, A. Hoffmann, Blowing magnetic skyrmion bubbles. *Science* **349**, 283–286 (2015).

8. O. Boulle, J. Vogel, H. Yang, S. Pizzini, D. de Souza Chaves, A. Locatelli, T. O. Menteş, A. Sala, L. D. Buda-Prejbeanu, O. Klein, M. Belmeguenai, Y. Roussigné, A. Stashkevich, S. M. Chérif, L. Aballe, M. Foerster, M. Chshiev, S. Auffret, I. M. Miron, G. Gaudin, Room-temperature chiral magnetic skyrmions in ultrathin magnetic nanostructures. *Nat. Nanotechnol.* **11**, 449–454 (2016).

9. C. Moreau-Luchaire, C. Moutafis, N. Reyren, J. Sampaio, C. a. F. Vaz, N. V. Horne, K. Bouzehouane, K. Garcia, C. Deranlot, P. Warnicke, P. Wohlhüter, J.-M. George, M. Weigand, J. Raabe, V. Cros, A. Fert, Additive interfacial chiral interaction in multilayers for stabilization of small individual skyrmions at room temperature. *Nat. Nanotechnol.* **11**, 444–448 (2016).

10. K. Litzius, J. Leliaert, P. Bassirian, D. Rodrigues, S. Kromin, I. Lemesh, J. Zazvorka, K.-J. Lee, J. Mulkers, N. Kerber, D. Heinze, N. Keil, R. M. Reeve, M. Weigand, B. Van Waeyenberge, G. Schütz, K. Everschor-Sitte, G. S. D. Beach, M. Kläui, The role of temperature and drive current in skyrmion dynamics. *Nat. Electron.* **3**, 30–36 (2020).

11. K. Zeissler, S. Finizio, C. Barton, A. J. Huxtable, J. Massey, J. Raabe, A. V. Sadovnikov, S. A. Nikitov, R. Brearton, T. Hesjedal, G. van der Laan, M. C. Rosamond, E. H. Linfield, G. Burnell, C. H. Marrows, Diameter-independent skyrmion Hall angle observed in chiral magnetic multilayers. *Nat. Commun.* **11**, 428 (2020).

12. A. A. Thiele, Theory of the Static Stability of Cylindrical Domains in Uniaxial Platelets. *J. Appl. Phys.* **41**, 1139–1145 (1970).

13. A. Hrabec, J. Sampaio, M. Belmeguenai, I. Gross, R. Weil, S. M. Chérif, A. Stashkevich, V. Jacques, A. Thiaville, S. Rohart, Current-induced skyrmion generation and dynamics in symmetric bilayers. *Nat. Commun.* **8**, ncomms15765 (2017).

14. F. Büttner, I. Lemesh, G. S. D. Beach, Theory of isolated magnetic skyrmions: From fundamentals to room temperature applications. *Sci. Rep.* **8**, 4464 (2018).

15. J. Barker, O. A. Tretiakov, Static and Dynamical Properties of Antiferromagnetic Skyrmions in the Presence of Applied Current and Temperature. *Phys. Rev. Lett.* **116**, 147203 (2016).

16. C. Jin, C. Song, J. Wang, Q. Liu, Dynamics of antiferromagnetic skyrmion driven by the spin Hall effect. *Appl. Phys. Lett.* **109**, 182404 (2016).

17. R. Tomasello, V. Puliafito, E. Martinez, A. Manchon, M. Ricci, M. Carpentieri, G. Finocchio, Performance of synthetic antiferromagnetic racetrack memory: domain wall versus skyrmion. *J. Phys. Appl. Phys.* **50**, 325302 (2017).

18. X. Zhang, Y. Zhou, M. Ezawa, Antiferromagnetic Skyrmion: Stability, Creation and Manipulation. *Sci. Rep.* **6**, 24795 (2016).

19. P. E. Roy, Method to suppress antiferromagnetic skyrmion deformation in high speed racetrack devices. *J. Appl. Phys.* **129**, 193902 (2021).

20. X. Zhang, Y. Zhou, M. Ezawa, Magnetic bilayer-skyrmions without skyrmion Hall effect. *Nat. Commun.* **7**, 10293 (2016).

21. H. Velkov, O. Gomonay, M. Beens, G. Schwiete, A. Brataas, J. Sinova, R. A. Duine, Phenomenology of current-induced skyrmion motion in antiferromagnets. *New J. Phys.* **18**, 075016 (2016).

22. A. Salimath, F. Zhuo, R. Tomasello, G. Finocchio, A. Manchon, Controlling the deformation of antiferromagnetic skyrmions in the high-velocity regime. *Phys. Rev. B* **101**, 024429 (2020).

23. J. Xia, X. Zhang, K.-Y. Mak, M. Ezawa, O. A. Tretiakov, Y. Zhou, G. Zhao, X. Liu, Current-induced dynamics of skyrmion tubes in synthetic antiferromagnetic multilayers. *Phys. Rev. B* **103**, 174408 (2021).

24. S. Komineas, N. Papanicolaou, Traveling skyrmions in chiral antiferromagnets. *SciPost Phys.* **8**, 086 (2020).

25. S. Gao, H. D. Rosales, F. A. G. Albarracín, V. Tsurkan, G. Kaur, T. Fennell, P. Steffens, M. Boehm, P. Čermák, A. Schneidewind, E. Ressouche, D. C. Cabra, C. Rüegg, O. Zaharko, Fractional antiferromagnetic skyrmion lattice induced by anisotropic couplings. *Nature*, doi: 10.1038/s41586-020-2716-8 (2020).

26. H. Jani, J.-C. Lin, J. Chen, J. Harrison, F. Maccherozzi, J. Schad, S. Prakash, C.-B. Eom, A. Ariando, T. Venkatesan, P. G. Radaelli, Antiferromagnetic half-skyrmions and bimerons at room temperature. *Nature* **590**, 74–79 (2021).

27. K. G. Rana, R. Lopes Seeger, S. Ruiz-Gómez, R. Juge, Q. Zhang, K. Bairagi, V. T. Pham, M. Belmeguenai, S. Auffret, M. Foerster, L. Aballe, G. Gaudin, V. Baltz, O. Boulle, Imprint from ferromagnetic skyrmions in an antiferromagnet via exchange bias. *Appl. Phys. Lett.* **119**, 192407 (2021).

28. W. Legrand, D. Maccariello, F. Ajejas, S. Collin, A. Vecchiola, K. Bouzehouane, N. Reyren, V. Cros, A. Fert, Room-temperature stabilization of antiferromagnetic skyrmions in synthetic antiferromagnets. *Nat. Mater.* **19**, 34–42 (2020).

29. R. Juge, N. Sisodia, J. U. Larrañaga, Q. Zhang, V. T. Pham, K. G. Rana, B. Sarpi, N. Mille, S. Stanescu, R. Belkhou, M.-A. Mawass, N. Novakovic-Marinkovic, F. Kronast, M. Weigand, J. Gräfe, S. Wintz, S. Finizio, J. Raabe, L. Aballe, M. Foerster, M. Belmeguenai, L. D. Buda-Prejbeanu, J. Pelloux-Prayer, J. M. Shaw, H. T. Nembach, L. Ranno, G. Gaudin, O. Boulle, Skyrmions in synthetic antiferromagnets and their nucleation via electrical current and ultra-fast laser illumination. *Nat. Commun.* **13**, 4807 (2022).

30. Y. Hirata, D.-H. Kim, S. K. Kim, D.-K. Lee, S.-H. Oh, D.-Y. Kim, T. Nishimura, T. Okuno, Y. Futakawa, H. Yoshikawa, A. Tsukamoto, Y. Tserkovnyak, Y. Shiota, T. Moriyama, S.-B. Choe, K.-J. Lee, T. Ono, Vanishing skyrmion Hall effect at the angular momentum compensation temperature of a ferrimagnet. *Nat. Nanotechnol.*, 1 (2019).

31. S. Woo, K. M. Song, X. Zhang, Y. Zhou, M. Ezawa, X. Liu, S. Finizio, J. Raabe, N. J. Lee, S.-I. Kim, S.-Y. Park, Y. Kim, J.-Y. Kim, D. Lee, O. Lee, J. W. Choi, B.-C. Min, H. C. Koo, J. Chang, Current-driven dynamics and inhibition of the skyrmion Hall effect of ferrimagnetic skyrmions in GdFeCo films. *Nat. Commun.* **9**, 959 (2018).

32. T. Dohi, S. DuttaGupta, S. Fukami, H. Ohno, Formation and current-induced motion of synthetic antiferromagnetic skyrmion bubbles. *Nat. Commun.* **10**, 1–6 (2019).

33. Y. Quessab, J.-W. Xu, E. Cogulu, S. Finizio, J. Raabe, A. D. Kent, Zero-Field Nucleation and Fast Motion of Skyrmions Induced by Nanosecond Current Pulses in a Ferrimagnetic Thin Film. *Nano Lett.* **22**, 6091 (2022).

34. S. S. P. Parkin, N. More, K. P. Roche, Oscillations in exchange coupling and magnetoresistance in metallic superlattice structures: Co/Ru, Co/Cr, and Fe/Cr. *Phys. Rev. Lett.* **64**, 2304–2307 (1990).

35. J. U. Larrañaga, N. Sisodia, V. T. Pham, I. Di Manici, A. Masseboeuf, K. Garello, F. Disdier, B. Fernandez, S. Wintz, M. Weigand, M. Belmeguenai, S. Pizzini, R. Sousa, L. Buda-Prejbeanu, G. Gaudin, O. Boulle, Electrical detection and nucleation of a magnetic skyrmion in a magnetic tunnel junction observed via operando magnetic microscopy. arXiv arXiv:2308.00445 [Preprint] (2023). https://doi.org/10.48550/arXiv.2308.00445.

36. See supplementary materials.

37. I. Gross, W. Akhtar, A. Hrabec, J. Sampaio, L. J. Martínez, S. Chouaieb, B. J. Shields, P. Maletinsky, A. Thiaville, S. Rohart, V. Jacques, Skyrmion morphology in ultrathin magnetic films. *Phys. Rev. Mater.* **2**, 024406 (2018).

38. R. Juge, S.-G. Je, D. de Souza Chaves, S. Pizzini, L. D. Buda-Prejbeanu, L. Aballe, M. Foerster, A. Locatelli, T. O. Menteş, A. Sala, F. Maccherozzi, S. S. Dhesi, S. Auffret, E.

Gautier, G. Gaudin, J. Vogel, O. Boulle, Magnetic skyrmions in confined geometries: Effect of the magnetic field and the disorder. *J. Magn. Magn. Mater.* **455**, 3–8 (2018).

39. L. Ranno, M. A. Moro, Design Rules for DMI-Stabilised Skyrmions. *ArXiv210700767 Cond-Mat* (2021).

40. J.-V. Kim, M.-W. Yoo, Current-driven skyrmion dynamics in disordered films. *Appl. Phys. Lett.* **110**, 132404 (2017).

41. C. Reichhardt, C. J. O. Reichhardt, M. V. Milosevic, Statics and Dynamics of Skyrmions Interacting with Pinning: A Review. *Rev. Mod. Phys.* **94**, 035005 (2022).

42. A. A. Thiele, Steady-State Motion of Magnetic Domains. *Phys. Rev. Lett.* **30**, 230 (1973).

43. J. Masell, D. R. Rodrigues, B. F. McKeever, K. Everschor-Sitte, Spin-transfer torque driven motion, deformation, and instabilities of magnetic skyrmions at high currents. *Phys. Rev. B* **101**, 214428 (2020).

44. H. Vakili, J.-W. Xu, W. Zhou, M. N. Sakib, M. G. Morshed, T. Hartnett, Y. Quessab, K. Litzius, C. T. Ma, S. Ganguly, M. R. Stan, P. V. Balachandran, G. S. D. Beach, S. J. Poon, A. D. Kent, A. W. Ghosh, Skyrmionics—Computing and memory technologies based on topological excitations in magnets. *J. Appl. Phys.* **130**, 070908 (2021).

45. R. E. Troncoso, Á. S. Núñez, Brownian motion of massive skyrmions in magnetic thin films. *Ann. Phys.* **351**, 850–856 (2014).

46. K. Yu. Guslienko, K.-S. Lee, S.-K. Kim, Dynamic Origin of Vortex Core Switching in Soft Magnetic Nanodots. *Phys. Rev. Lett.* **100**, 027203 (2008).

47. Z. Chen, X. Zhang, Y. Zhou, Q. Shao, Skyrmion Dynamics in the Presence of Deformation. *Phys. Rev. Appl.* **17**, L011002 (2022).

48. E. A. Tremsina, G. S. D. Beach, Atomistic simulations of distortion-limited high-speed dynamics of antiferromagnetic skyrmions. *Phys. Rev. B* **106**, L220402 (2022).

49. A. Thiaville, S. Rohart, É. Jué, V. Cros, A. Fert, Dynamics of Dzyaloshinskii domain walls in ultrathin magnetic films. *EPL Europhys. Lett.* **100**, 57002 (2012).

50. B. Jinnai, C. Zhang, A. Kurenkov, M. Bersweiler, H. Sato, S. Fukami, H. Ohno, Spin-orbit torque induced magnetization switching in Co/Pt multilayers. *Appl. Phys. Lett.* **111**, 102402 (2017).

51. K.-F. Huang, D.-S. Wang, H.-H. Lin, C.-H. Lai, Engineering spin-orbit torque in Co/Pt multilayers with perpendicular magnetic anisotropy. *Appl. Phys. Lett.* **107**, 232407 (2015).

52. J. Kim, Y. Otani, Orbital angular momentum for spintronics. *J. Magn. Magn. Mater.* **563**, 169974 (2022).

53. R. Xu, H. Zhang, Y. Jiang, H. Cheng, Y. Xie, Y. Yao, D. Xiong, Z. Zhu, X. Ning, R. Chen, Y. Huang, S. Xu, J. Cai, Y. Xu, T. Liu, W. Zhao, Giant orbit-to-charge conversion induced via the inverse orbital Hall effect. arXiv arXiv:2308.13144 [Preprint] (2023). https://doi.org/10.48550/arXiv.2308.13144.

54. S. Panigrahy, S. Mallick, J. Sampaio, S. Rohart, Skyrmion inertia in synthetic antiferromagnets. *Phys. Rev. B* **106**, 144405 (2022).

55. O. Boulle, Data of "Fast current induced skyrmion motion in synthetic antiferromagnets." doi: 10.17605/OSF.IO/8QHWR (2024).

56. S. Bandiera, R. C. Sousa, S. Auffret, B. Rodmacq, B. Dieny, Enhancement of perpendicular magnetic anisotropy thanks to Pt insertions in synthetic antiferromagnets. *Appl. Phys. Lett.* **101**, 072410 (2012).

57. P. J. Metaxas, J. P. Jamet, A. Mougin, M. Cormier, J. Ferré, V. Baltz, B. Rodmacq, B. Dieny, R. L. Stamps, Creep and Flow Regimes of Magnetic Domain-Wall Motion in Ultrathin $\mathrm{Pt}/\mathrm{Co}/\mathrm{Pt}$ Films with Perpendicular Anisotropy. *Phys. Rev. Lett.* **99**, 217208 (2007).

58. S.-G. Je, D.-H. Kim, S.-C. Yoo, B.-C. Min, K.-J. Lee, S.-B. Choe, Asymmetric Magnetic Domain-Wall Motion by the Dzyaloshinskii-Moriya Interaction. *Phys. Rev. B* **88** (2013).

59. T. H. Pham, J. Vogel, J. Sampaio, M. Vaňatka, J.-C. Rojas-Sánchez, M. Bonfim, D. S. Chaves, F. Choueikani, P. Ohresser, E. Otero, A. Thiaville, S. Pizzini, Very large domain wall velocities in Pt/Co/GdOx and Pt/Co/Gd trilayers with Dzyaloshinskii-Moriya interaction. *EPL* **113**, 67001 (2016).

60. C. O. Avci, K. Garello, C. Nistor, S. Godey, B. Ballesteros, A. Mugarza, A. Barla, M. Valvidares, E. Pellegrin, A. Ghosh, I. M. Miron, O. Boulle, S. Auffret, G. Gaudin, P. Gambardella, Fieldlike and antidamping spin-orbit torques in as-grown and annealed Ta/CoFeB/MgO layers. *Phys. Rev. B* **89**, 214419 (2014).

61. H.-B. Braun, Fluctuations and instabilities of ferromagnetic domain-wall pairs in an external magnetic field. *Phys. Rev. B* **50**, 16485–16500 (1994).

62. N. Romming, A. Kubetzka, C. Hanneken, K. von Bergmann, R. Wiesendanger, Field-Dependent Size and Shape of Single Magnetic Skyrmions. *Phys. Rev. Lett.* **114**, 177203 (2015).

63. U. Ritzmann, S. von Malottki, J.-V. Kim, S. Heinze, J. Sinova, B. Dupé, Trochoidal motion and pair generation in skyrmion and antiskyrmion dynamics under spin–orbit torques. *Nat. Electron.* **1**, 451–457 (2018).

64. A. Vansteenkiste, J. Leliaert, M. Dvornik, M. Helsen, F. Garcia-Sanchez, B. V. Waeyenberge, The design and verification of MuMax3. *AIP Adv.* **4**, 107133 (2014).

65. L. Belliard, A. Thiaville, S. Lemerle, A. Lagrange, J. Ferré, J. Miltat, Investigation of the domain contrast in magnetic force microscopy. *J. Appl. Phys.* **81**, 3849–3851 (1997).

66. D. W. Abraham, F. A. McDonald, Theory of magnetic force microscope images. *Appl. Phys. Lett.* **56**, 1181–1183 (1990).

67. A. Hubert, W. Rave, S. L. Tomlinson, Imaging Magnetic Charges with Magnetic Force Microscopy. *Phys. Status Solidi B* **204**, 817–828 (1997).

68. O. Hellwig, A. Berger, J. B. Kortright, E. E. Fullerton, Domain structure and magnetization reversal of antiferromagnetically coupled perpendicular anisotropy films. *J. Magn. Magn. Mater.* **319**, 13–55 (2007).

69. S. Hamada, K. Himi, T. Okuno, K. Takanashi, MFM observation of perpendicular magnetization and antiferromagnetically coupled domains in Co/Ru superlattices. *J. Magn. Magn. Mater.* **240**, 539–542 (2002).

70. O. Hellwig, A. Berger, E. E. Fullerton, Domain Walls in Antiferromagnetically Coupled Multilayer Films. *Phys. Rev. Lett.* **91**, 197203 (2003).

71. A. Baruth, L. Yuan, J. D. Burton, K. Janicka, E. Y. Tsymbal, S. H. Liou, S. Adenwalla, Domain overlap in antiferromagnetically coupled [Co∕Pt]∕NiO∕[Co∕Pt] multilayers. *Appl. Phys. Lett.* **89**, 202505 (2006).

72. L. Berges, Size-dependent mobility of skyrmions beyond pinning in ferrimagnetic GdCo thin films. *Phys. Rev. B* **106** (2022).

73. C. A. Akosa, O. A. Tretiakov, G. Tatara, A. Manchon, Theory of the Topological Spin Hall Effect in Antiferromagnetic Skyrmions: Impact on Current-Induced Motion. *Phys. Rev. Lett.* **121**, 097204 (2018).

74. P. M. Buhl, F. Freimuth, S. Blügel, Y. Mokrousov, Topological spin Hall effect in antiferromagnetic skyrmions. *Phys. Status Solidi RRL – Rapid Res. Lett.* **11**, 1700007 (2017).

75. B. Göbel, A. Mook, J. Henk, I. Mertig, Antiferromagnetic skyrmion crystals: Generation, topological Hall, and topological spin Hall effect. *Phys. Rev. B* **96**, 060406 (2017).